\documentclass[fleqn,usenatbib]{mnras}
\usepackage{newtxtext,newtxmath}
\usepackage[T1]{fontenc}
\pdfoutput=1
\DeclareRobustCommand{\VAN}[3]{#2}
\let\VANthebibliography\thebibliography
\def\thebibliography{\DeclareRobustCommand{\VAN}[3]{##3}\VANthebibliography}

\usepackage{graphicx}	% Including figure files
\usepackage{amsmath}	% Advanced maths commands
\usepackage{color}
\title{Cluster Assembly Times as a Cosmological Test}
\author[Y. Amoura et al.]{Yuba Amoura,$^{1,2}$\thanks{E-mail: ayuba@uwaterloo.ca} 
Nicole E.~Drakos,$^{3}$
Anael Berrouet$^{2}$, 
James E.~Taylor$^{1,2}$\thanks{E-mail: taylor@uwaterloo.ca}
\\
$^{1}$Waterloo Centre for Astrophysics, University of Waterloo, Waterloo, Ontario N2L 3G1 Canada\\
$^{2}$Department of Physics and Astronomy, University of Waterloo, 200 University Avenue West, Waterloo, Ontario N2L 3G1, Canada\\
$^{3}$Department of Astronomy and Astrophysics, University of California, Santa Cruz, 1156 High Street, Santa Cruz, CA 95064 USA
}

\date{October 2020}

\begin{document}
\label{firstpage}
\pagerange{\pageref{firstpage}--\pageref{lastpage}}
\maketitle
\begin{abstract}
The abundance of galaxy clusters in the low-redshift universe provides an important cosmological test, constraining a product of the initial amplitude of fluctuations and the amount by which they have grown since early times. The degeneracy of the test with respect to these two factors remains a limitation of abundance studies. Clusters will have different mean assembly times, however, depending on the relative importance of initial fluctuation amplitude and subsequent growth. Thus, structural probes of cluster age such as concentration, shape or substructure may provide a new cosmological test that breaks the main degeneracy in number counts. We review analytic predictions for how mean assembly time should depend on cosmological parameters, and test these predictions using cosmological simulations. Given the overall sensitivity expected, we estimate the cosmological parameter constraints that could be derived from the cluster catalogues of forthcoming surveys such as {\it Euclid}, the {\it Nancy Grace Roman Space Telescope}, {\it eROSITA}, or CMB-S4. We show that by considering the structural properties of their cluster samples, such surveys could easily achieve errors of $\Delta \sigma_8 = 0.01$ or better.
\end{abstract}


%%%%%%%%%%%%%%%%%%%%%%%%%%%%%%%%%%%%%%%%%%%%%%%%%%%%%%%%%%%%%%%%%%%%%%

% Section 1
\section{Introduction}

The `concordance' $\Lambda$-Cold Dark Matter ($\Lambda$CDM) cosmological model is now well established as a single theoretical framework that is consistent with many different observational tests. The present-day abundance of dark matter and dark energy and the statistical properties of the matter distribution are increasingly well constrained, as expressed by cosmological parameters with gradually decreasing uncertainties \citep[e.g.][]{Planck.2020}. As the uncertainties in parameter values shrink; however, they reveal tension in several places in the model. Most notably, the Hubble parameter $H_0$ appears to differ significantly between high-redshift and low-redshift tests, with the tension in independent measurements of this parameter now exceeding $4\sigma$ (\citealt{Riess2019}; see \citealt{Verde.2019} for a review). In addition to the $H_0$ tension, there is also growing evidence that the amplitude of the matter fluctuations (typically expressed as $\sigma_8$, the r.m.s.~of fluctuations in the matter density on a scale of $8 h^{-1}$Mpc) may display a similar tension at the 2--3$\sigma$ level \cite[or $\sim0.1$ in this parameter, e.g.][]{Battye.2015, Douspis.2019, To.2020, Heymans2020}. More generally, the fundamental natures of dark energy and dark matter remain unknown, raising the possibility of new, exotic physics not yet included in the standard cosmological model. 

Given the evidence for tension in the current results, multiple, independent tests of the standard cosmological model are needed, on different mass and length scales and at different redshifts, to either reconcile the current results, or to reveal the physical origin of the disagreements. Current and forthcoming space missions, including \emph{Euclid}, the \emph{Nancy Grace Roman Space Telescope} (\emph{Roman}), and \emph{eROSITA} \citep{Pillepich2012}, together with data from large ground-based surveys such as UNIONS \citep{UNIONS2020}, DESI \citep{DESI2016} or Rubin LSST \citep{LSST2009}, or experiments such as CMB-S4 \citep{CMBS4}, will provide remarkable new datasets, mapping out structure over a significant fraction of the observable universe, out to redshifts of a few. Given, on the one  hand, the enormous potential of this data, and, on the other hand, the exacting precision required to resolve current parameter tensions, there is a need for new tests of the cosmological model that make full use of our growing understanding of structure formation. 

The measured abundance of massive galaxy clusters is a classic cosmological test. Cluster abundance has been estimated using samples detected in the X-ray \citep[e.g.][]{Henry.2009,Mantz.2010, Bohringer.2014}, via the Sunyaev-Zel'dovich (SZ) effect \citep[e.g.][]{deHaan.2016,Planck.2020}, in optical galaxy redshift surveys \citep[e.g.][]{Abdullah2020}, in weak lensing surveys \citep{Kacprzak2016}, or using combinations of these techniques \citep[e.g.][]{Abbott2020,Costanzi2020}. 
These rare objects evolve from peaks in the matter distribution present at early times, and their present-day abundance places a tight constraint on the function $S=\sigma_8\Omega_{\rm m}^{\gamma}$, where $\Omega_{\rm m}$ is the present-day matter density parameter, and $\gamma\approx 0.5$ is the growth index. Cluster abundance measurements alone do not place very tight constraints on $\Omega_{\rm m}$ or $\sigma_8$ individually, due to the degeneracy between them. Simply counting clusters does not leverage the full potential of the underlying datasets, however. The structural properties of clusters --- their projected shape, central concentration, substructure and non-axisymmetry --- are all related to their degree of dynamical relaxation, which in turn traces their formation history \citep[see][for a review]{Taylor.2011}. Thus, measurements of these properties provide a separate constraint on the growth rate. While measurements of structural parameters in individual clusters may be noisy, the sheer number of systems expected in forthcoming surveys should allow us to make robust measurements of the average trends, using the expertise developed in fields like weak lensing.

The idea of using the structural properties of clusters to constrain cosmological parameters is not new \cite[e.g.][]{Richstone1992,Evrard1993,Mohr1995}, but the context for these tests has changed radically in the 30 years since the idea was first proposed. First, the size of the datasets has grown enormously, giving better statistics. Second, our understanding of the systematics in individual structural measurements has developed considerably. Third, there is increasing sophistication in understanding and exploiting large, complex datasets. In particular, fields such as cosmic shear have illustrated how it is possible to extract parameter constraints from large sets of noisy measurements, even when the link between parameters and observables is indirect and non-linear. Finally, simulations of structure formation have progressed dramatically, allowing us to calibrate some aspects of non-linear structure formation at the per cent level, even if other aspects remain uncertain. Thus, it seems high time to reconsider cosmological tests based on the internal structure of haloes.

In this paper, we consider the possibility of estimating from their structural properties the mean assembly time or formation epoch for a large sample of galaxy clusters. This measurement of mean `age' would leverage the same data already collected for cluster abundance studies, providing an independent constraint on the cosmological parameters. We focus in particular on the parametric dependence of halo age, and its sensitivity to the parameters $\Omega_{\rm m}$ and $\sigma_8$; in a subsequent paper, we will consider the (non-trivial) path to developing practical observational tests based on age estimates. The outline of the paper is as follows. In Section~\ref{sec:analytic}, we review theoretical models of cluster abundance and age, and use them to predict how these properties vary as a function of the cosmological parameters. Given the approximate nature of the theoretical estimates, in Section~\ref{sec:simulations} we test these predictions using catalogues from several different N-body simulations. We show that with some careful analysis, we can reconcile the analytic and numerical results to reasonable accuracy. In Section~\ref{sec:observations}, we estimate the sensitivity a realistic observational program could achieve, using concentration as a proxy for age. Finally, in Section~\ref{sec:conc}, we review and summarize our results. The details of the analytic calculations and the dependence of several important quantities on the cosmological parameters are discussed in the appendices.
We consider a range of cosmologies throughout the paper, but assume flatness ($\Omega_{\rm m} + \Omega_\Lambda = 1$), and take a model with $\Omega_{\rm m} = 0.3$ as the fiducial case.

%%%%%%%%%%%%%%%%%%%%%%%%%%%%%%%%%%%%%%%%%%%%%%%%%%%%%%%%%%%%%%%%%%%

%Section 2 
\section{Cosmological Sensitivity of Halo Abundance and Halo Age}
\label{sec:analytic}

We will begin by estimating the potential sensitivity of age tests, using theoretical predictions of how cluster abundance and age depend on the cosmological parameters. We use analytic models of abundance and age based on the extended Press \& Schechter formalism, and calculated using standard tools and techniques summarized in Appendix \ref{sec:Appendix_A}.

\subsection{Analytic Models of the Halo Mass Function}
\label{subsec:2.1}

The Press--Schechter \citep[PS --][]{Press.Schechter,Bond.1991}
and extended Press--Schechter \citep[EPS -- ][]{lacey.cole} formalisms provide a convenient analytic framework for computing the number density of dark matter haloes and their growth rate, given a background cosmological model. The basic expression for the halo mass function, derived assuming spherical collapse, is 
\begin{equation}
    n(M)dM = \sqrt{\frac{2}{\pi}}\frac{\rho_0}{M}\frac{\delta_c}{\sigma^2}\exp\left(-\frac{\delta_c^2}{2\sigma^2}\right)\left|\frac{d\sigma}{dM}\right|dM\,,
    \label{eq:1}
\end{equation}
where $\rho_0$ is the matter density at the redshift of interest, $\sigma = \sigma(M)$ is the r.m.s.~amplitude of fluctuations in the density field smoothed on mass scale $M$, and $\delta_c$ is the threshold for collapse to a virialized halo. Although fluctuations grow in amplitude as $z$ decreases to zero, the condition for collapse by redshift $z$ can also be considered at some fixed, early redshift, taking $\sigma(M)$ to be a function of mass only, and $\delta_c = \delta_c(z)$ to be a function of the collapse redshift. 

The mass function can also be written in a more compact form as
\begin{equation}
    n(M,t)dM = \frac{\rho_0}{M}f_{PS}(\nu)\left|\frac{d\nu}{dM}\right|dM\,,
    \label{eq:2}
\end{equation}
 where $\nu(M, z) \equiv \delta_c(z)/\sigma(M)$ is the height of the collapse threshold at redshift $z$, relative to typical fluctuations on mass scale $M$, and
\begin{equation}
    f_{\rm PS}(\nu) = \sqrt{\frac{2}{\pi}}\exp\left(-\frac{\nu^2}{2}\right)
    \label{eq:fps}
\end{equation}
is the mass fraction that has collapsed per unit interval of $\nu$\footnote{Note the mass fraction is often defined per unit $\ln\nu$; the expressions for $f_{\rm PS}$ and $f_{\rm ST}$ (Eqs.~\ref{eq:fps} \&\ \ref{eq:fst}) then contain an extra factor of $\nu$.}. 

It is well known, however, that this basic form of the mass function fails to reproduce the halo abundance found in N-body simulations, particularly for low-mass haloes \citep{Sheth.Tormen.1999, Sheth.Tormen.2002, Jenkins.2001}. This failure is due to several simplifying assumptions made in the model, the most important one being a fixed threshold for (spherical) collapse $\delta_c$ that is independent of mass and environment. To solve this problem, \citet[][ST hereafter]{Sheth.Tormen.1999} considered a mass-dependent collapse threshold (or `moving barrier'), to derive a functional form that better fits the mass function from simulations 
\begin{equation}
    f_{ST}(\nu) = A\sqrt{\frac{2a}{\pi}}\left(1 + \frac{1}{\left(\sqrt{a}\nu\right)^{2p}}\right) \exp\left(-\frac{a\nu^2}{2}\right)\,,
    \label{eq:fst}
\end{equation}
with $A=0.322$, $a=0.707$ and $p=0.3$. 

A number of subsequent studies have improved our understanding of the mass function. \cite{Tinker.2008} demonstrated that the HMF is not completely universal, but evolves with redshift; allowing the parameters $A$, $a$, and $p$ in the fit to evolve as a power-law of $1+z$ provides a better match to simulations. This non-universality has since been confirmed by other groups \citep[e.g.][]{Watson.2013}. \cite{Despali.2016}, argued that it is in fact, an artifact of the halo mass definition, and that the common choices of overdensity of 200 or 178 times the critical density induce much of the non-universality. Finally, several authors \citep{Velliscig.2014, Bocquet.2016, Castro2021} have studied the impact of baryonic effects on the HMF by measuring halo masses, profiles, and abundance in hydrodynamic simulations. These improvements to the HMF fit are required in precision applications, but are generally secondary in importance ($\le 20$\% -- \citealt{Tinker.2008, Velliscig.2014, Bocquet2020}), relative to the large variations in abundance with cosmology shown below. Thus, for simplicity in what follows, we will assume the ST form of the collapsed fraction (Eq.~\ref{eq:fst}), in order to study how abundance and age depend on cosmology. We discuss the possible effect of baryons on the {\it internal} structure of haloes in Section \ref{subsec:4.4} below.

\subsection{Cosmological Dependence of Halo Abundance}
\label{subsec:2.2}

Cluster abundance depends on the cosmological parameters both through the cluster mass function and through the survey volume. Within a survey volume subtending a solid angle $\Delta\Omega$, the expected number of clusters in the mass bin $i$: $[M^i, M^{i + 1}]$ and redshift bin $j$: $[z_j, z_{j+1}]$ is 
\begin{equation}
    N(M_i, z_j) = \frac{\Delta \Omega}{4\pi} \int_{z_j}^{z_{j+1}} dz\frac{dV}{dz}\int_{M_i}^{M_{i+1}}
    \frac{dn}{dM}\,dM \,,
\end{equation}
where $dn/dM$ is the HMF given above and $dV/dz$ is the volume element per unit solid angle and per unit redshift. As discussed previously, the HMF is calculated as a fraction of the material in a region that has collapsed to form haloes on some mass scale. Thus, rather than relating the halo abundance to the volume probed by the survey, we can
express it in terms of the total mass $M_V$ of material in the survey volume:
\begin{equation}
    N(M_i, z_j) = \frac{\Delta \Omega}{4\pi} \int_{z_j}^{z_{j+1}}dz \frac{dM_V}{dz}\int_{M_i}^{M_{i+1}} f(\nu) \frac{d\nu }{dM}dM\,.
\end{equation}
The advantage of this form is that we can now separate the cosmological dependence of the first factor, the total mass $M_V$ probed by the survey in a given redshift range $\Delta z$, from that of the second factor, which is the fraction $f(\nu)\Delta\nu$ of that mass that has collapsed to form haloes in the mass range $\Delta\nu = (d\nu/dM)\Delta M$ by that redshift.  

To make explicit the cosmological dependence of the HMF, in Appendices \ref{sec:Appendix_B} and \ref{sec:Appendix_C} we consider each of these factors separately. Over a realistic range of ($\Omega_{\rm m}$, $\sigma_8$), the survey mass $M_V$ varies by a factor of $\lesssim 2$,  while the collapsed fraction can vary by several orders of magnitude, and thus dominates the parametric dependence of the HMF.

\begin{figure*}
    \centering
    \includegraphics[width=0.9\textwidth]{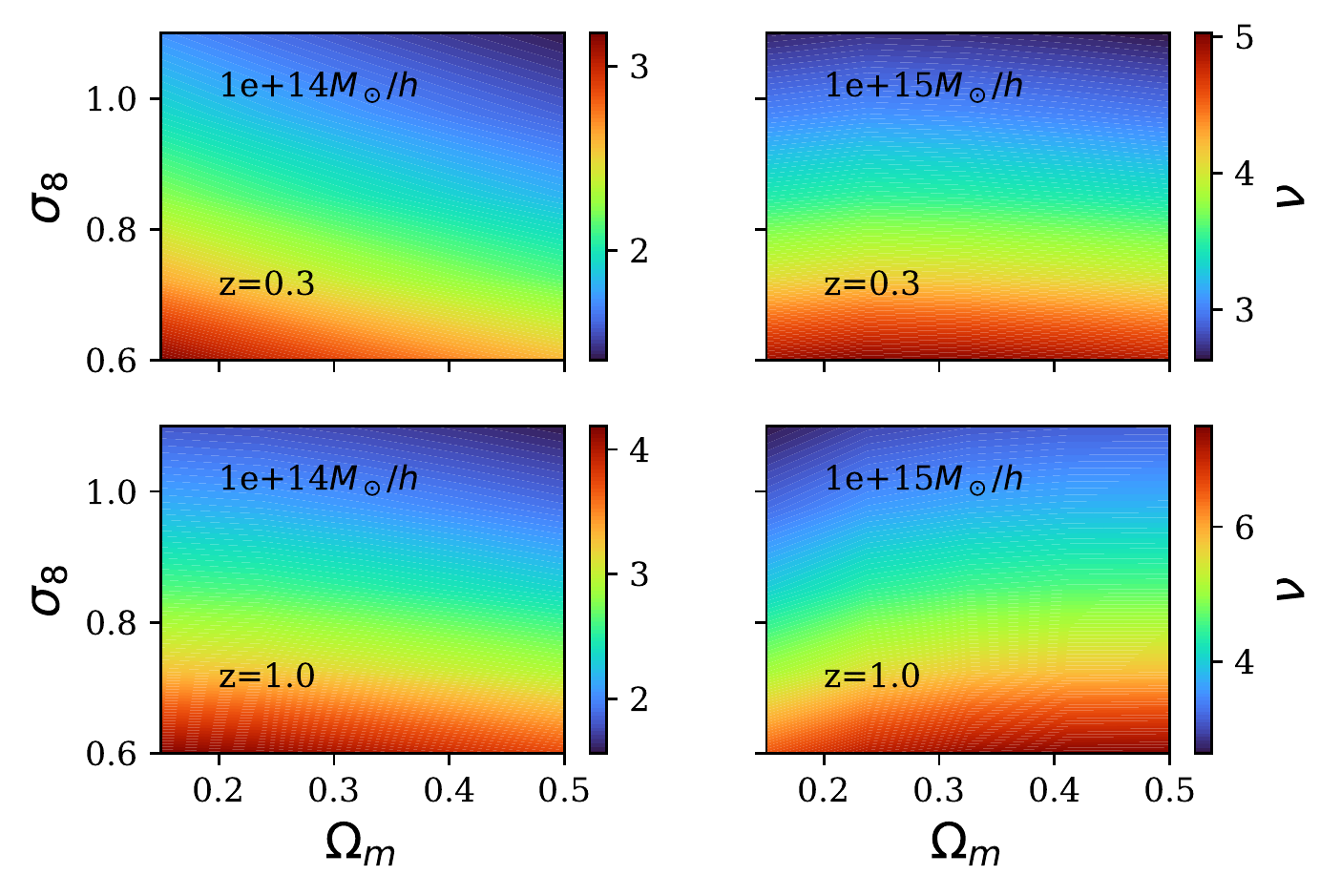}

    \caption{Variation of the peak height $\nu$, for a halo of the mass and redshift indicated, in the $\Omega_{\rm m}$--$\sigma_8$ plane.}
    \label{fig:cosmo_peak_height_om_s8_plane}
\end{figure*}

As demonstrated in Appendix \ref{sec:Appendix_C}, the peak height $\nu$ varies approximately as 
$\nu(M, z) \propto (\sigma_8)^{-1}\Omega_{\rm m}^{\alpha(z)}\Omega_{\rm m}^{-\beta (M)}\,$. The resulting behaviour in the $\Omega_{\rm m}$--$\sigma_8$ plane is shown in Fig.~\ref{fig:cosmo_peak_height_om_s8_plane}, for several mass/redshift combinations. We see that peak height depends mainly on $\sigma_8$; variations in $\Omega_{\rm m}$ introduce a slight tilt in the contours, that goes from negative at low mass/redshift, where $\beta(M) > \alpha(z)$, to slightly positive at high mass/redshift (and low $\Omega_{\rm m}$), where $\alpha(z) > \beta(M)$ (see also Appendix \ref{sec:Appendix_C}, and the lower right panel of Fig.~\ref{fig:g_ofz_gamma}, which shows the dependence on $\Omega_{\rm m}$ for fixed $\sigma_8$, at several different masses and redshifts).

% Fig 1

In both spherical collapse (PS -- Eq. \ref{eq:fps}) and ellipsoidal collapse (ST -- Eq. \ref{eq:fst}) models, the collapsed fraction exhibits power-law growth for $\nu<1$, and 
an exponential decay for $\nu>1$. Thus, there are two main regimes, the first where the abundance of haloes increases with $\nu$, and the second where it decreases rapidly. The effect of the cosmological parameters on the two regimes is shown, for instance, in Fig.~\ref{fig:dimensionless_hmf}. The amplitude of fluctuations $\sigma_8$ controls the mass at which the transition between regimes occurs, while $\Omega_{\rm m}$ controls the sharpness of the transition. In the case of clusters, we are generally in the second regime, where increases in $\nu$ produce an exponential decrease in abundance. Thus, the parametric dependence of the collapsed fraction (shown in the right panel of Fig.~\ref{fig:mass_omega_m}) is very similar to the corresponding figure for peak height, Fig.~\ref{fig:cosmo_peak_height_om_s8_plane}, but with an inverted and logarithmic colour scale, since $f(\nu) \propto \exp (-\nu^2/2)$ implies that in a log scale, $\ln(f) \propto -\nu^2/2$. 

\begin{figure*}
    \centering
\includegraphics[width=0.9\textwidth]{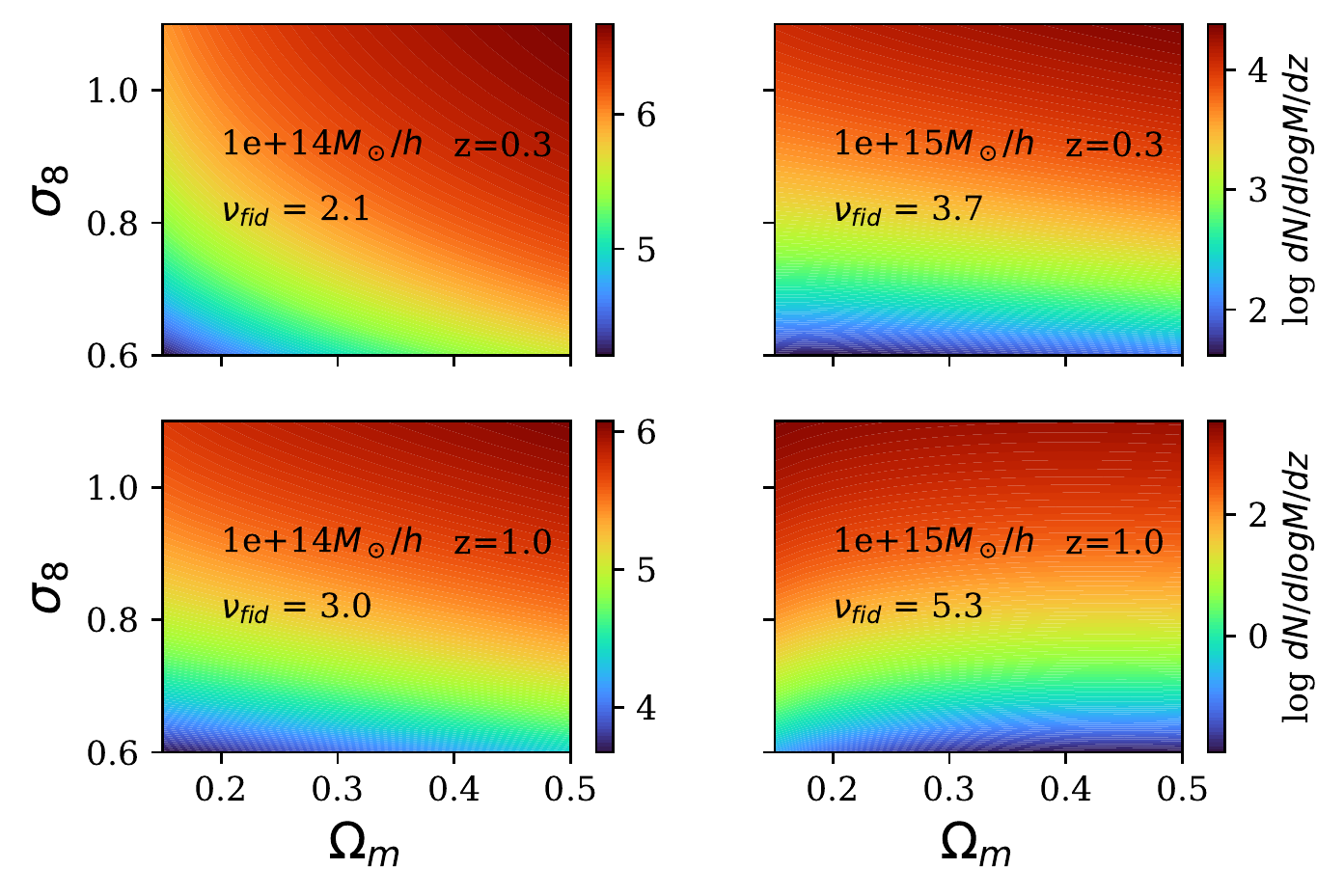}
  \caption{The cosmological dependence of the differential cluster number counts, per unit log mass and per unit redshift. The colour scale is logarithmic. The value of the peak height at different masses and redshifts for our fiducial cosmology ($\Omega_m=0.3$, $\sigma_8=0.8$) is shown to illustrate the dependence of the contour shapes on peak height.} %Note the similarity to Fig.~\ref{fig:cosmo_fnu_om_s8_plane}}
     \label{fig:om_s8_diff_counts}
\end{figure*}

Finally, we can combine the parametric dependence of the survey volume and the collapsed fraction (shown in the left and right panels of Fig.~\ref{fig:mass_omega_m} respectively) to plot the dependence of number counts on $\Omega_{\rm m}$ and $\sigma_8$. This is shown in Fig.~\ref{fig:om_s8_diff_counts}, for the same mass/redshift combinations considered previously. We note that the variation of the total mass within the survey volume has a minimal effect, and aside from the change in the overall scale, the contours are almost identical to those for the collapsed fraction.

% Fig 2

\subsection{Theoretical Estimates of Halo Assembly Time}
\label{subsec:2.3}

In CDM cosmologies, dark matter haloes grow through repeated, stochastic mergers, gradually assembling their mass from a large number of smaller progenitors. Thus, deciding when a given halo has `formed' is rather arbitrary. Most definitions in the literature are based on the Mass Accretion History (MAH) \citep{VandenBosch.2002}. This is calculated by tracing the growth of the halo backwards in time and selecting the largest progenitor of each merger, to produce a single monotonic growth sequence $M(z)$; the MAH is then defined as the relative value $M(z)/M(0)$. Given a MAH, the formation epoch is often defined as the redshift by which a halo has reached some fixed fraction $f$ of its final mass (e.g.~$z_{50}$ for $f=0.5$ -- \citealt{lacey.cole}).

As for the HMF, the EPS formalism provides an analytic framework for exploring the cosmological dependence of halo age. We will first consider the predicted $z_{50}$ distribution derived by \cite{lacey.cole} assuming spherical collapse, and then give two different models of the ellipsoidal collapse equivalent, derived by \cite{Sheth.Tormen.2002} and \cite{Zhang.2008} respectively.

Given a halo of mass $M_0$ at redshift $z_0$, the fraction of its mass that was in progenitor haloes of mass $M_1 \pm dM_1/2$ at redshift $z_1$, is given by the conditional probability
\begin{equation}
\begin{gathered}
\label{eq:fsc}
    f_{SC}(M_1, z_1 |M_0, z_0)dM_1 = \frac{1}{\sqrt{2\pi}}\frac{\delta_c(z_1) - \delta_c(z_0)}{(S(M_1)-S(M_0))^{3/2}} \\ \times \exp{\left(-\frac{\left(\delta_c(z_1) - \delta_c(z_0)\right)^2}{2(S(M_1)-S(M_0))}\right)}dS_1\,,
\end{gathered}
\end{equation}
where $S(M) = \sigma^2(M)$, and the other variables are as in Section \ref{subsec:2.1} 

Multiplying by the factor $M_0/M_1$, we get the progenitor mass function (PMF), that is the number of progenitors of mass $M_1$
\begin{equation}
\label{eq:PMF}
    PMF(M_1, z_1|M_0, z_o)dM_1 = \frac{M_0}{M_1}f(M_1, z |M_0, z_0)dM_1\,.
\end{equation}

Following \cite{lacey.cole}, if we integrate the PMF from masses $M_0/2$ to $M_0$, we are calculating the average number of progenitors at redshift $z_1$ that have more than half the final halo mass at $z_0$. Since the halo cannot have more than one progenitor with more than half of its final mass, this quantity is also the probability that the halo had built up at least half of its mass in a single progenitor by redshift $z_1$. Thus, it gives the cumulative distribution of the formation redshift $z_{50}$:
\begin{equation}
\label{eq:z50}
    P(z_{50}>z|M_0, z_0) \equiv \int_{M_0/2}^{M_0}\frac{M_0}{M}f(M, z |M_0, z_0)dM\,.
\end{equation}
(We note that this approach only works for formation redshifts $z_f$ with $f \ge 0.5$; there is no simple analytic way to obtain the distribution of $z_f$ for $f<0.5$.) 

As with the HMF, this estimate of halo formation redshift is limited by the assumption of spherical collapse. \cite{Sheth.Tormen.2002} provided a version of the conditional mass function using a Taylor expansion of their moving barrier from \citep{Sheth.Tormen.1999} that can be used to calculate $z_{50}$ \citep[e.g.][]{Giocoli.2007}. Their conditional probability is
\begin{equation}
\begin{gathered}
    \label{eq:sheth2002}
    f_{c,ST}(M_1, z_1 |M_0, z_0) =\frac{\left|T(M_1, z_1 |M_0, z_0)\right|}{\sqrt{2\pi \left(S(M1) - S(M_0)\right)^3}}\\
    \times\exp{\left(-\frac{\left(B(M_1,z_1)-B(M_0, z_0)\right)^2}{2(S(M_1)-S(M_0)}\right)}\,, 
\end{gathered}
\end{equation}
where $B$ is the moving barrier 
\begin{equation}
    \label{eq:barrier}
    B(M, z) = \sqrt{a}\delta_c(z)\left[1 + \beta \left( \frac{S(M)}{a(\delta_c(z))^2}\right)^\gamma\right]\,,
\end{equation}
with parameters $a = 0.7$, $\beta = 0.485$ and $\gamma = 0.615$, while T is the first terms of a Taylor expansion of the function $B$
\begin{equation}
    \label{eq:taylor}
    T(M_1, z_1 |M_0, z_0)= \sum_0^5 \frac{(S_0-S_1)^n}{n!}\frac{\partial^n\left[B(M_1, z_1)-B(M_0, z_0)\right]}{\partial S^n(M_1)}\,.
\end{equation}
Inspired by the ellipsoidal collapse model, \cite{Zhang.2008} also developed a fitting function for the conditional probability based on ellipsoidal collapse: 

\begin{equation}
\begin{gathered}
\label{eq:fec}
    f_{EC}(M_1, z_1 |M_0, z_0)dM_1 = \frac{A_0}{\sqrt{2\pi}}\frac{\delta_c(z_1) - \delta_c(z_0)}{(S(M_1)-S(M_0))^{3/2}} \exp{\left(-\frac{A_1^2}{2}\tilde S\right)} \\ \times \left\{\exp{\left(-A_3\frac{\left(\delta_c(z_1) - \delta_c(z_0)\right)^2}{2(S(M_1)-S(M_0))}\right)} + A_2\tilde S^{3/2}\left(1+2A\sqrt{\frac{\tilde S}{\pi}} \right) \right\}dM_1\,,
\end{gathered}
\end{equation}
where $A_0 = 0.8661(1-0.133\nu_0^{-0.615})$, $A_1 = 0.308\nu_0^{-0.115}$, $A_2 = 0.0373\nu_0^{-0.115}$, $A_3 = A_0^2 + 2A_0A_1\sqrt{\Delta S \tilde S}/\Delta \omega$, $\nu_0 = \omega_0^2/S(M_0)$, $\tilde S = \Delta S/S(M_0)$, $\Delta S = S(M_1) - S(M_0)$ and $\omega_i \equiv \delta_c(z_i)$.

\begin{figure}
    \centering
    \includegraphics[width=0.45\textwidth]{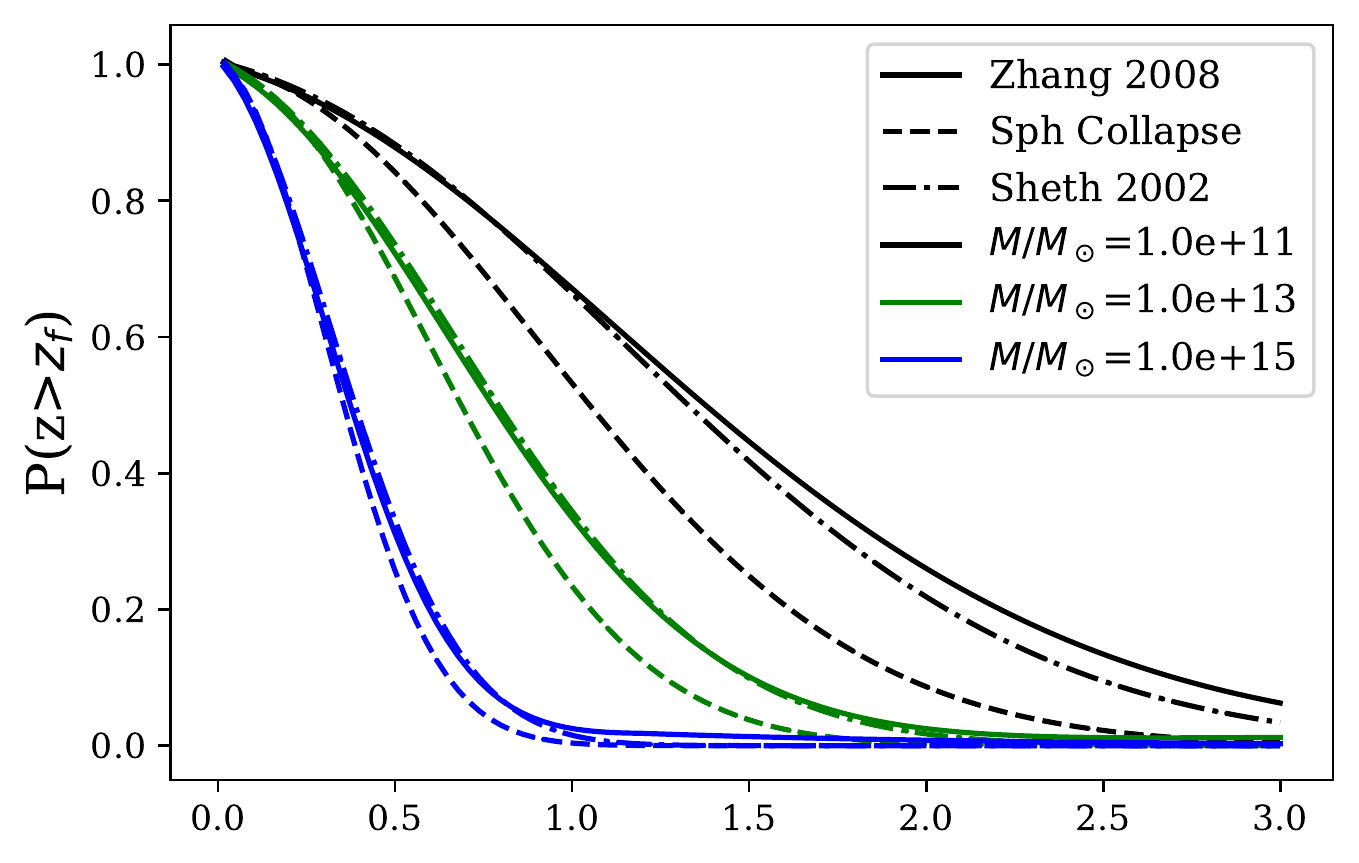}
    \includegraphics[width=0.45\textwidth]{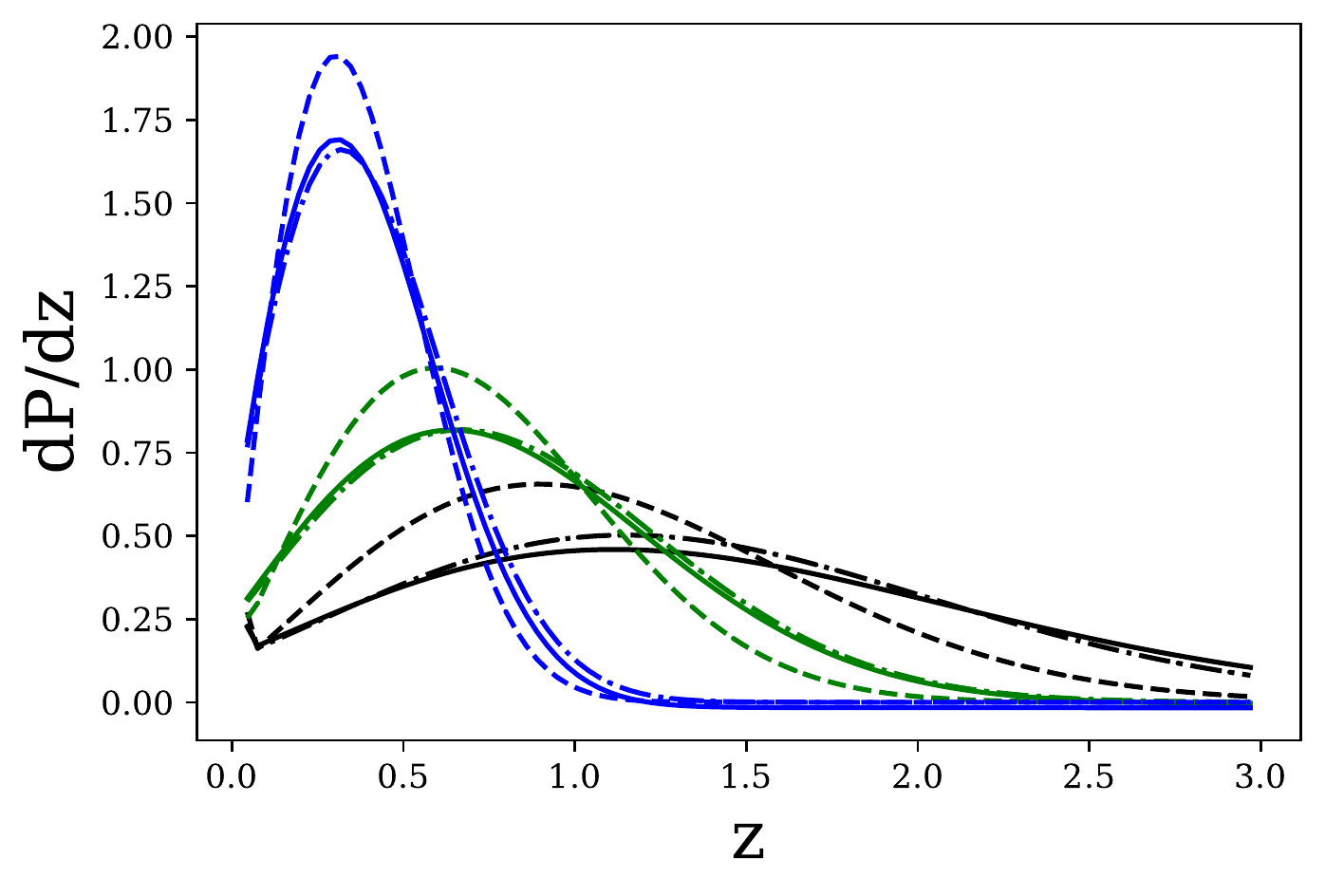}
    \caption{Cumulative (top panel) and differential (bottom panel) distributions of the formation redshift $z_{50}$ in our fiducial cosmology, for the SC model \citep[][dashed lines]{lacey.cole}, and the EC models from \citet[][dot-dashed lines]{Sheth.Tormen.2002} and  \citet[][solid lines]{Zhang.2008}.}
    \label{fig:ages}
\end{figure}

The probability distributions obtained using the three models (Eqs.~\ref{eq:fsc}, ~\ref{eq:sheth2002} and \ref{eq:fec}) in Eq.~\ref{eq:z50} are shown in Fig.~\ref{fig:ages}. The ellipsoidal collapse models predict earlier formation times $z_{50}$ at all masses, although the difference is largest at low mass. The figure also illustrates a well-known feature of hierarchical structure formation, that massive haloes have formed more recently. The predictions of the two ellipsoidal collapse models are very similar, so we will use the model from \cite{Zhang.2008} as our base model in what follows, as it is slightly faster to calculate.

% Fig 3

%subsection 2.4
\subsection{Cosmological Dependence of Halo Assembly Time}
\label{subsec:2.4}

Given a prediction for the distribution of halo formation redshifts, we can study how it varies with cosmological parameters. We have tested the dependence of three quantities in particular: 
\begin{itemize}
    \item The median formation redshift, defined as $\left<z_{50}\right> = z_m$ such that $P(Z > z_m) = 0.5$
    \item The peak of the differential probability distribution, $z_p = \rm{max}\left[\frac{dP}{dz}(z)\right]$
    \item The average formation redshift, $z_a = \int_{z_{obs}}^\infty z\, \frac{dP}{dz}dz$\,.
\end{itemize}
Of these three, we will focus on the median formation redshift, noting that the average formation redshift is slightly higher. 

\begin{figure*}
    \centering
    \includegraphics[width=0.45\textwidth]{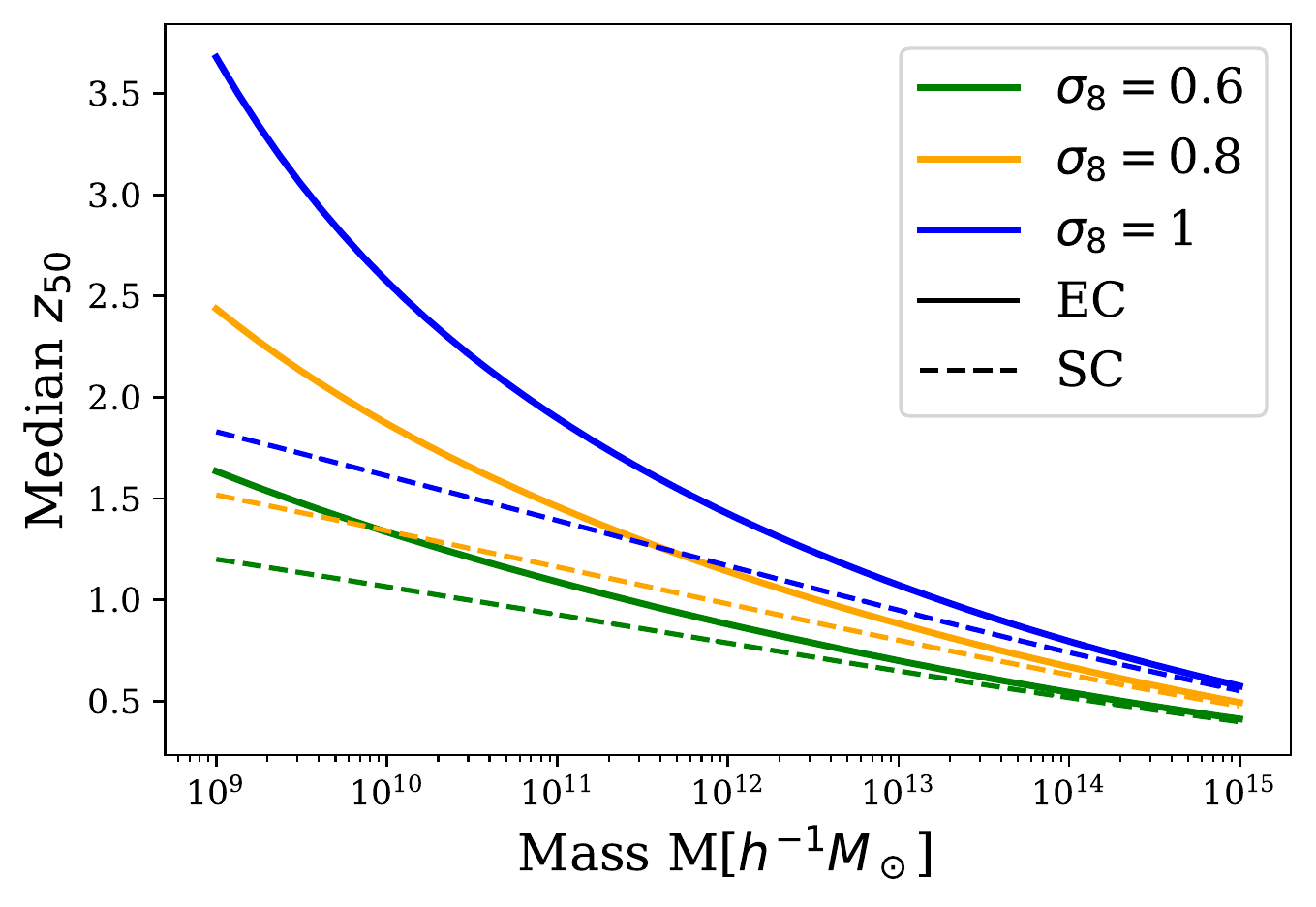}
    \includegraphics[width=0.45\textwidth]{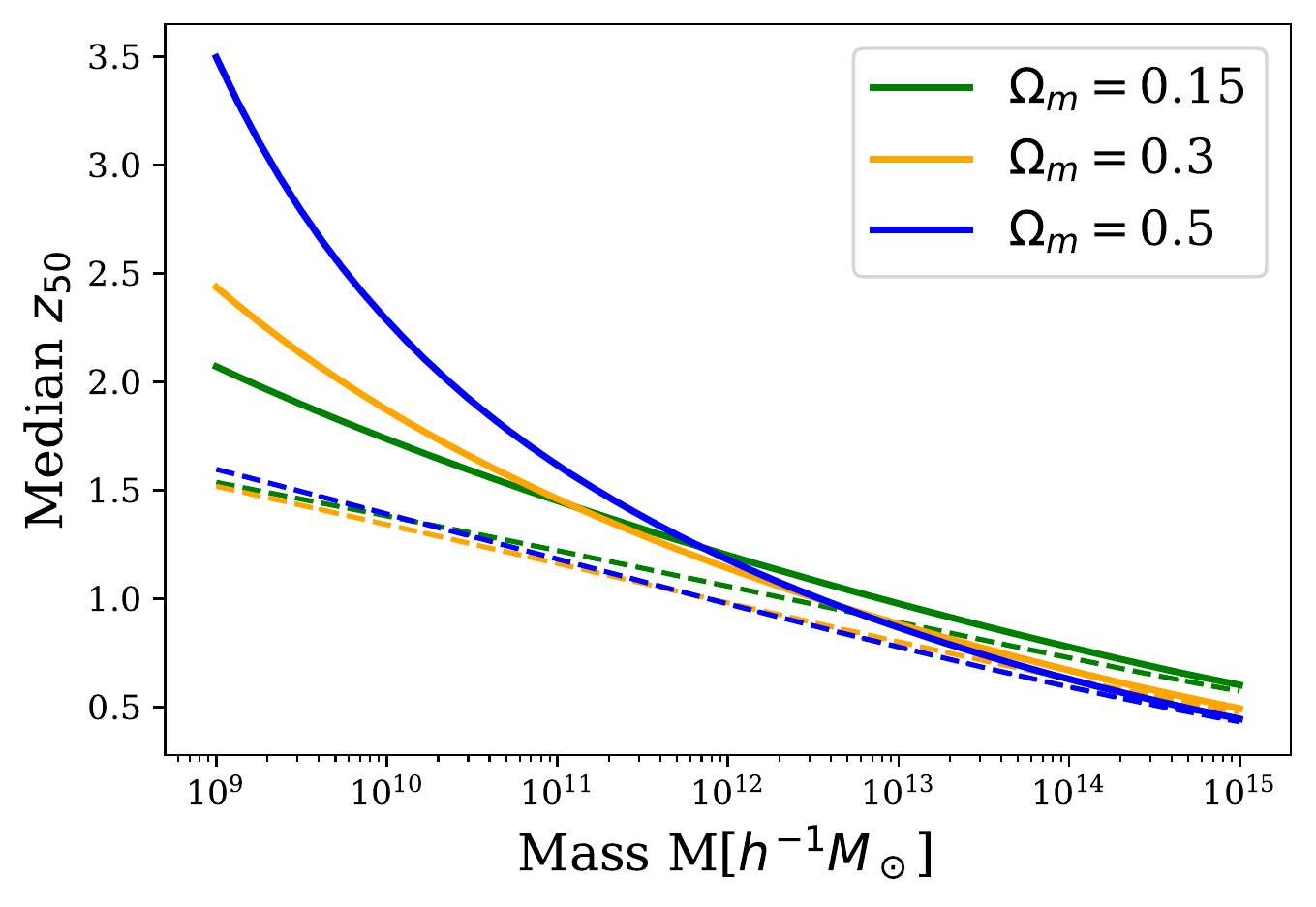}
    \caption{Mass dependence of the median formation redshift $\left<z_{50}\right>$, for haloes at $z=0.15$, and for different values of $\sigma_8$ (left panel) and $\Omega_{\rm m}$ (right panel). Solid lines show the ellipsoidal collapse predictions; dashed lines show the spherical collapse predictions.} 
    \label{fig:median_mass}
\end{figure*}

Both spherical and ellipsoidal collapse models predict the same behaviour of the median formation time, as shown in Fig.~\ref{fig:median_mass}. Haloes form earlier in high--$\sigma_8$ cosmologies, since a greater mean fluctuation amplitude causes typical peaks in the density field to cross the threshold for collapse earlier in the process of structure formation.
The $\Omega_{\rm m}$-dependence is less trivial and differs between low-mass and high-mass haloes. As explained in Appendix \ref{sec:Appendix_C}, high-$\Omega_{\rm m}$ cosmologies have more power on small scales relative to large ones. Thus at fixed $\sigma_8$, low-mass haloes form earlier in higher $\Omega_{\rm m}$ universes, while high mass haloes form slightly later. This agrees with the previous findings of \cite{Giocoli.2012}.

% Fig 4

\begin{figure*}
    \centering
    \includegraphics[width=0.45\textwidth]{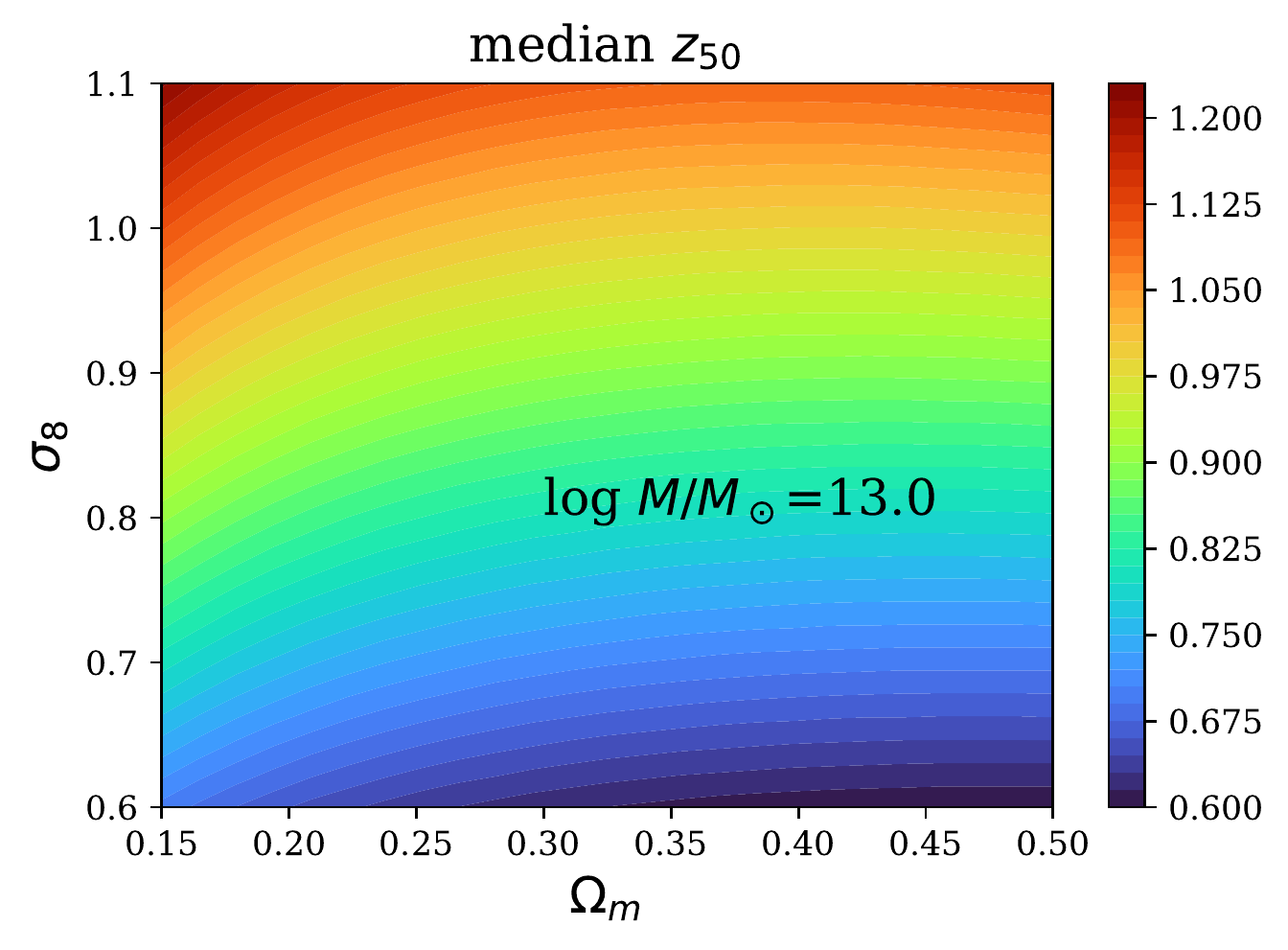}
    \includegraphics[width=0.45\textwidth]{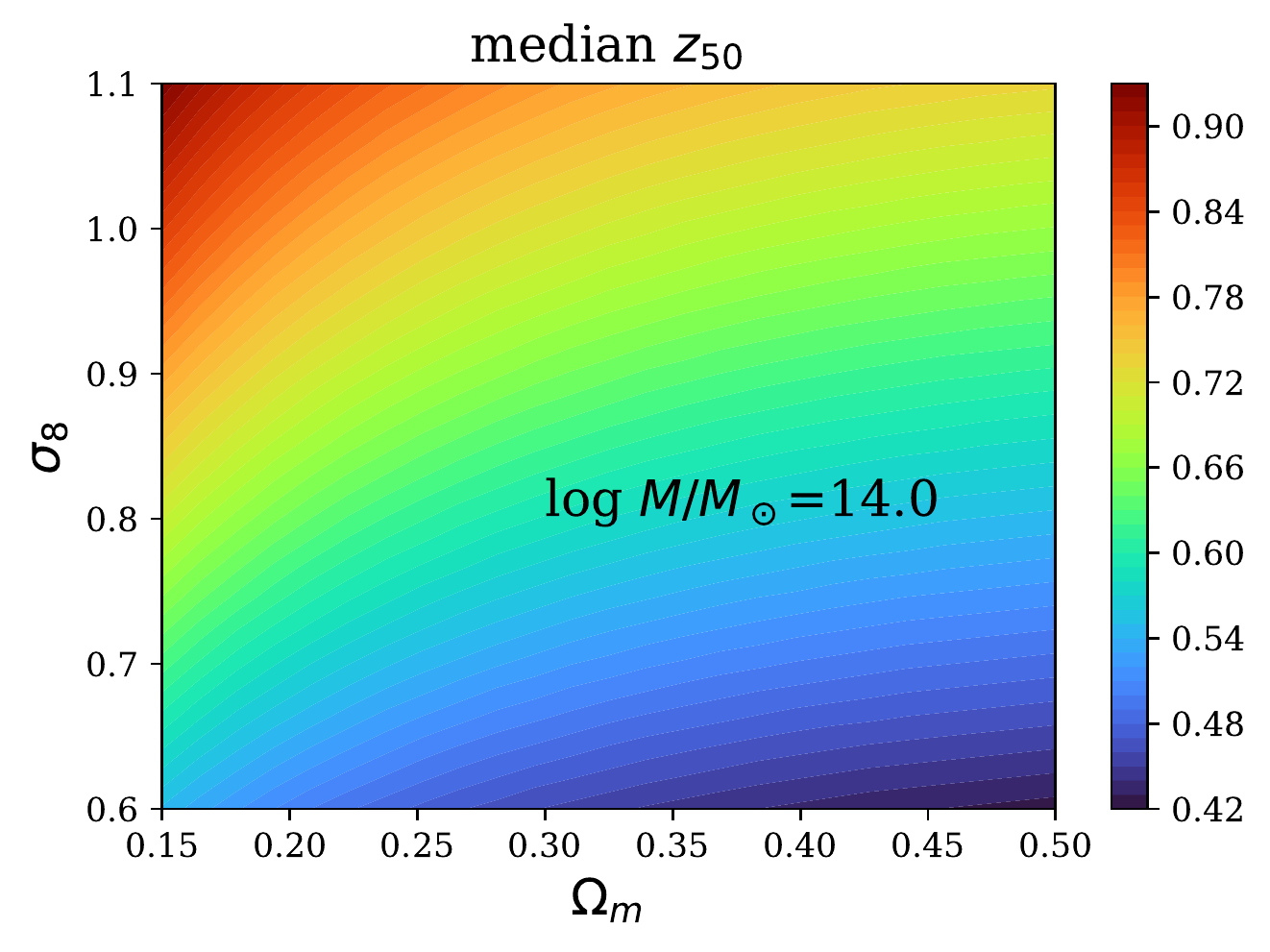}\\
    \includegraphics[width=0.45\textwidth]{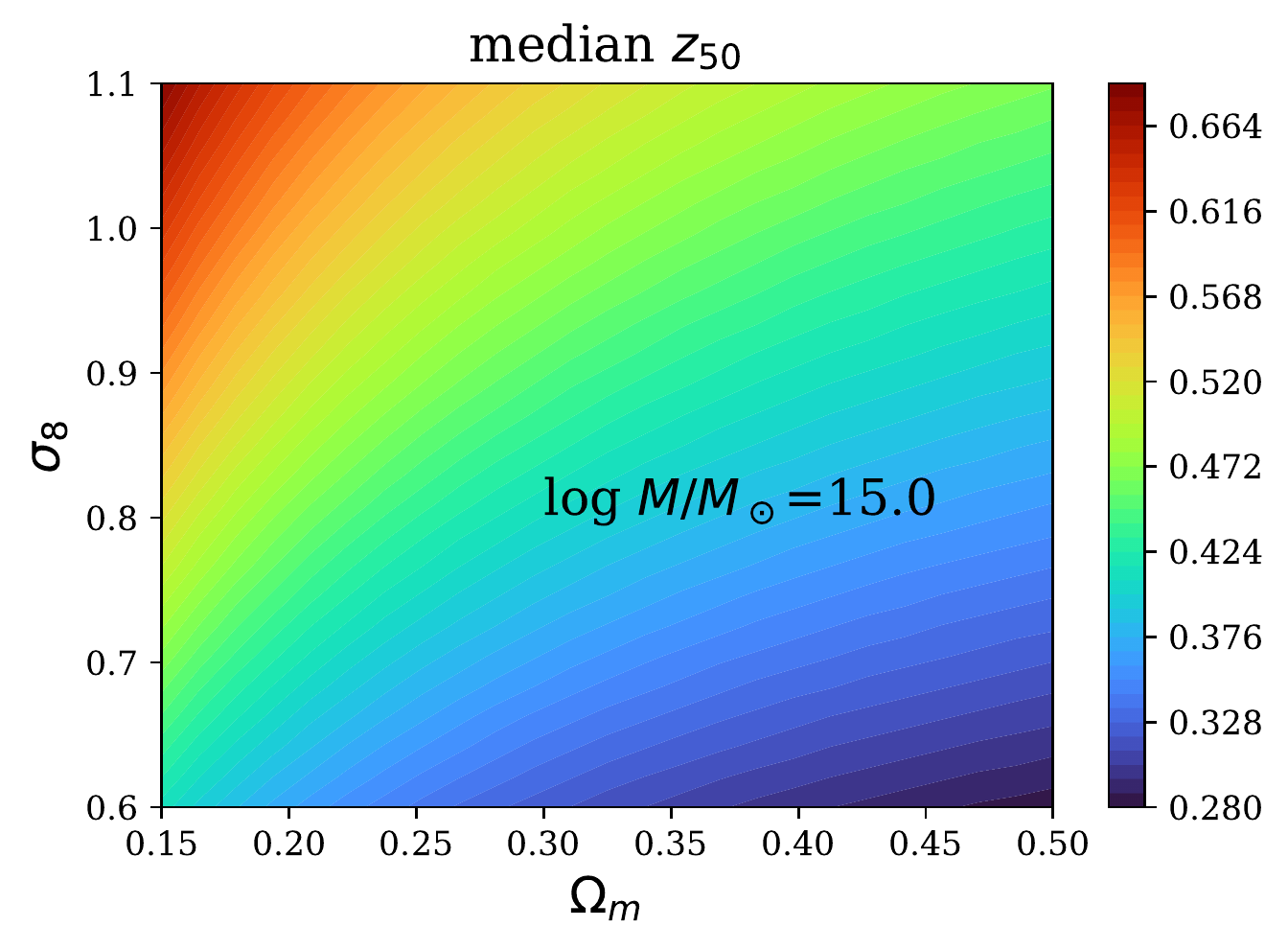}
    \caption{Median $z_{50}$ for present-day haloes of the mass indicated, as a function of $\Omega_{\rm m}$ and $\sigma_8$. Calculations assume the EC model (cf.~Sec.\ref{subsec:2.3}).} 
    \label{fig:anti-banana}
\end{figure*}

The general dependence of formation epoch $z_{50}$ on $\Omega_{\rm m}$ and $\sigma_8$ is shown in Fig.~\ref{fig:anti-banana}. The main trend is for age to increase with $\sigma_8$; since the masses shown here are all beyond the cross-over point in Fig.~\ref{fig:median_mass}, median age also decreases slightly with $\Omega_{\rm m}$, particularly at large masses. We explore how the dependence of $z_{50}$ on $\Omega_{\rm m}$ and $\sigma_8$ arises in more detail in Appendix \ref{sec:Appendix_D}. 

% Fig 5

\begin{figure*}
    \centering
    \includegraphics[width=0.45\textwidth]{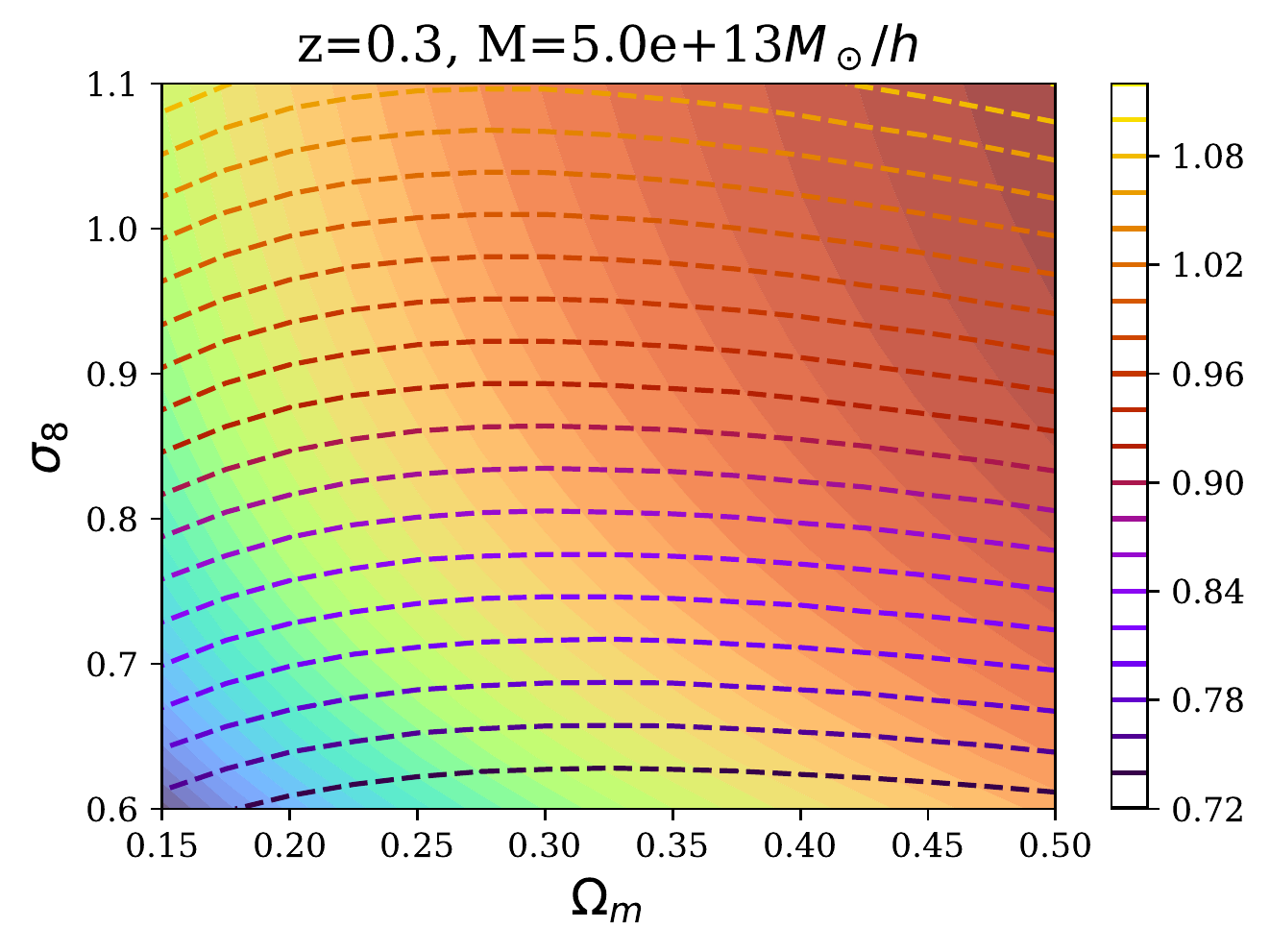}
    \includegraphics[width=0.45\textwidth]{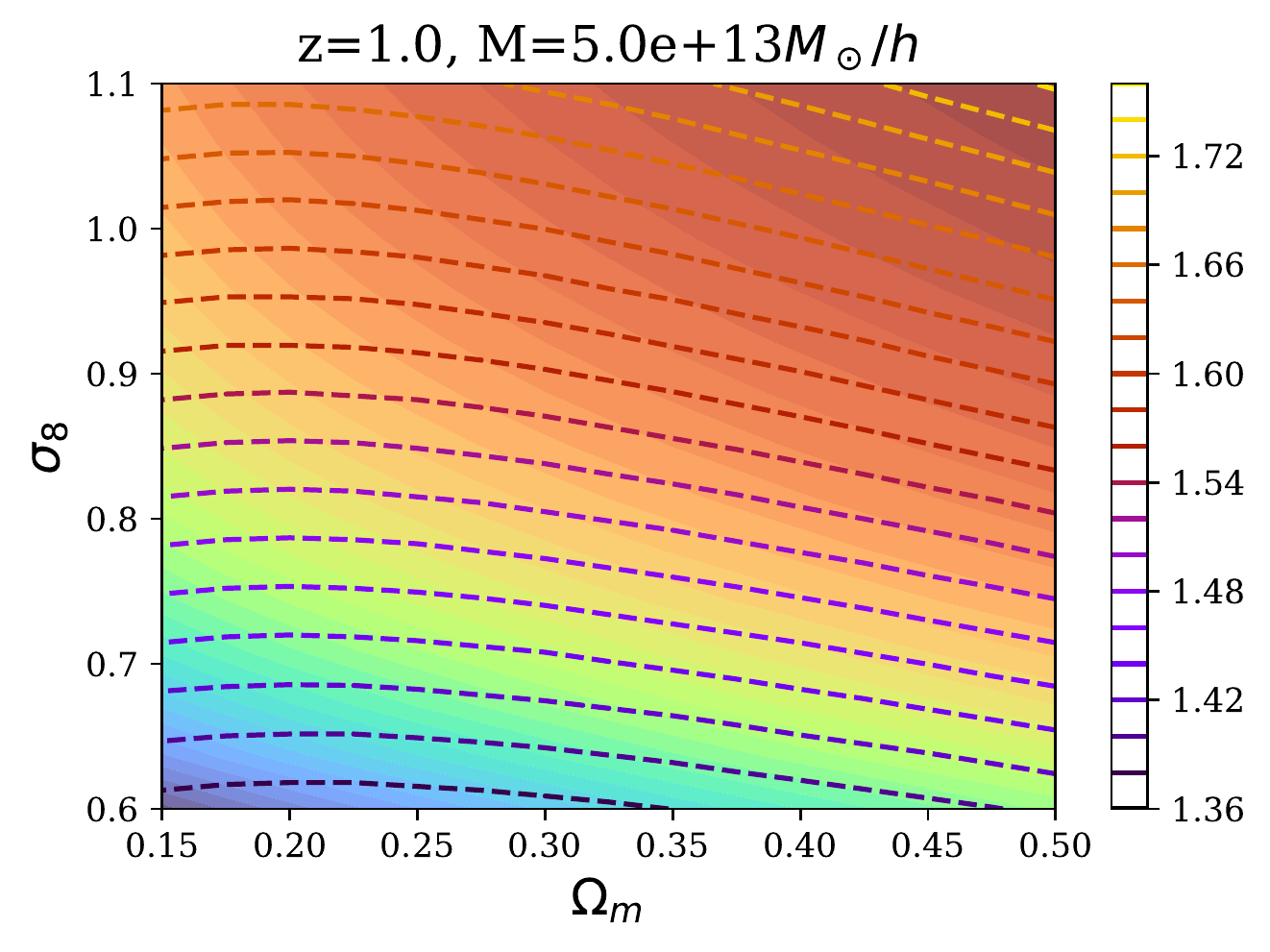}
    \includegraphics[width=0.45\textwidth]{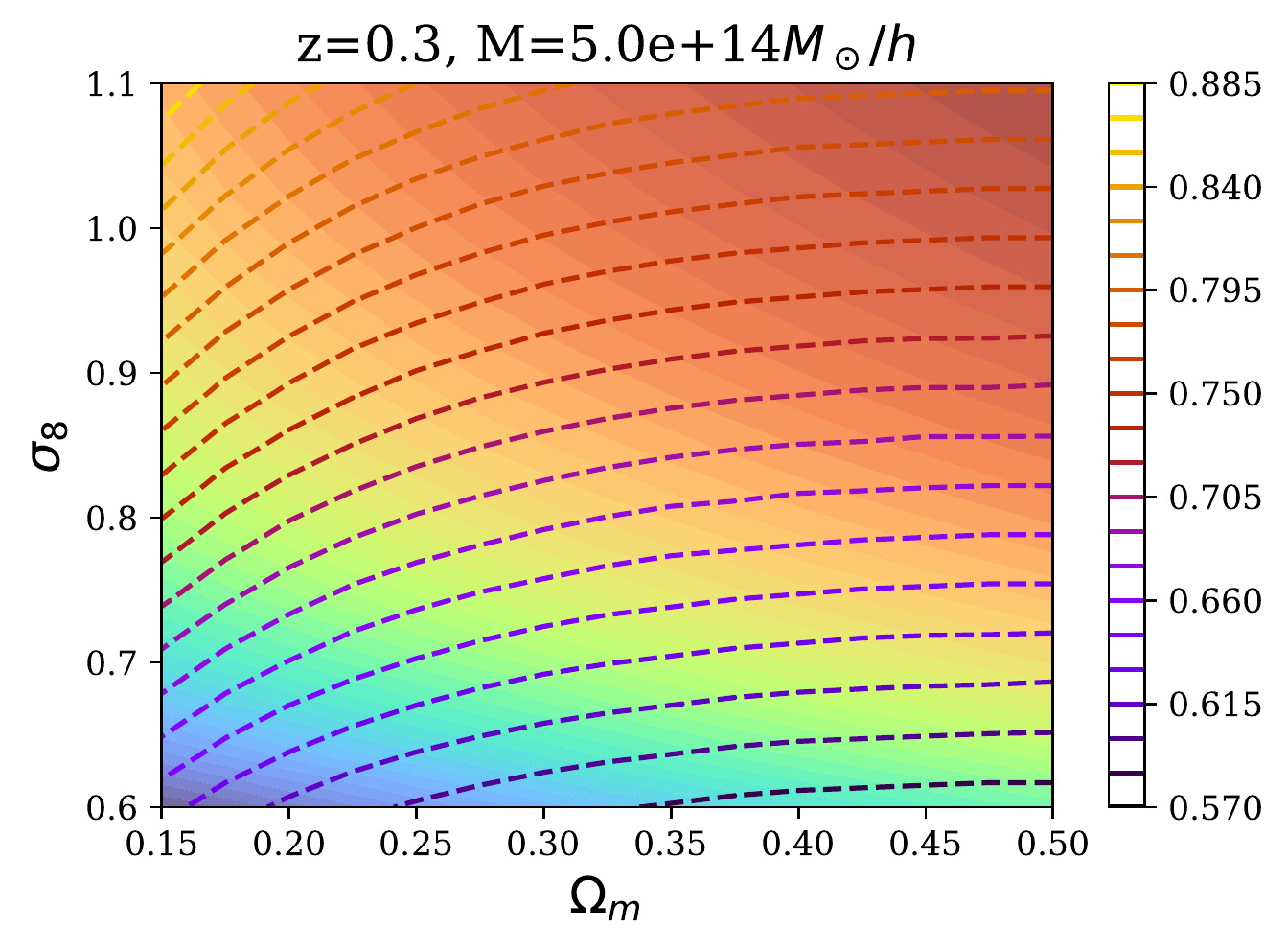}
    \includegraphics[width=0.45\textwidth]{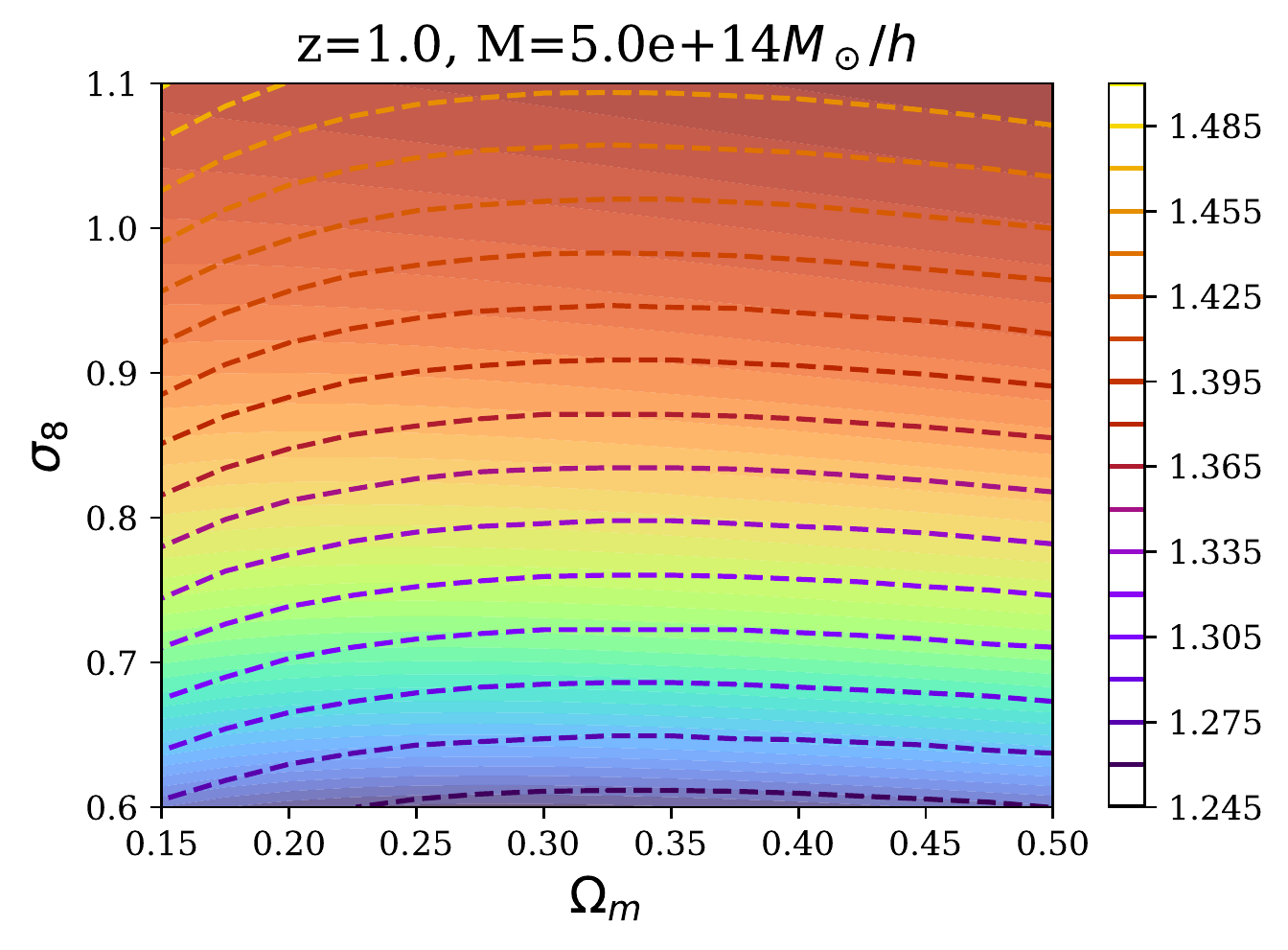}
    \includegraphics[width=0.45\textwidth]{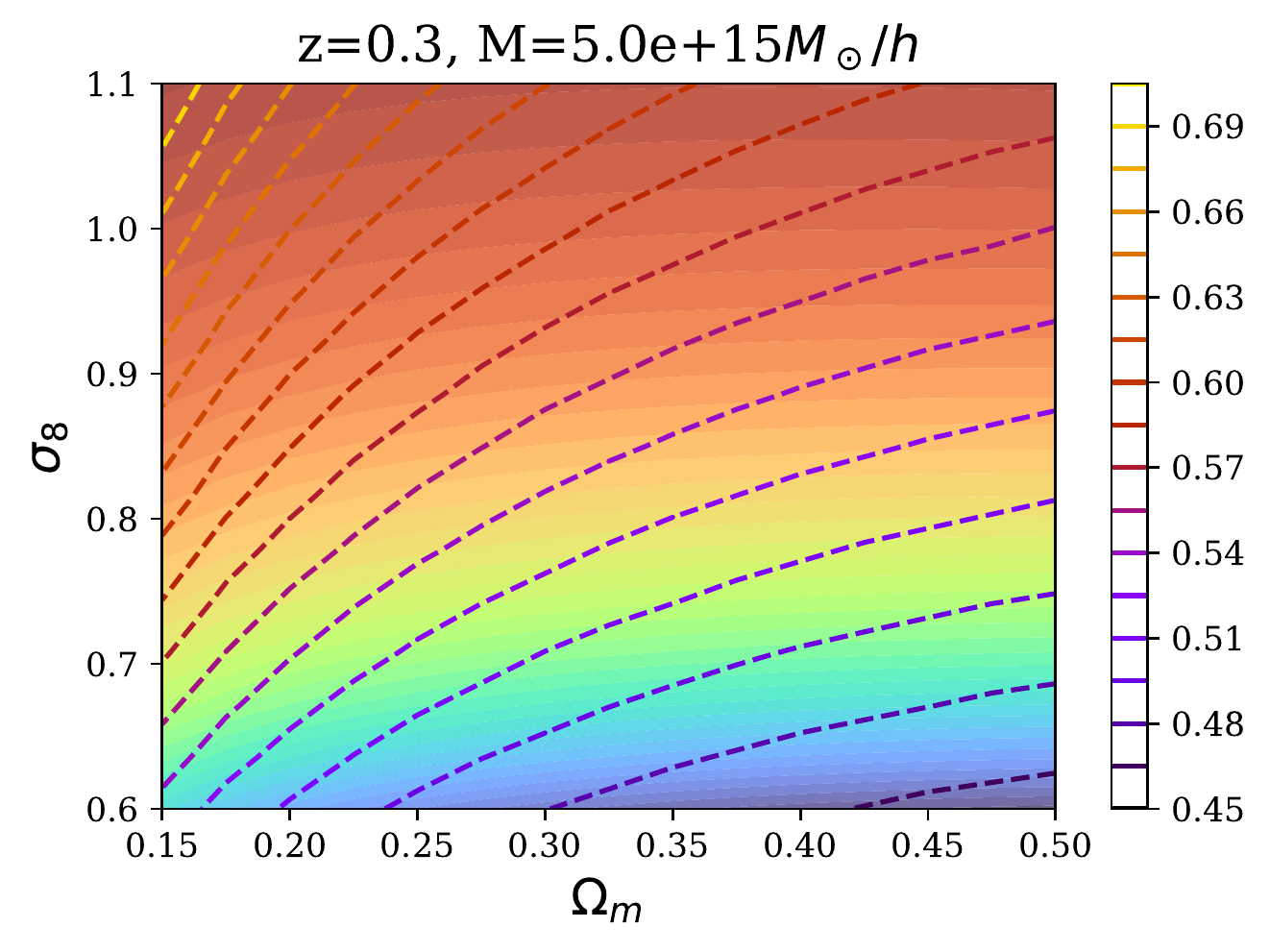}
    \includegraphics[width=0.45\textwidth]{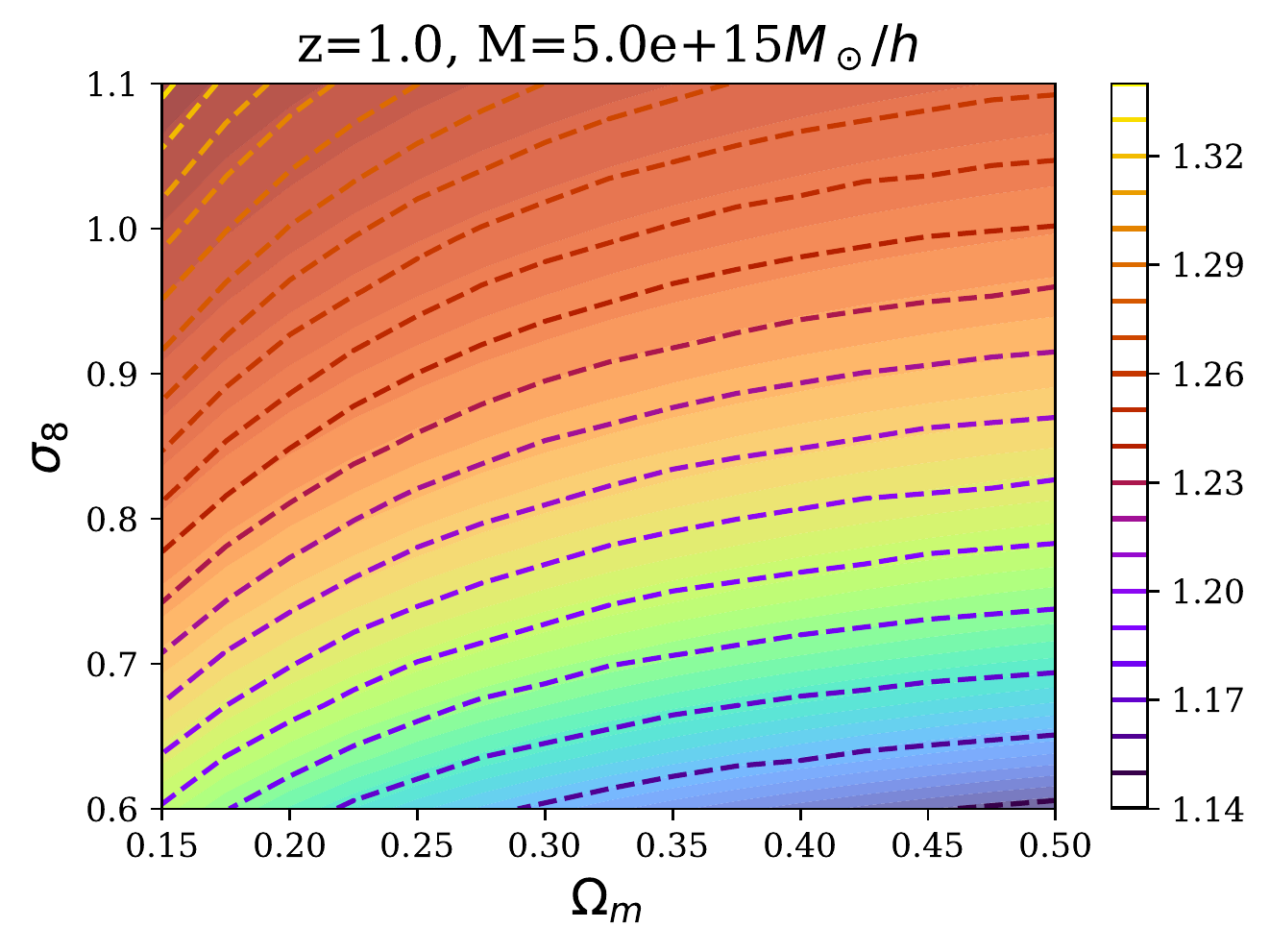}
    \caption{A comparison of age and abundance contours in the $\Omega_{\rm m}$--$\sigma_8$ plane, for the halo mass and redshift indicated at the top of each panel. Filled contours show curves of constant number density per unit log\,$M$. Dashed lines show curves of constant $z_{50}$, with the scale given by the colorbar.} 
    \label{fig:cross_constraint}
\end{figure*}

Comparing Figs.~\ref{fig:om_s8_diff_counts} and \ref{fig:anti-banana} closely, we note an important feature of halo age relative to halo abundance: for lower mass haloes and/or at lower redshift, the contours for the two are fairly orthogonal over much of the $\Omega_{\rm m}$--$\sigma_8$ plane. To highlight this point, Fig.~\ref{fig:cross_constraint} shows the two sets of contours superimposed, for the ranges of mass and redshift accessible to large cluster surveys. In the region of particular interest, around the concordance model ($\Omega_{\rm m} = 0.3$, $\sigma_8 = 0.8$), the two sets of contours are almost exactly orthogonal for lower masses and/or redshifts, where $\nu \sim$2--3 (top and middle left hand panels). They only become similar for the most massive clusters, at $z\ge 1$, where $\nu \sim 4$--6 (bottom right panel).
This implies that age or age proxies, measured for clusters with masses $M < 5\times 10^{14}M_\odot$ at $z < 1$, can potentially break the main degeneracy in cluster abundance measurements. 

% Fig 6

%%%%%%%%%%%%%%%%%%%%%%%%%%%%%%%%%%%%%%%%%%%%%%%%%%%%%%%%%%%%%%%%%%%

%Section 3
\section{Comparison to simulations} \label{sec:simulations}

\subsection{Simulation Data}

In Section~\ref{subsec:2.3}, we provided analytic EPS estimates of how the halo formation time $z_{50}$ depends on $\Omega_{\rm m}$ and $\sigma_8$. Given the approximations made in these models, it is worth testing the accuracy of their predictions in N-body simulations.

Previous work has demonstrated that different merger tree algorithms can produce significantly different MAHs \citep{avila.2014.sussing, srisawat.2013}.
The exact value of $z_{50}$ may be particularly sensitive to these differences, as discussed in \citet{srisawat.2013}. In particular, some merger tree algorithms allow fragmentation events, where haloes lose mass with time, such that MAHs are not always monotonic. Our previous EPS estimates assume strictly
hierarchical growth, and thus we anticipate that the numerical results may disagree with them to some degree. To test the effect of different methods of analysis, we consider three sets of simulations (two public, and one of our own), that employed three different merger tree algorithms:
\begin{itemize}
    \item The Illustris TNG simulation, \citep{nelson2019illustristng} using the Sublink merger tree algorithm \citep{Sublink.Rodriguez_Gomez.2015}.
    \item The Bolshoi/BolshoiP simulation, \citep{Bolshoi.Klypin_2011} using the Rockstar halo finder \citep{Rockstar.Behroozi_2012}  and the Consistent Trees merger tree code \citep{Consistent_trees.Behroozi_2012}. (Note that Bolshoi uses WMAP cosmological parameters, whereas BolshoiP uses Planck ones.) 
    \item Our own set of 9 cosmological simulations, spanning a range of cosmological parameters, and analyzed using the AHF halo finder and merger tree code \citep{AHF,Gill2004.AHF}. These will be labelled MxSy, where $x  = 25/3/35$ indicates the value of $\Omega_{\rm m}$, and $y = 7/8/9$ indicates the value of $\sigma_8$. 
\end{itemize}

The simulation parameters are summarized in Table~\ref{table:sims}. Data from the TNG and Bolshoi simulations were obtained directly from their respective websites. In particular, we used the Rockstar merger tree data available for the Bolshoi simulation. For the TNG and MxSy simulations, mass accretion histories were calculated using the Sublink and AHF codes, respectively. For the Bolshoi simulations, they were generated by following the main progenitor sequence in the Rockstar files. 
\begin{table*}
\begin{tabular}{ |c|c|c|c|c|c|c|c| } 
 \hline
 Simulation & $\Omega_{\rm m}$ & $\sigma_8$ & particle mass [$M_\odot /h$]& $N_{\rm part}$ & merger tree & $N_{\rm snap}$  \\ 
 \hline
Illustris TNG & 0.31 & 0.81 & $3 \times 10^9$ & $625^3$ & Sublink & 100 \\ 
 Bolshoi & 0.27 & 0.82 & $1.35 \times 10^8$ & $2048^3$ & Consistent Trees & 181 \\
 BolshoiP & 0.31 & 0.82 & $1.55 \times 10^8$ & $2048^3$ & Consistent Trees & 178  \\
 MxSy & 0.25/0.3/0.35 & 0.7/0.8/0.9 & $4 \times 10^9$ & $512^3$ & Amiga Halo Finder & 44 \\
 \hline
\end{tabular}
\caption{Summary of the simulations used and their main parameters, including cosmological parameters, particle mass, total number of particles $N_{\rm part}$, merger tree code, and the number of snapshots $N_{\rm snap}$ used to make the merger trees. The MxSy simulations are a set of 9 of our own simulations that span a range of different values of $\Omega_{\rm m}$ and $\sigma_8$.}
\label{table:sims}
\end{table*}

\subsection{Formation Time Distributions Compared}

\begin{figure*}
    \centering
    \includegraphics[width=0.32\textwidth]{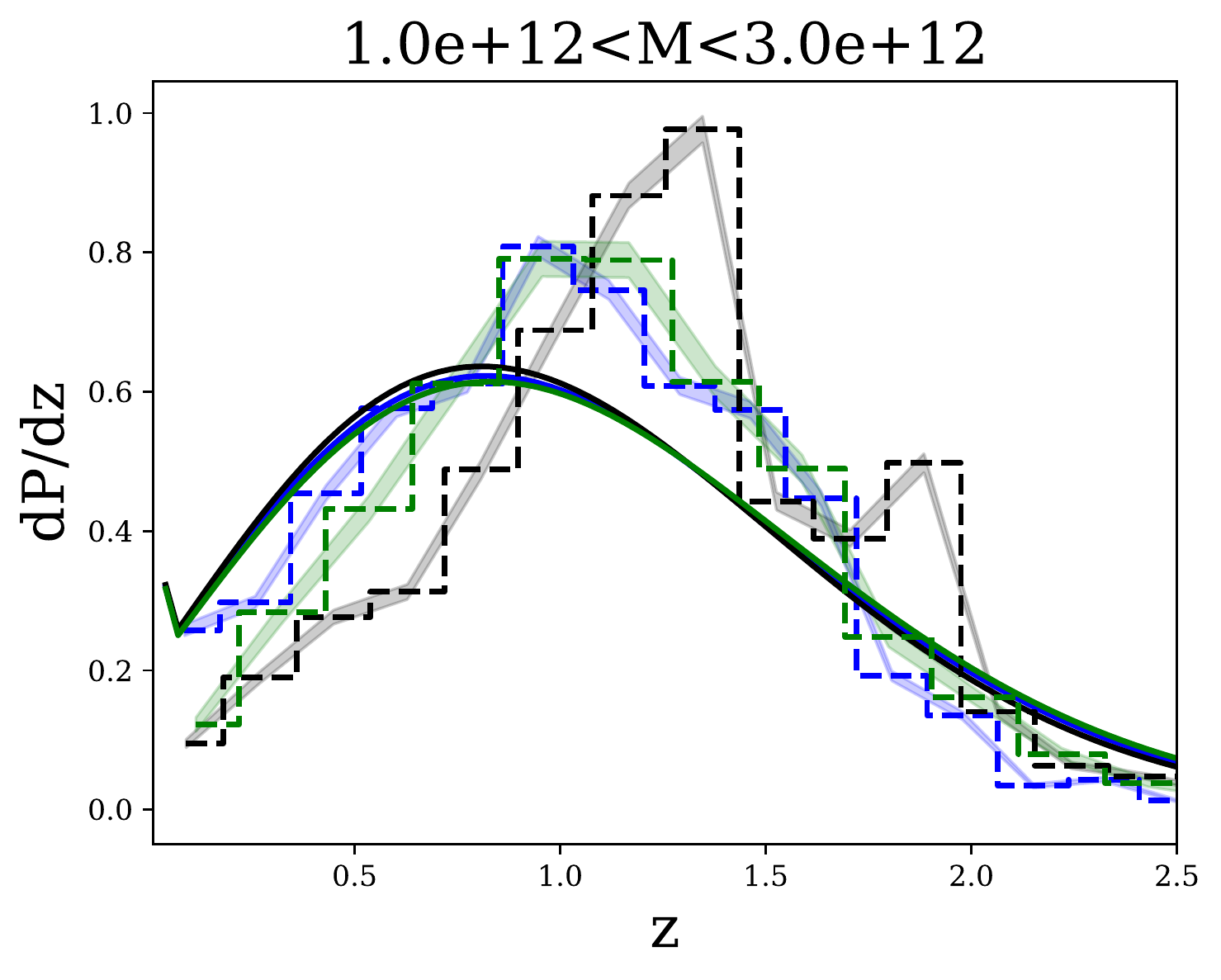}
    \includegraphics[width=0.3\textwidth]{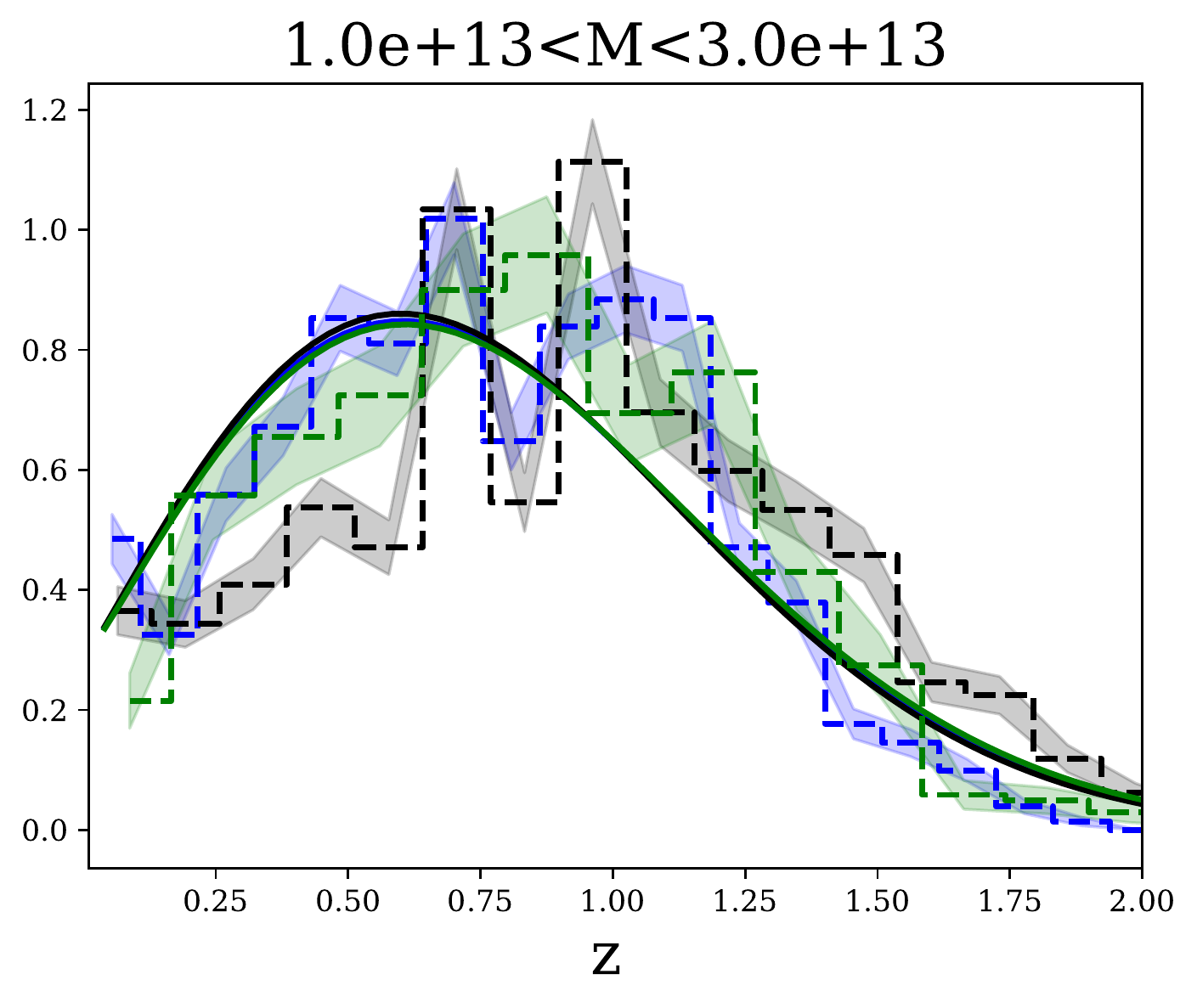}
    \includegraphics[width=0.3\textwidth]{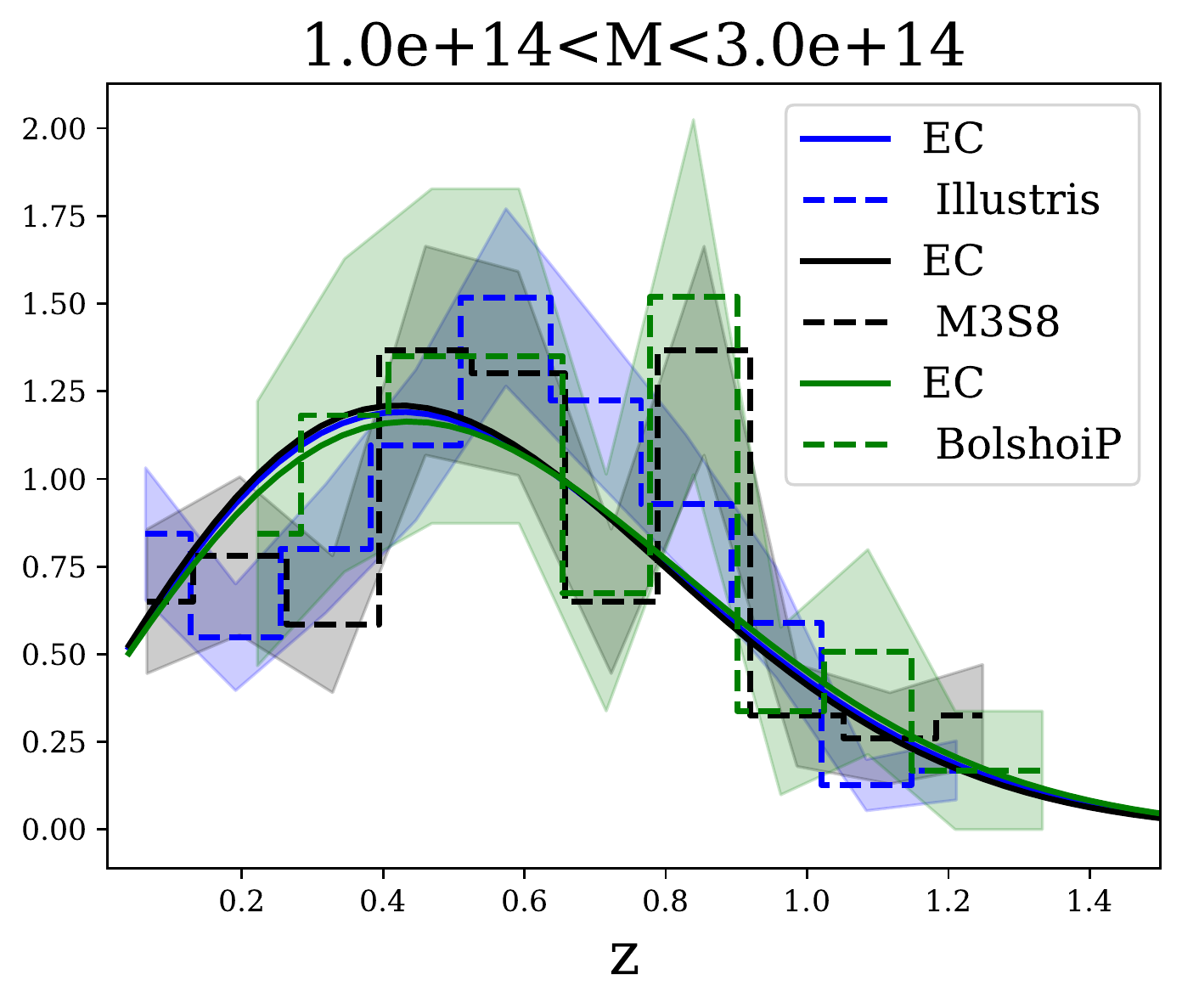} \\
    \includegraphics[width=0.32\textwidth]{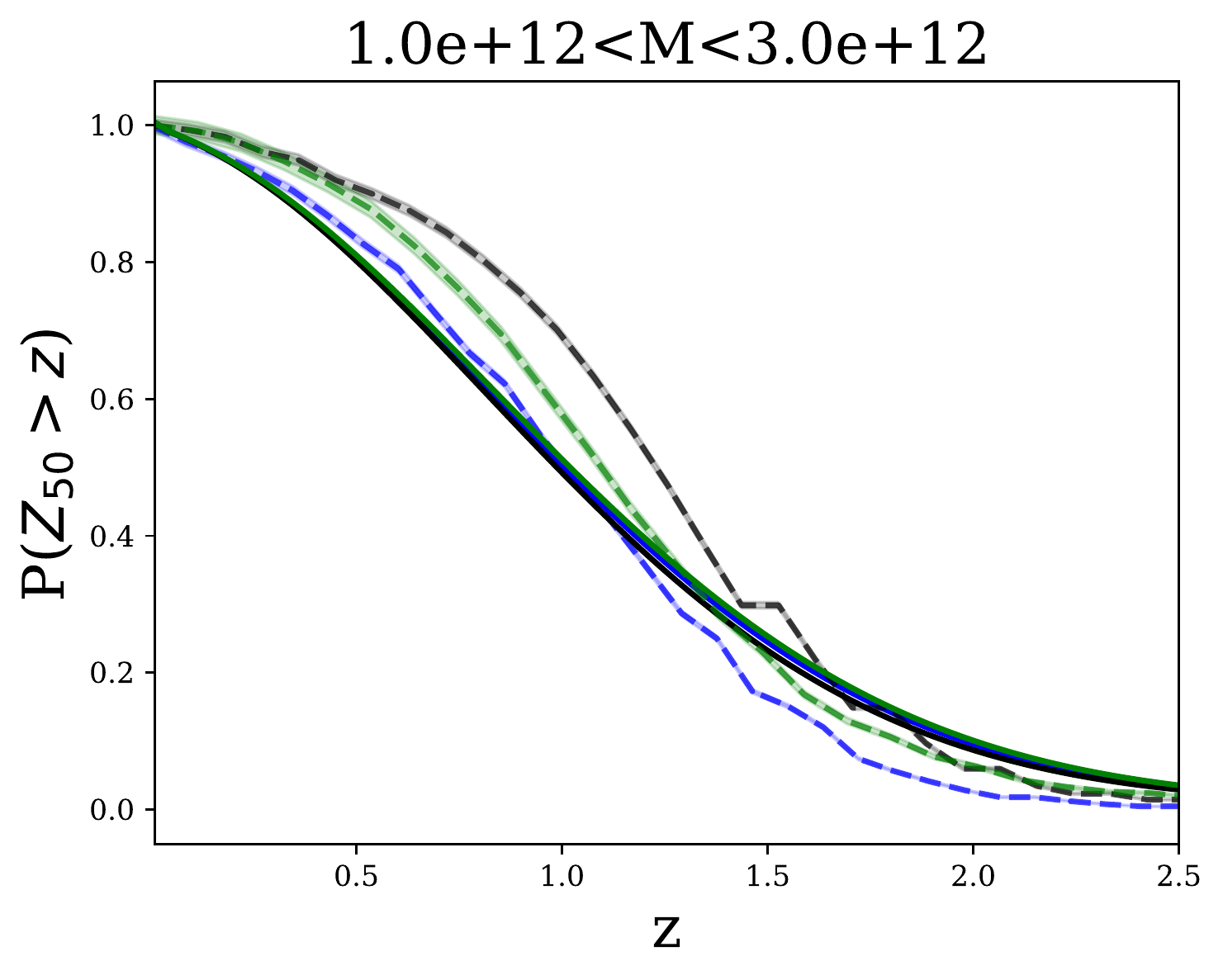}
    \includegraphics[width=0.3\textwidth]{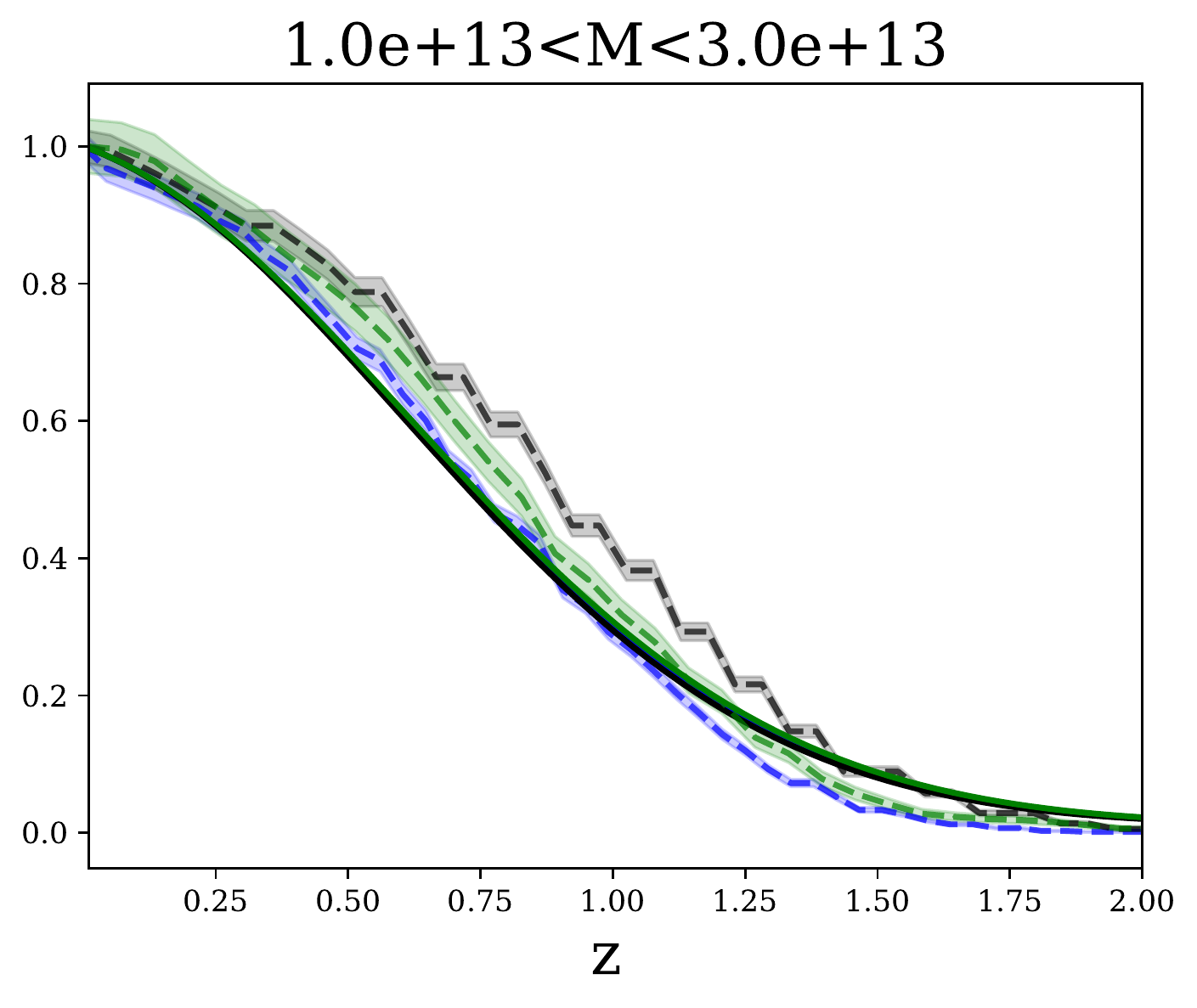}
    \includegraphics[width=0.3\textwidth]{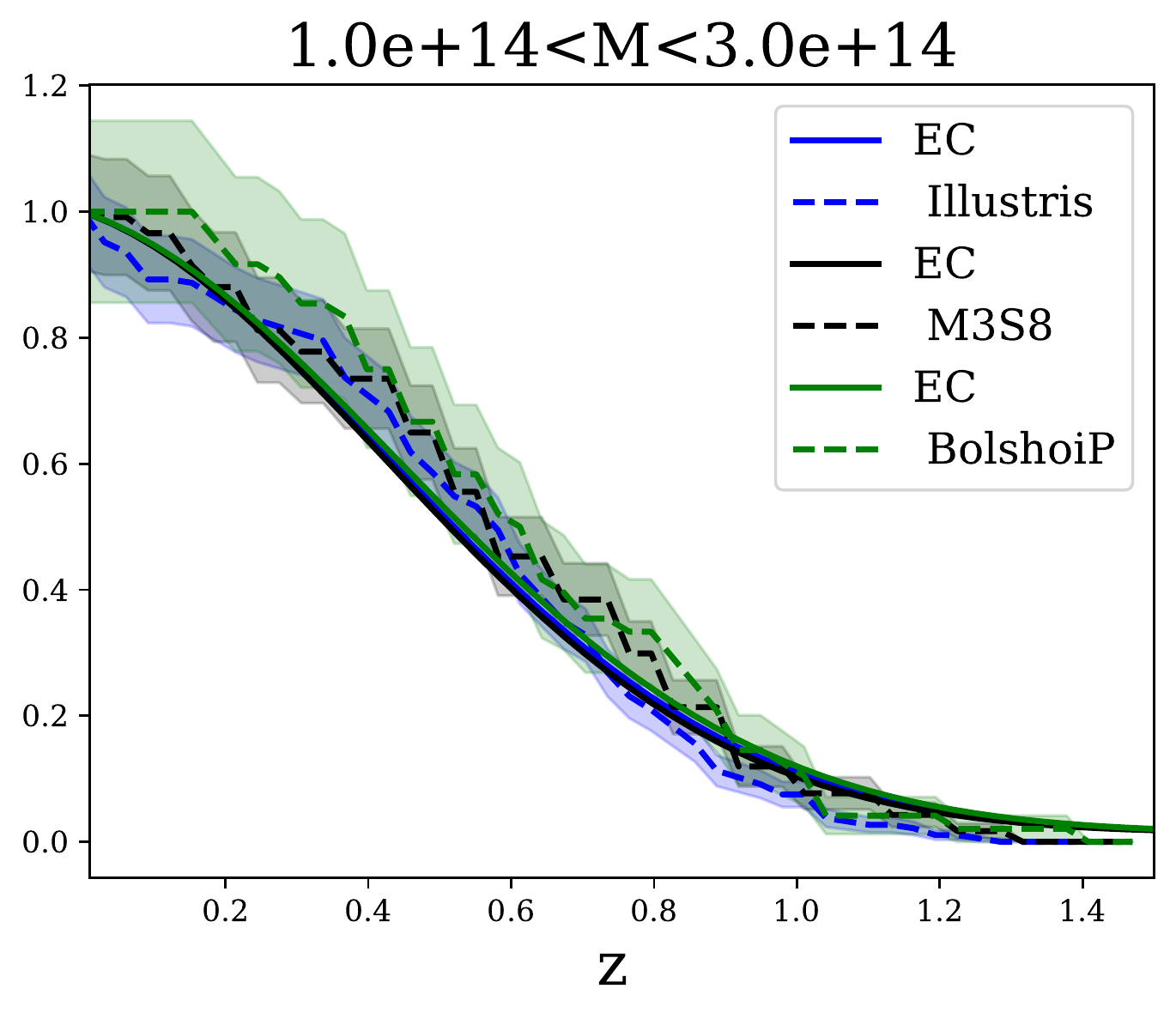}
    \caption{Differential (top) and cumulative (bottom) distributions of $z_{50}$ in our fiducial cosmology, for three different ranges of halo mass. Smooth curves show the (EC) analytic prediction, while dashed lines with shading show the numerical results and associated Poisson uncertainties.}
    \label{fig:problem}
\end{figure*}

In each numerical MAH, we define $z_{50}$ to be the lowest redshift at which the mass of the halo has dropped to less than half of the mass at $z=0$. Fig.~\ref{fig:problem} compares the distribution of these formation redshifts to the analytic (EC) predictions. For all three simulations considered, but particularly for the Bolshoi and M3S8 simulations, we see a clear offset between the numerical results and the EC predictions, that is largest at small masses. In general, the numerical formation redshifts are larger than predicted, by up to 0.1--0.2 on average.

% Fig 7

\begin{figure*}
    \centering
    \includegraphics[width=0.32\textwidth]{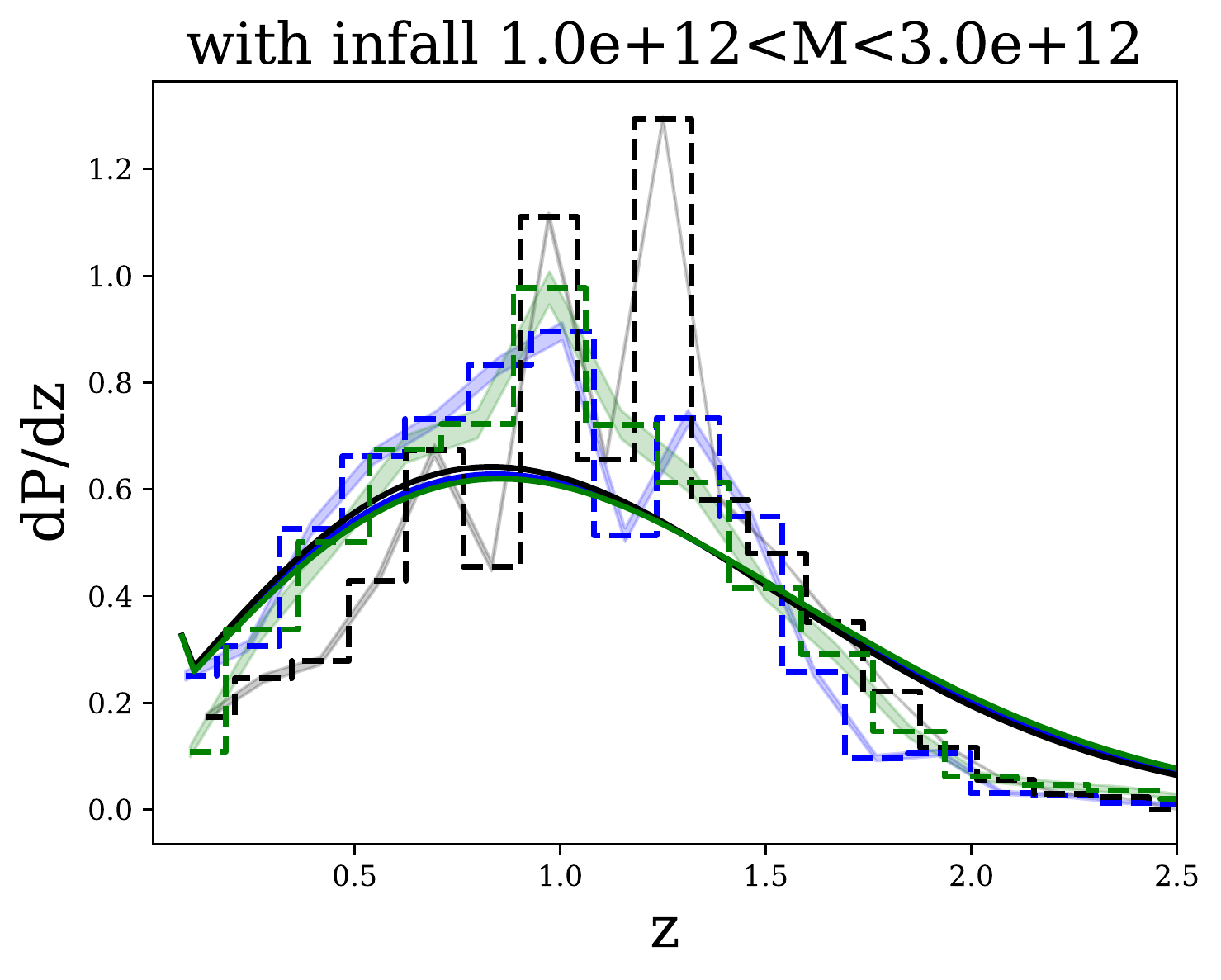}
    \includegraphics[width=0.3\textwidth]{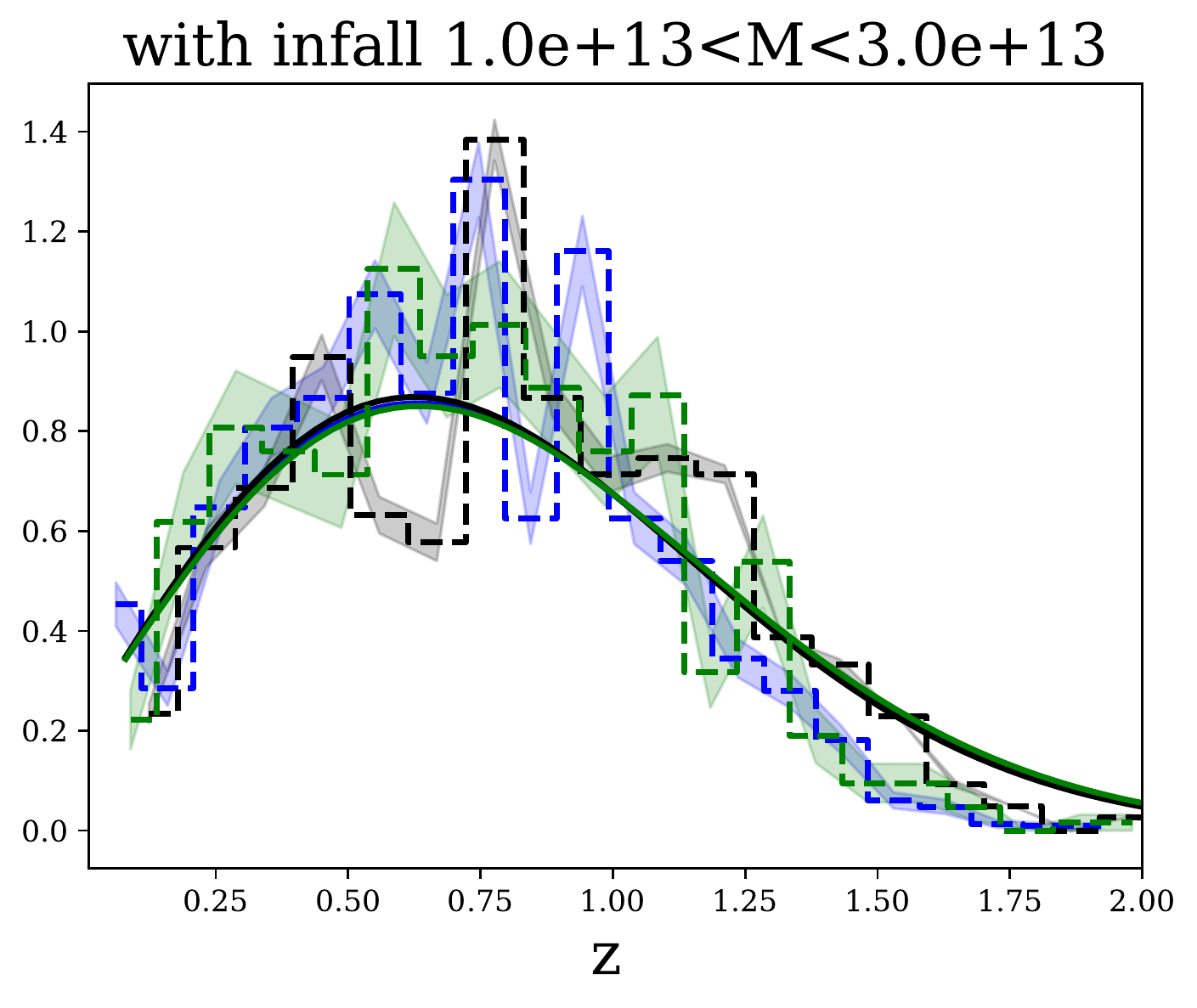}
    \includegraphics[width=0.3\textwidth]{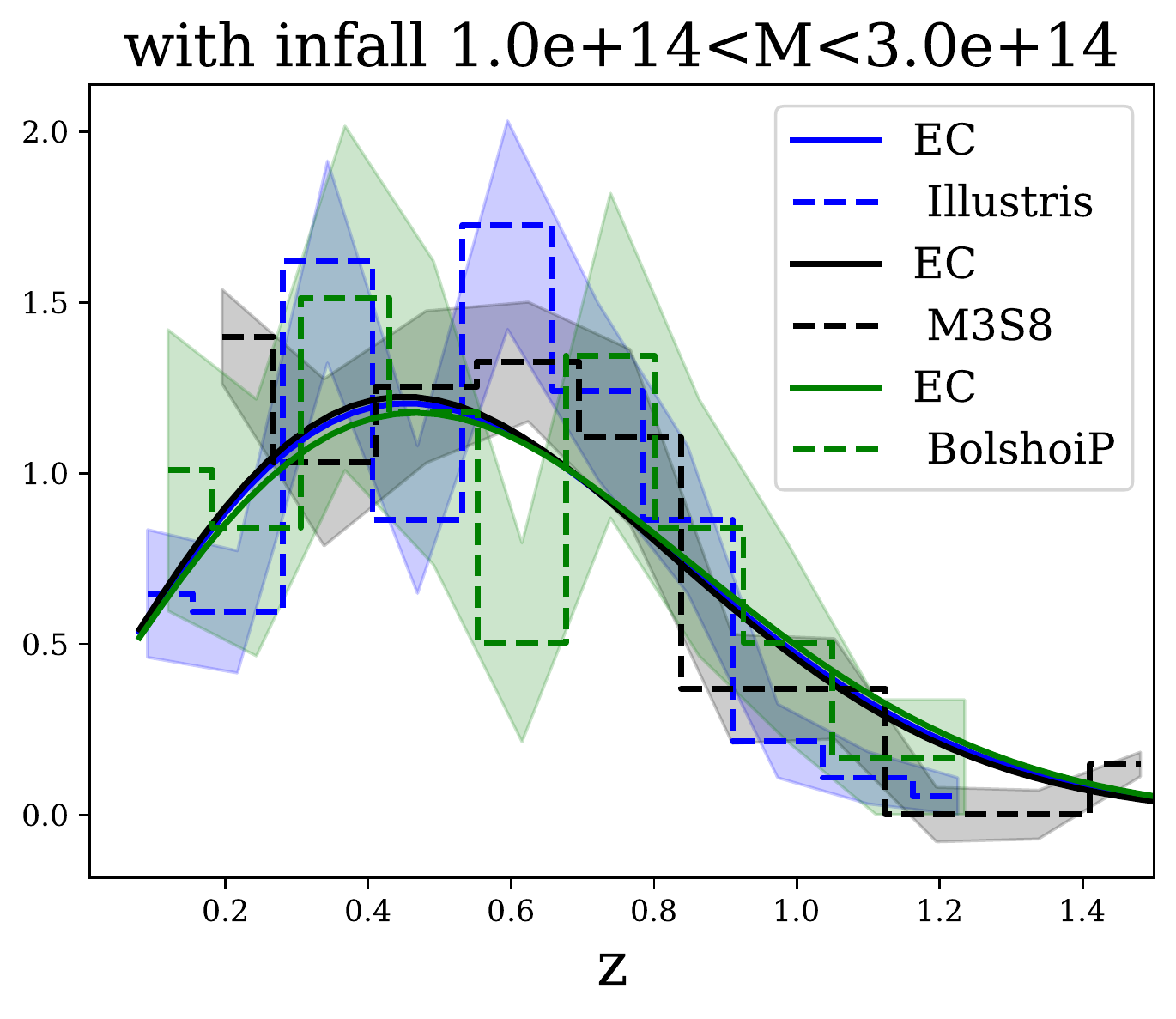} \\
    \includegraphics[width=0.32\textwidth]{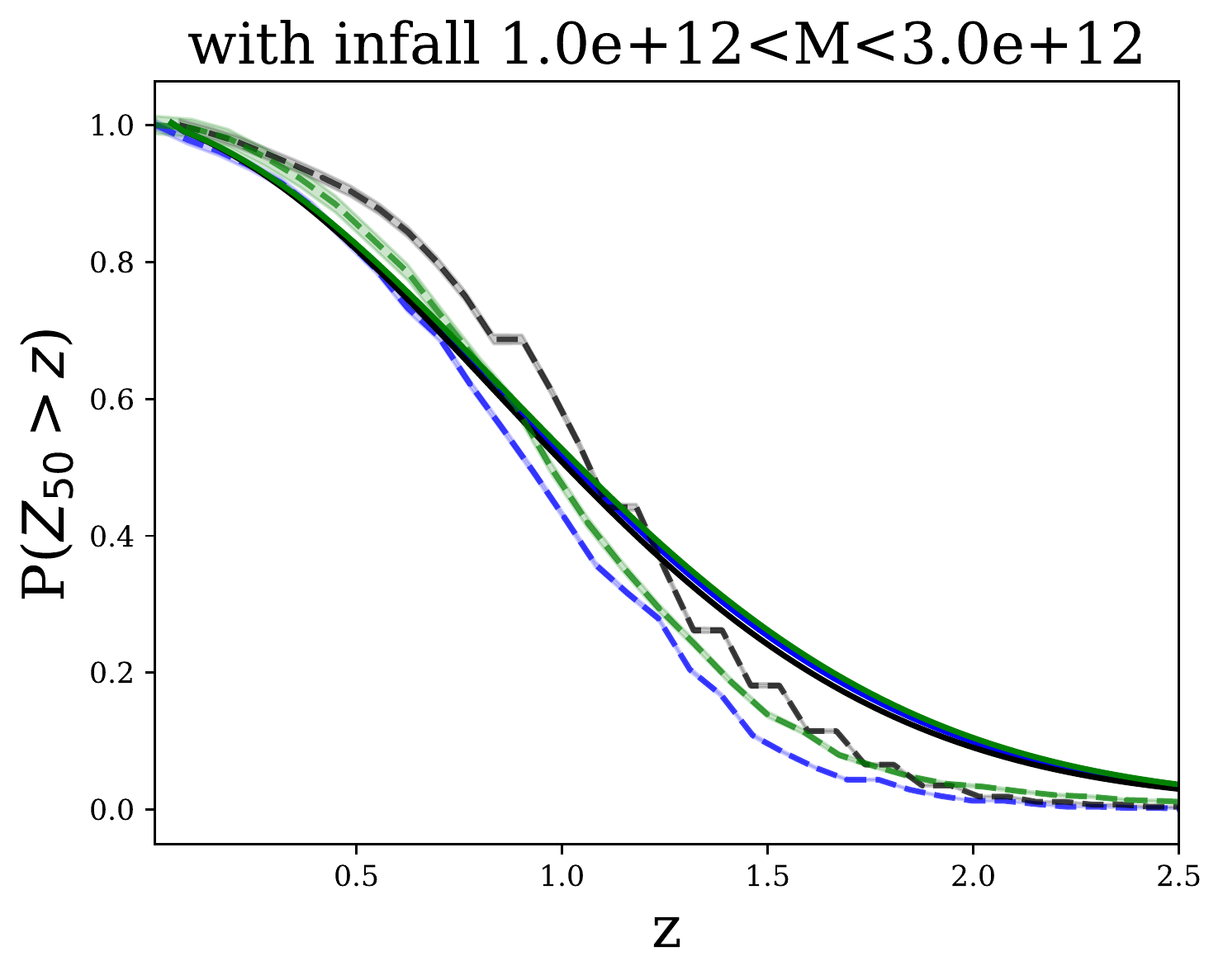}
    \includegraphics[width=0.3\textwidth]{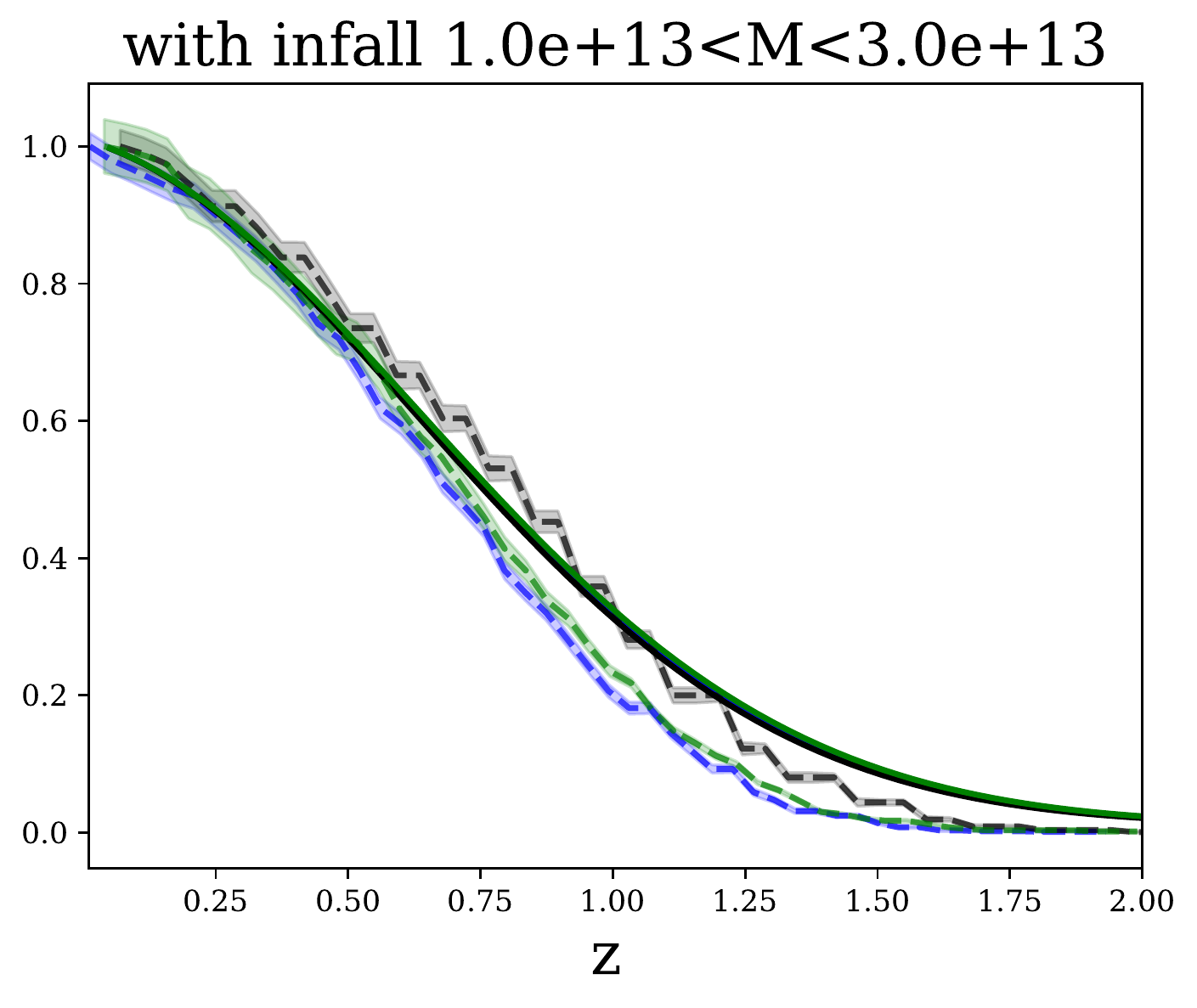}
    \includegraphics[width=0.3\textwidth]{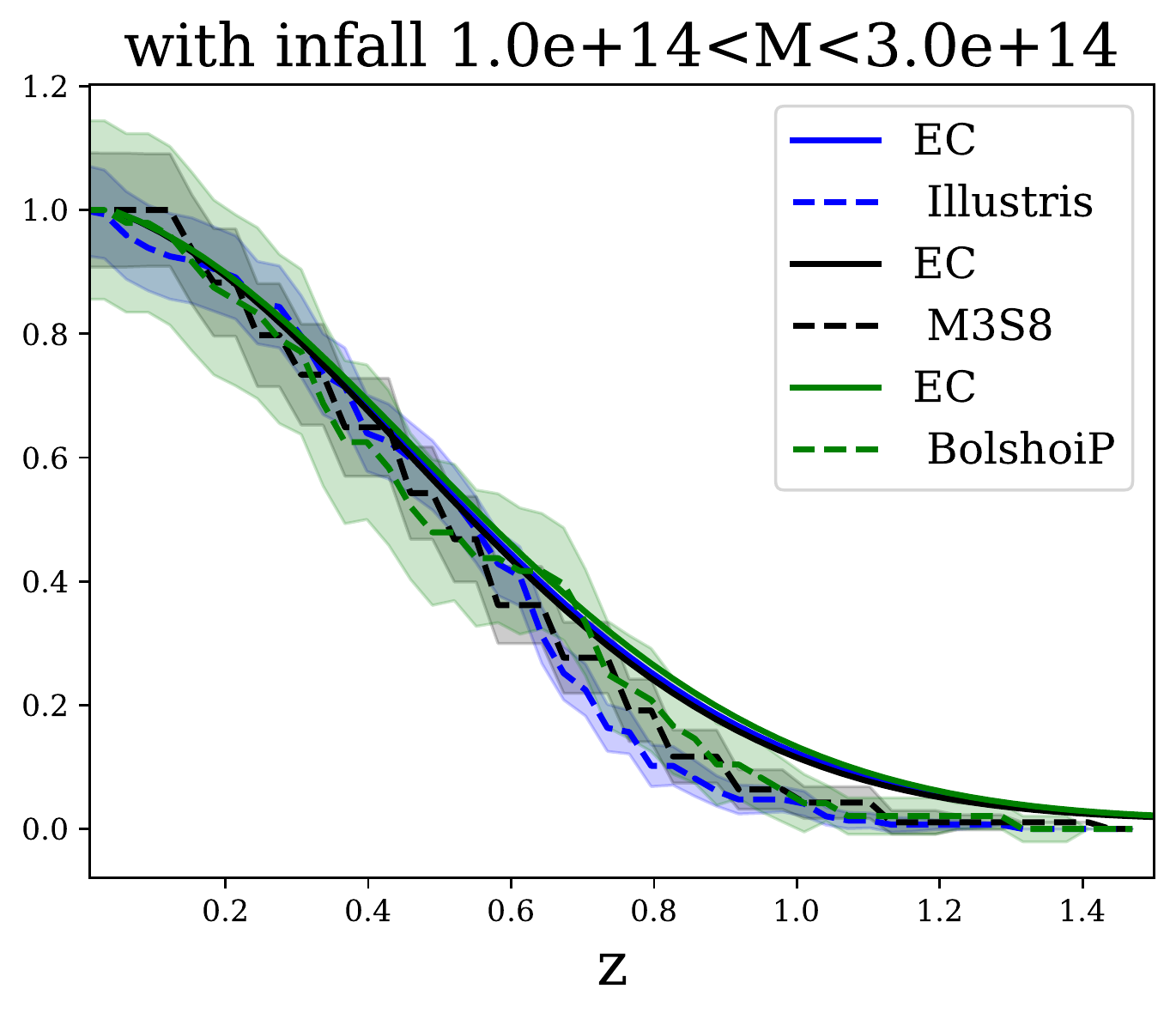}
    \caption{Differential and cumulative $z_{50}$ distributions, as in  Fig.~\ref{fig:problem}, after taking into account the delay due to infall. }
    \label{fig:with_infall}
\end{figure*}

One possible explanation for this shift lies in the different definitions of merger time assumed. Given a particular merger event, EPS theory takes the corresponding collapse redshift (that is, roughly, the time by which newly-accreted mass has first fallen to the centre of the halo) to be the moment at which a halo's virial mass is said to increase. In contrast, numerical group finders may link haloes when their outer virial surfaces first touch. Thus, numerical mergers may occur up to one infall time earlier than analytic ones. Adding a delay equal to the infall time to the numerical results, we obtain the $z_{50}$ distributions in Fig.~\ref{fig:with_infall}. The discrepancy between the numerical and analytic results is greatly reduced, although some differences remain, as seen most clearly in the cumulative distributions.

% Fig 8

The remaining differences may have several possible explanations. There are slight offsets between the distributions for the three simulations, suggesting 
 that the different halo finders and merger tree algorithms used to analyze them affect the results. A detailed comparison of halo finders and merger tree algorithms, including AHF, Rockstar/ConsistentTrees and Subfind/Sublink, was presented in \cite{Sussing_Knebe_2011}. They highlight a number of significant differences between methods, notably in how they treat fragmentation events and non-monotonic MAHs. We note that the discrepancy between numerical and analytic results is greatest at low mass, so resolution may also play a role. Finally, even the revised EC version of EPS remains an approximate theory, so its predictions may be inaccurate at some level.  
 
\begin{figure*}
    \centering
    \includegraphics[width=0.45\textwidth]{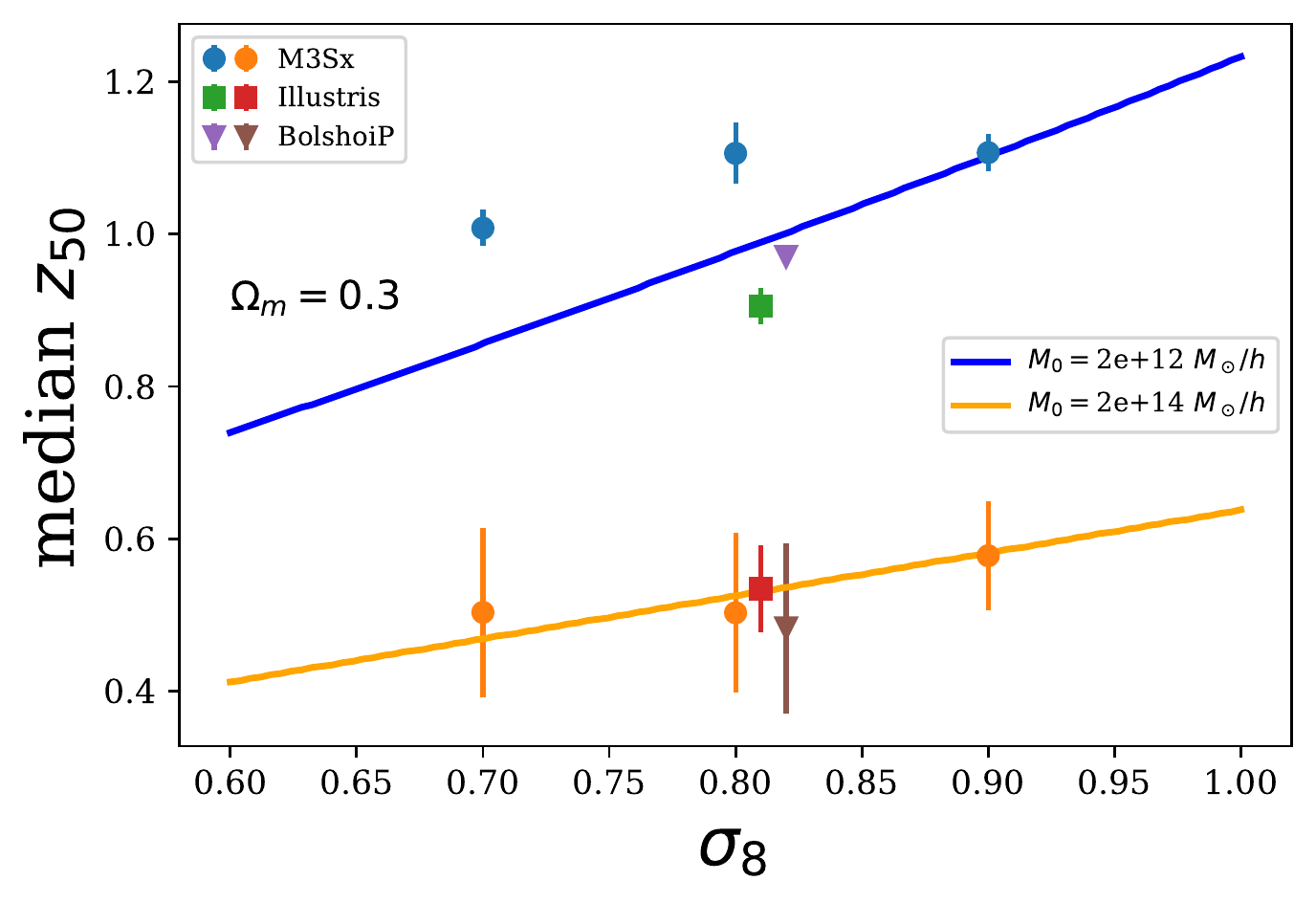}
    \includegraphics[width=0.45\textwidth]{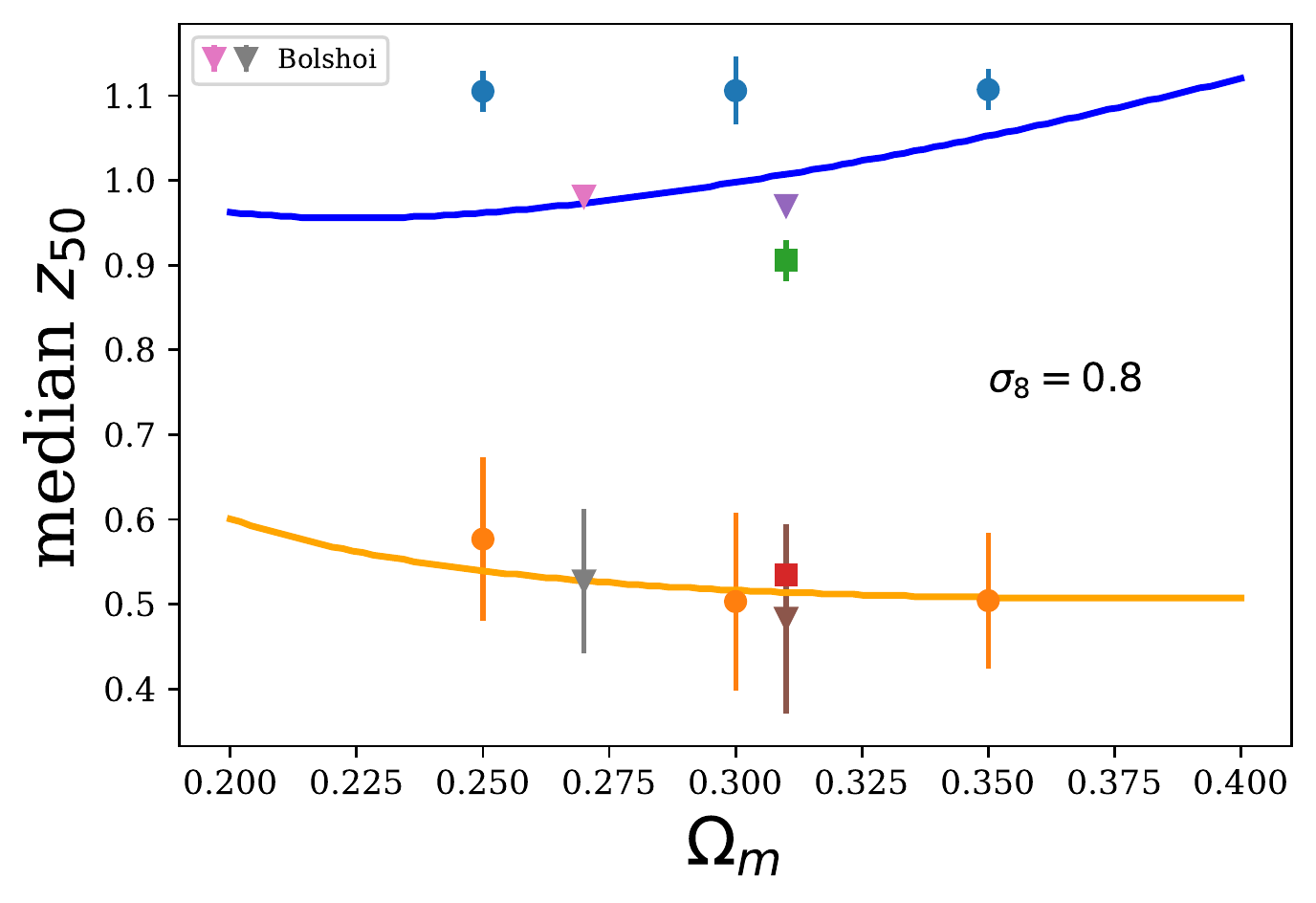}
    \caption{Median value of $z_{50}$ for present-day haloes, as a function of $\sigma_8$ (left panel) and $\Omega_{\rm m}$ (right panel). Smooth curves show the EC prediction; points with error bars show the numerical results and Poisson uncertainties, corrected for infall time. Each symbol represents a different set of simulations and different point colors represent different masses. Note that on the right panel two versions of Bolshoi are shown, one with Planck cosmology and one with WMAP cosmology. See Table \ref{table:sims}.}
    \label{fig:med_z50}
\end{figure*}

We summarize the comparison between numerical and analytic results in Fig.~\ref{fig:med_z50}, which shows the median $z_{50}$ of each of the simulations, together with the uncertainty (points with errorbars),
compared to the analytical predictions from the EC model of \cite{Zhang.2008}. As discussed in Section \ref{subsec:2.4}, age is most sensitive to the amplitude of fluctuations $\sigma_8$, and depends only weakly on $\Omega_{\rm m}$. At high mass, the numerical results agree well with the analytic prediction, at least in terms of the median value of $z_{50}$. Since this mass range is the one relevant for cluster surveys, we conclude that our previous analytic estimates are reasonably valid, although the details of the halo age distribution require further study in future work. 

% Fig 9

%%%%%%%%%%%%%%%%%%%%%%%%%%%%%%%%%%%%%%%%%%%%%%%%%%%%%%%%%%%%%%%%%%%

%Section 4
\section{Observational prospects} \label{sec:observations}

A number of ongoing and future surveys are expected to produce very large samples of galaxy clusters, with $O(10^5)$ significant detections, out to redshifts of $z=1$ or higher \cite[e.g.][]{Pillepich2012,sartoris.2016,CMBS4}.
Supposing such a sample were available, with age information based on one or more observational proxies, we can ask what sensitivity this dataset would have to the cosmological parameters, or equivalently how large a sample would be needed to provide significant improvement on parameter constraints.

To estimate age observationally, we need a structural proxy for age (as expressed, say, by the formation epoch $z_{50}$). There are several known examples of structural properties that correlate with $z_{50}$ \cite[][]{Wong2012}, including concentration \citep{Zhao.2003, Wang2020}, shape (as a product of major mergers -\citealt{Drakos2019a}), substructure \citep[e.g.][]{Gao2004,TB2005,Diemand2007}, or overall degree of relaxation, as measured by a centre-of-mass offset \citep{Maccio2007,Power2012}. We will take concentration as an example here, as its age dependence is the best studied. We note that on some mass scales and at some redshifts, baryons may have an important effect on halo structure, and on concentration specifically. We will start by discussing concentration measurements ignoring these possible effects, but then consider them separately in Section \ref{subsec:4.4} below.

\subsection{Mean Concentration versus \texorpdfstring{$z_{50}$}{z50}}

Since the discovery of the universal density profile \cite[][NFW hereafter]{NFW.1997}, the value of the concentration parameter $c=r_{\rm vir}/r_{\rm s}$ has been linked to the halo's formation history. In NFW, concentration depends on how early a critical fraction of the final mass was first assembled into (any number of) progenitors. Subsequent models \citep[e.g.][]{Bullock.2001, Wechsler.2002, Zhao.2003, Zhao.2009, Ludlow2014, Correa2015a} 
related concentration instead to the growth history of a single main progenitor, as expressed by the mass accretion history (MAH). In the simplest picture \citep{Wechsler.2002, Zhao.2003}, $c \simeq$ $c_0 (a_0/a_r)$ where $a_r$ is the scale factor at the end of the period of rapid growth in the MAH, and $c_0 \sim 3$--4 is the concentration of newly-formed systems at this time. In these models, there is, therefore, a direct correlation between $c$ and $z_{50}$, or any similar estimate of the formation epoch $z_{\rm f}$ \citep{Wong2012}. 

All of these models focus on the relation between the average growth rate and the mean concentration of a sample of haloes of a given range of mass and redshift (although \citealt{Ludlow.2013} does consider the connection between individual MAHs and concentration values.) Major mergers can lead to large variations in concentration, however, depending on the net input of (orbital) energy \citep{Drakos2019b}. Most recently, \cite{Wang2020} have shown that the measured value of the concentration parameter oscillates during major mergers, and that these fluctuations may dominate the statistics of the average values measured for large ensembles. Clearly, the subject is complicated and requires further study; we will not consider it in further detail here, but will assume a correlation between $c$ and $z_{50}$, that makes mean concentration measurements sensitive to mean age.

\begin{figure}
    \centering
    \includegraphics[width=0.48\textwidth]{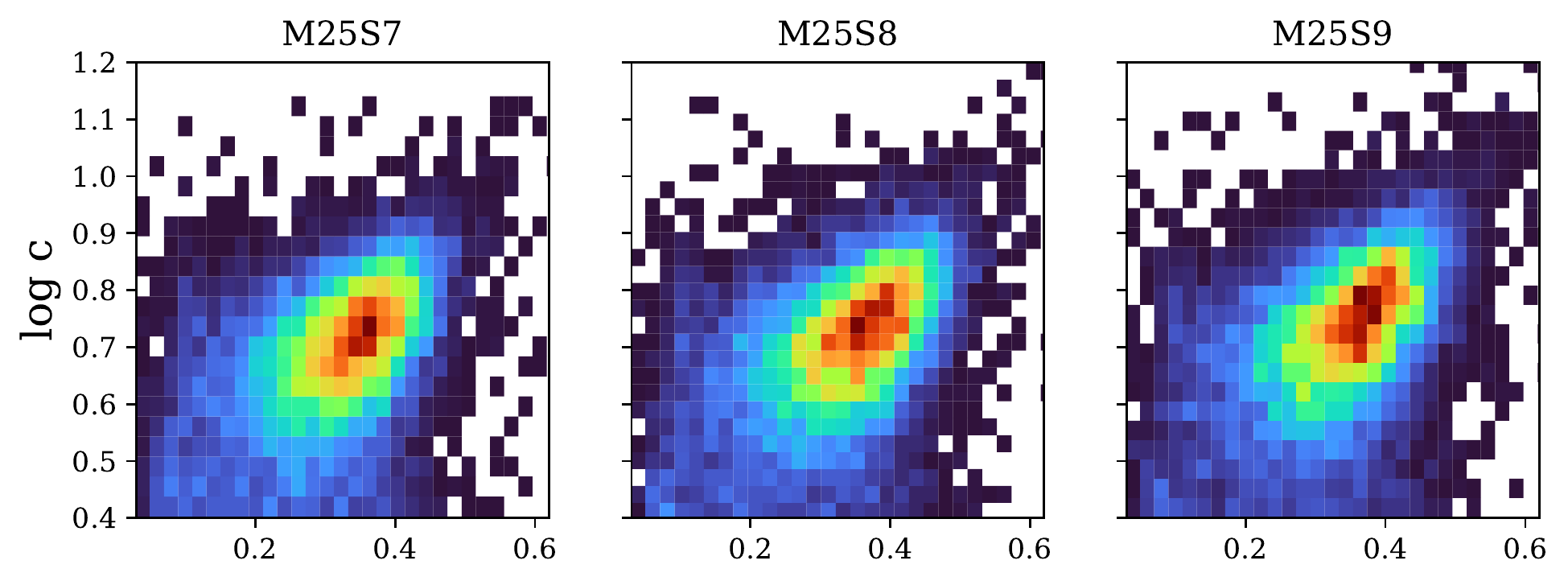}
    \includegraphics[width=0.48\textwidth]{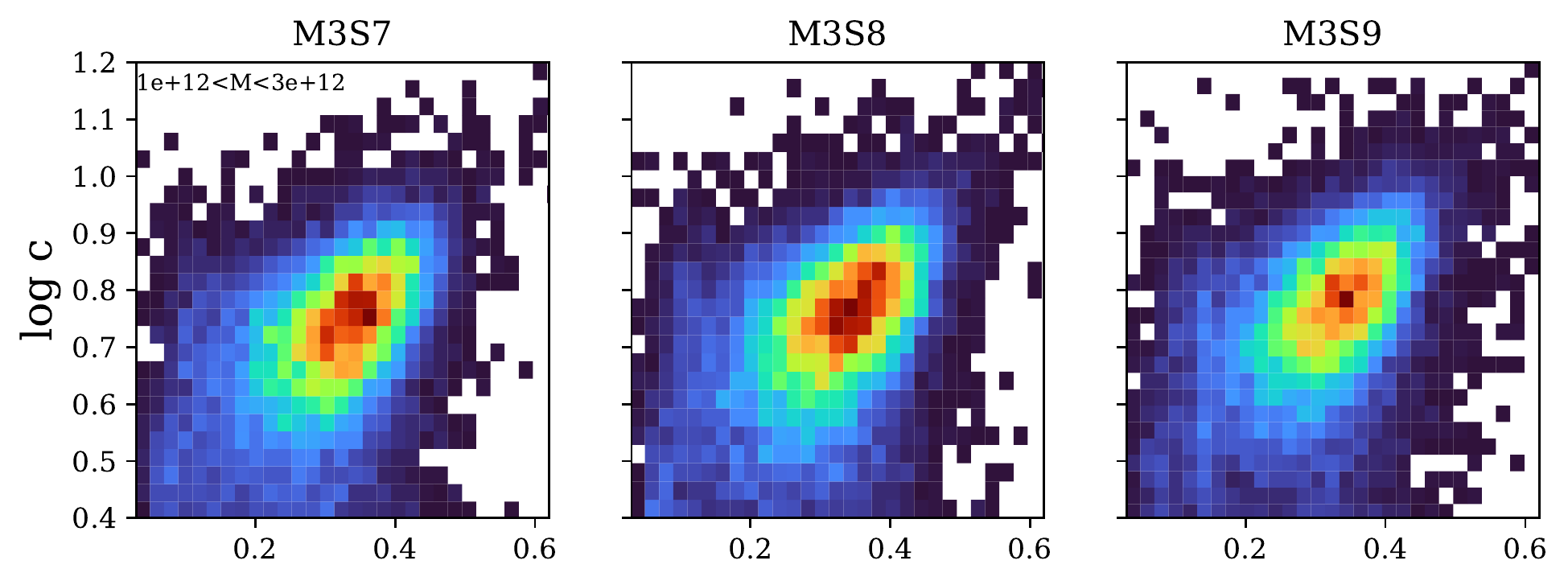}
    \includegraphics[width=0.48\textwidth]{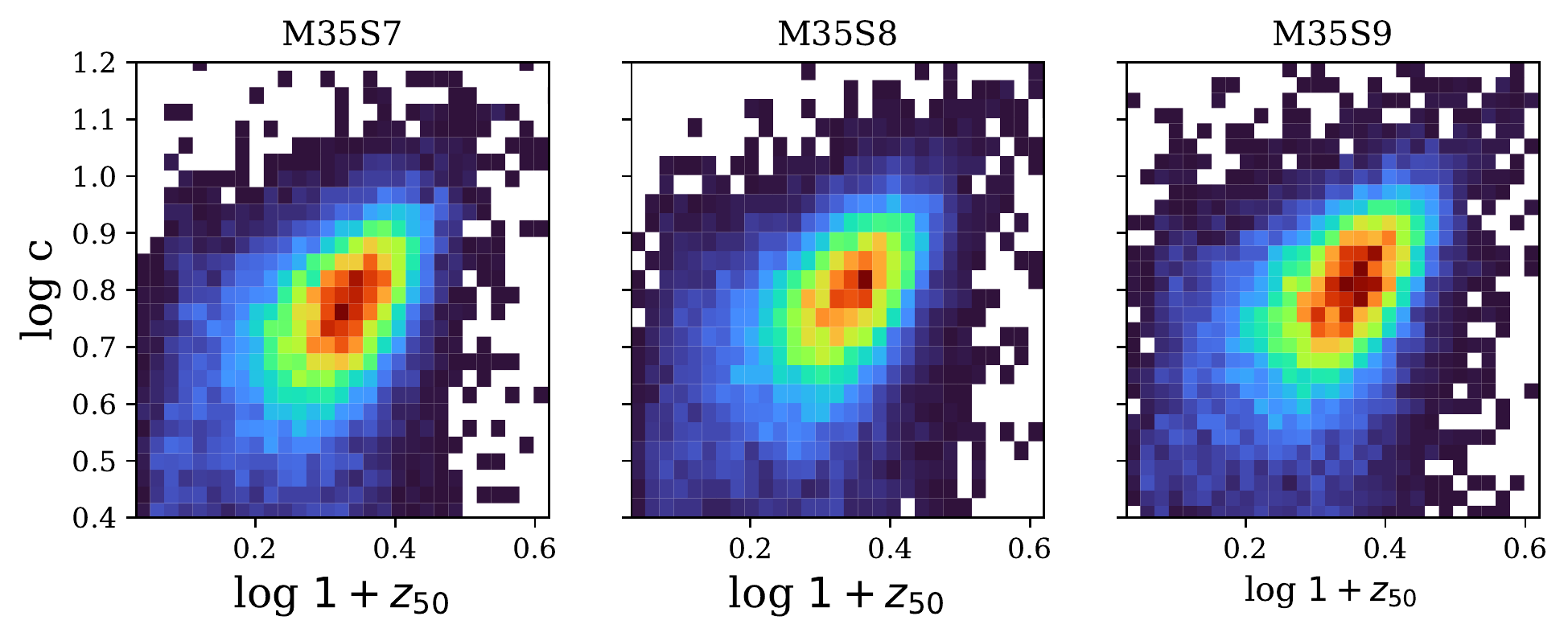}
    \caption{Concentration versus $z_{50}$ for present-day haloes with masses between $10^{12}M_\odot$ and $3\times 10^{12}M_\odot$, in the 9 MxSy simulations.}
    \label{fig:logc_logz50_m12}
\end{figure}

Fig.~\ref{fig:logc_logz50_m12} shows this correlation in practice, as measured from our grid of MxSy simulations for different cosmological parameters (we have chosen a lower halo mass range, 1--3$\times 10^{12}M_\odot$, 
to reduce Poisson noise in the figure). The basic pattern is similar for each set of cosmological parameters, and has been explored extensively in the literature \citep[e.g.][]{Wechsler.2002, Zhao.2003, Zhao.2009, Giocoli.2012, correa.c.2015, Ludlow.2013}, though interestingly, there is also a slight change in the mean relation over the range of parameters explored. In particular, the intercept of the linear regression relation between $\log c$  and $\log (1+z)$ increases monotonically, both with $\Omega_m$, and with $\sigma_8$. This indicates that the concentration has an additional cosmological sensitivity, beyond its main dependence on formation history.
Here too, there is clearly further complexity to explore in the concentration-mass-redshift relation; in future work, we will focus on understanding and calibrating the mean $c(z,M)$ and $c $--$z_{50}$ relationships, and consider more generally the links between concentration, mass accretion history, and cosmology. For the purpose of our present calculations, we will assume a power-law correlation with a fiducial scatter of 30\%, which provides a reasonable fit to the results from all nine cosmologies.  

% Fig 10

\subsection{Measured Concentration versus Mean Concentration}

From the preceding results, the actual 3D concentration of an individual halo should scatter by $\sim30$\%\ for a given value of $z_{50}$. This actual concentration can be estimated in various ways, including weak lensing convergence, detailed modelling of the lensing potential in strong lensing systems, X-ray or SZ emission, or even the galaxy distribution within a group or cluster. Each of these techniques will add observational errors and biases; see, for instance, \cite{Groener2016}, which provides a fairly recent review of individual concentration estimates in clusters. Generally, the observational errors are 0.1 or 0.2 dex, i.e.~25--60\%, for each individual system. There are also systematic uncertainties, both identified and unidentified, associated with each method. In principle, future work with large samples, dedicated simulations, and comparison between observational modalities may help reduce these. Overall, we will assume typical errors of either 30\%\ or 50\%\ in going from an actual 3D concentration to an observational estimate. 

Combining these errors with the intrinsic scatter in the $c$--$z_{50}$ relation, we expect a net scatter of $\epsilon_{z_{50}} = $\,40--60\%\ in the relation between an observational estimate of concentration and the formation epoch $z_{50}$. This large uncertainty makes individual measurements relatively uninteresting; we can compare the situation to weak gravitational lensing, however, where shape measurements for individual galaxies are extremely noisy, but careful averaging extracts the mean value in an unbiased way.

\subsection{\texorpdfstring{$z_{50}$}{z50} versus Cosmology}
\label{subsec:4.3}

The remaining factor in our calculation is the connection between an estimate of the mean value of $z_{50}$ and the values of the cosmological parameters. From Fig.~\ref{fig:med_z50}, to achieve a nominal precision of 0.01 in $\sigma_8$, we need 0.55\%\ precision in the estimate of $\langle z_{50} \rangle$. Assuming unbiased averaging over a sample of $N$ clusters, $\sigma \langle z_{50}\rangle = \epsilon_z/\sqrt{N}$. Solving, we get $N = (0.40/0.0055)^2$--$(0.60/0.0055)^2 \sim$ 5,000--12,000. Thus, with low-precision but unbiased concentration measurements for $O$(10,000) clusters, we could obtain constraints on the value of $\sigma_8$ that 
correspond to 1/10 or less of the current range of uncertainty in this parameter.

\subsection{Baryonic Corrections}
\label{subsec:4.4}

A major uncertainty in the preceding calculations is the net effect of baryons on cluster concentrations. Baryons 
may alter the halo density profile, increasing halo concentration via adiabatic contraction, or reducing it through outflows driven by stellar or AGN feedback. These effects are complex, and depend on mass, redshift, and radius within the halo. Overall, these processes can affect halo concentration and masses significantly \citep{Debackere.2021}. Although simulations suggest that baryonic effects are largest in galaxy-scale haloes \cite[e.g.][]{Velliscig.2014}, they may still be significant when measuring concentration or internal structure in clusters -- \cite{Debackere.2021}, for instance, find a 10\% bias in estimates of the scale length $r_s$ at masses of 5$\times 10^{14} h^{-1}M_\odot$.

Baryonic effects are not yet well enough understood to include in our predictions; in particular, their detailed dependence on the cosmological parameters is not yet known. We can point out a few possible avenues, however, to calibrating and correcting for their effects on structural measurements. First, simulations now model these effects with increasing accuracy, allowing the potential for calibration of any net bias in structural properties. Second, observations of nearby, well-studied systems allow verification of the simulations, independent of the samples used for cosmological tests. Third, as pointed out in Section \ref{subsec:2.4}, over some mass and redshift ranges, cluster age and abundance are predicted to vary almost identically with the cosmological parameters $\Omega_{\rm m}$ and $\sigma_8$. This should provide an independent test of any assumed concentration-mass relationships, at least for this range of mass and redshift, as the constraint in the  $\Omega_{\rm m}$--$\sigma_8$ plane derived from age/concentration measurements must agree with the one derived from abundance, and since the contours for the two are parallel, there is little room for bias in one relative to the other.

A final, and basic, reason for optimism is the differential nature of structural tests, whether based on concentration or on other structural properties. These would depend on the relative distribution of structural properties, measured across a population of systems. A simple proxy for the mean age of the sample, for instance, might be the number of high-concentration systems, relative to low-concentration ones. Thus, to lowest order, a net shift in concentration for the whole population would largely cancel out, reducing the bias in the final results. At the same time, the {\it shape} of the measured concentration distribution for the whole sample would provide another test of the consistency of the method, and any biases or effects due to sample selection.

Overall, it is clear that the impact of baryonic effects on halo structure requires much more detailed study, to see whether and to what degree they would compromise structural tests of cosmology, for a given survey and methodology.

\subsection{What is Achievable?}

In summary, in previous sections, we have shown that with low-precision but unbiased concentration estimates for $O(10^4)$ clusters, one could obtain excellent constraints on $\sigma_8$, assuming the net effect of baryons is small, or can be corrected using simulations. While measuring concentration observationally is challenging, our method does not require particularly accurate measurements for individual systems -- given the intrinsic scatter in the $c$--$z_{50}$ relation, the errors in individual estimates need only be accurate at the $\sim$30\%\ level.

 We can consider this goal in more detail for a particular survey. The {\it Euclid} mission is a 1.2m space telescope, operating in the visible and near-infrared (NIR). As part of its wide survey, it will image 15,000 deg$^2$ of the sky in one optical and three NIR bands, detecting galaxies down to an AB magnitude of 24 or fainter. Photo-zs will be derived for these objects, using ancillary data from ground-based surveys such as UNIONS \citep{UNIONS2020}. The {\it Euclid} wide survey should provide large catalogues of clusters detected photometrically (that is by clustering in projection and in photo-z space), and also as peaks in weak lensing maps. Based on the forecasts of \citep{sartoris.2016}, the photometric detections should include all clusters with masses $M\gtrsim 10^{14}M_\odot$ out to redshift $z\sim 2$. Below redshift z = 0.5, the wide survey is expected to detect 1.5 million clusters at 3$\sigma$ or greater significance, and 200,000 clusters at 5$\sigma$ significance. Extrapolating from these predictions, the number of 7$\sigma$ detections is in excess of 20,000 (with half that number at redshifts $z < 0.7$). Admittedly, these objects may be slightly more massive than the example considered above in section \ref{subsec:4.3} ($3.5\times10^{14} M_\odot/h$, versus $2\times10^{14} M_\odot/h$) but they are close enough that the slope of the $\langle z_{50}\rangle$--$\sigma_8$ relation should be similar. 
 
{\it The Roman Space Telescope} mission is a 2.4m wide-field space telescope, with optical/NIR imaging and slitless spectroscopy capabilities. Its High Latitude Survey will image $\sim$2000 deg$^2$ of sky in four NIR bands, reaching depths 1--2 AB magnitudes deeper than Euclid Wide, as well as providing slitless spectroscopy of brighter targets.
 This deeper data over a smaller area should produce a cluster sample that extends to lower masses and higher redshifts, and thus provides an interesting counterpoint to Euclid data. In particular, we note that at high redshift, the complementarity of age and abundance is reduced (cf.~the left-hand panels of Fig.~\ref{fig:cross_constraint}). This could provide an important consistency check on age estimates, as discussed previously. Finally, a number of other forthcoming experiments expect to detect large numbers of clusters, including \emph{eROSITA} \citep{Pillepich2012} in the X-ray, CMB-S4 \citep{CMBS4} in the mm, and the ground-based UNIONS \citep{UNIONS2020}, DESI \citep{DESI2016}, and Rubin LSST \citep{LSST2009} surveys.
 
Overall, we conclude that multiple samples of $O(10^4)$ clusters with sufficient signal-to-noise ratio (SNR) to allow structural measurements should become available in the near future. One could imagine using a large, uniform survey such as \emph{Euclid} for the low-redshift sample selection,  together with other complementary observations to make structural measurements on individual clusters. A high-redshift sample could then be used to test and calibrate age proxies, as mentioned above. It might also be possible to stack clusters to use a mean projected density profile, measured at high SNR, to derive constraints. We will consider these and other approaches in future work.

 As in weak lensing studies, the challenge of averaging over large numbers of low SNR measurements will be in controlling for and reducing systematics. Beyond the baryonic effects discussed in the previous section, systematics related to basic structure formation could also affect sample selection (e.g. by preferentially highlighting or neglecting disturbed systems); they could bias individual mass estimates (although the slope of the concentration-mass relation is fairly shallow, so accurate masses are less important than in abundance studies); or they could bias concentration measurements, e.g.~by biasing the sample selection towards objects with a particular 3D shape, or with a disturbed IGM (if the confirmation or structural measurements are performed in the X-ray).  In lensing-based studies, false peaks and projections could be a particular problem, as these may look less regular and have lower concentrations, biasing the average. Environment can also have an impact on formation time through assembly bias, as simulations have shown that haloes form earlier in dense environments \citep{gao.white.springel.2005, Wechsler.2006, Harker.2006}, so unbiased sampling of large volumes is important. Here again, there is much future work to be done considering the the possible biases for different observational modalities and survey strategies.

%%%%%%%%%%%%%%%%%%%%%%%%%%%%%%%%%%%%%%%%%%%%%%%%%%%%%%%%%%%%%%%%%%%

%Section 5
\section{Conclusion} \label{sec:conc}

The enormous success of CMB analyses and large cosmological surveys over the last few decades has been driven, for the most part, by a robust and detailed understanding of structure formation in the linear regime. Cluster number counts are an important exception, but they only probe one limited aspect of non-linear structure formation. As cluster catalogues grow by several orders of magnitude in size over the next decade, it is worth considering what other cosmological information we might extract from them. Measurements of internal halo structure can, in principle, tell us about the rate of non-linear structure formation, and are worth considering as a next-generation cosmological test. 

Previous work has established that as haloes grow through hierarchical merging, this process leaves structural signatures that can last for many dynamical times, that is, for many Gyr at low redshift. As a result, structural measurements provide several different avenues to estimate cluster assembly times or ``ages''. In this paper, we have shown that for typical clusters at $z < 1$, age varies almost orthogonally to abundance in the space of the cosmological parameters $\Omega_{\rm m}$ and $\sigma_8$. The same datasets that provide abundance constraints could be used to estimate mean values for structural parameters, and thus age, providing significantly tighter parameter constraints from a single set of observations. 

Of course, given the accuracy of current constraints from the CMB, it may seem less interesting to invest further in other techniques. A survey of current results hints at tension between the different measurements, however, emphasizing the importance of redundant cosmological tests, over different ranges of redshift, mass and/or spatial scale. To resolve the deep mysteries of dark energy and dark matter, and to rule out yet-undiscovered variations on the current cosmological model, we need to test it as sensitively as possible, across as broad as possible a range of parameter space. In pursuit of this goal, our growing understanding of non-linear structure formation will open up many exciting possibilities for new tests and new tools. 

\section*{Acknowledgements}

JET acknowledges support from the Natural Sciences and Engineering Research Council (NSERC) of Canada through a Discovery Grant. NED acknowledges support from NASA, through contract NNG16PJ25C. We thank the authors of the Illustris TNG and Bolshoi/BolshoiP simulations, and the halo finders and merger tree codes cited in section 3.1, for making their data and codes publicly available. We thank the scientific editor and reviewer for suggestions that improved the final paper.

\section*{Data Availability}
N-body simulation data from the Bolshoi simulation is publicly available (after registration) at \url{https://www.cosmosim.org/}. N-body simulation data from the Illustris TNG simulation is publicly available (after registration) at \url{https://www.tng-project.org/}. The rest of the data underlying this article will be shared on reasonable request to the corresponding author.
\bibliographystyle{mnras}
\bibliography{mybib}

\appendix

%Appendix A
\section{Details of the Analytic Calculations}
\label{sec:Appendix_A}
We use standard tools and techniques for the analytic calculations in Section~\ref{sec:analytic}. In particular, the growth factor $g(z)$ is calculated using the approximation given by \cite{Carroll.1992}, which is accurate to a few percent: 
\begin{equation}
    \label{growth approx}
   g(z) \approx \frac{5\Omega_m(z)}{2\left[\Omega_m^{4/7}(z) -\Omega_\Lambda(z) + (1+\Omega_m(z)/2)(1 + \Omega_\Lambda(z) /70)\right]}\,. 
\end{equation}

The critical overdensity $\delta_c$ is the
value a linearly-extrapolated density perturbation needs to reach to collapse and form a virialized object, and can be estimated using the spherical collapse model. Its present-day value varies weakly with $\Omega_m$ \citep{mbw2010}: 
%§ 5.1 \citep{lacey.cole}
\begin{equation}
\label{critical overdensity eq}
    \delta_c = \frac{3}{5}\left(\frac{3\pi}{2}\right)^{2/3}\Omega_m^{0.0055} \approx 1.686\Omega_m^{0.0055}\,.
\end{equation}

The linear matter power spectrum 
\begin{equation}
    P(k) = Ak^{n_s}T(k)
\end{equation}
is computed using the approximation to the transfer function $T(k)$ given by Equation 16 in \cite{Eisenstein.Hu.1998}.
We take the primordial amplitude to be A=1 initially, and then adjust this value retroactively to set the correct value for $\sigma_8$. The index of the primordial spectrum is taken to be $n_s = 0.965$.

The variance of the density fluctuation field, $\sigma^2$, is computed numerically by convolving the power spectrum $P(k)$ with a top hat smoothing filter: 
\begin{equation}
    \sigma^2(R) = \frac{1}{2\pi^2} \int_0^\infty k^2 P(k) \widetilde{W}_R^2(k)dk\,,
\end{equation}
where 
\begin{equation}
    \widetilde{W}_R(k) = 3\frac{\sin(kR) - kR\cos (kR)}{(kR)^3}
\end{equation}

We have compared our derived values of $P(k)$ and $\sigma (M)$ to values calculated using the Colossus python package \citep{Diemer2018}, and find good agreement. 

%Appendix B
\section{Survey mass versus \texorpdfstring{$\Omega_{\rm m}$}{Om}}
\label{sec:Appendix_B}

\begin{figure*}
    \centering
    \includegraphics[width=0.45\textwidth]{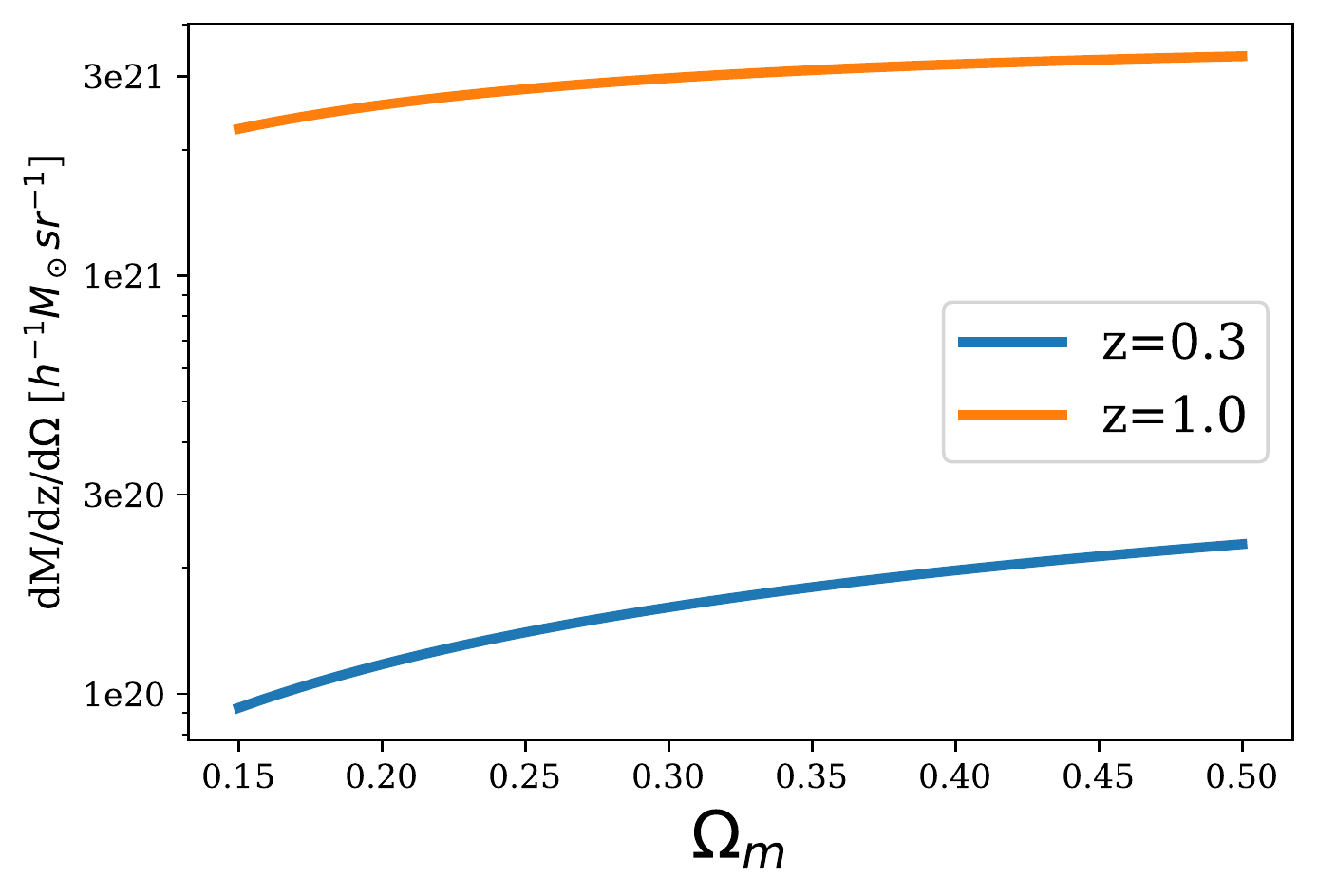}
    \includegraphics[width=0.45\textwidth]{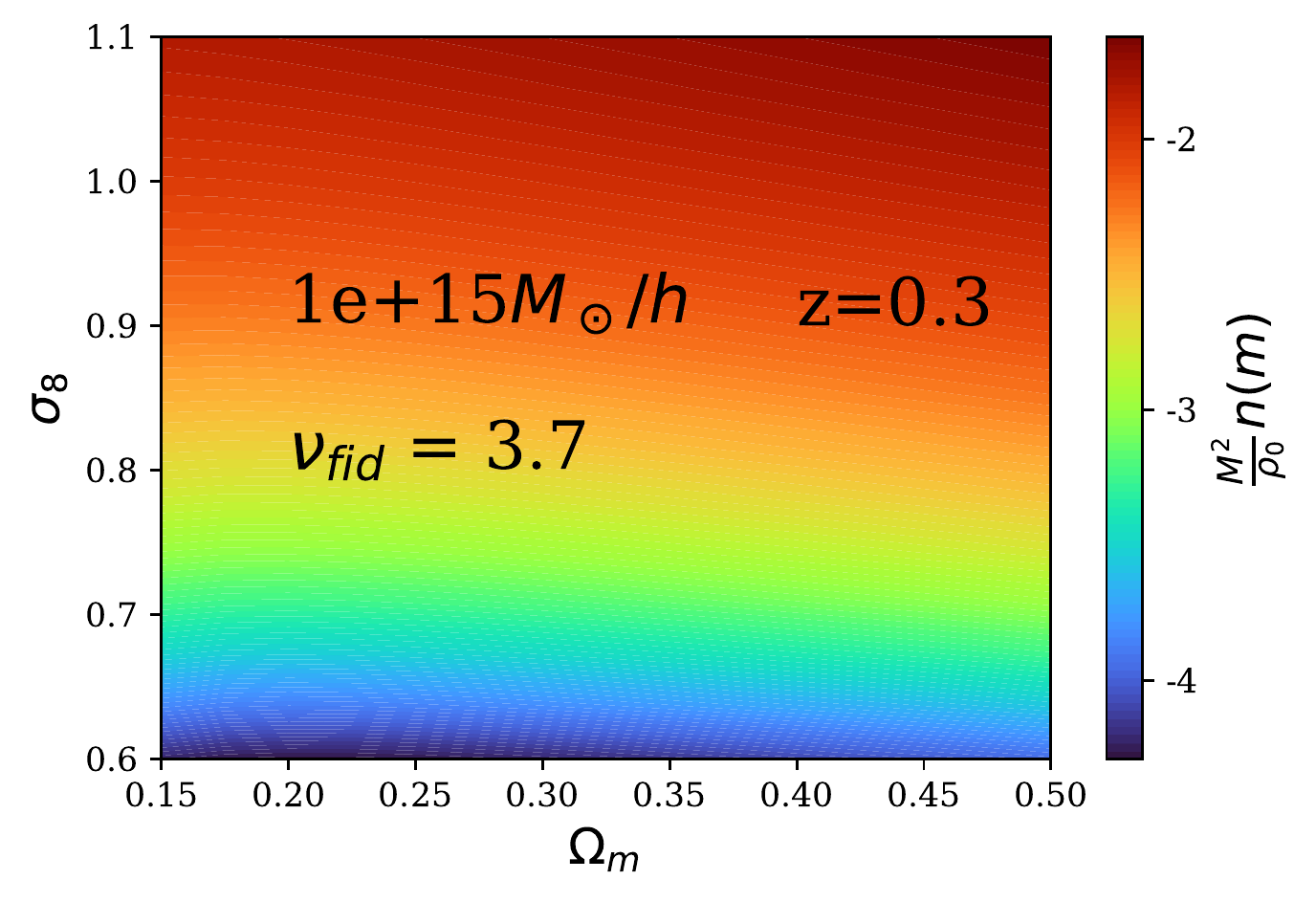} 
    \caption{Left : total mass $M_V$ contained within a survey volume, per unit solid angle and per unit redshift interval, as a function of $\Omega_{\rm m}$, at redshifts 0.3 and 1 (note a flat $\Lambda$CDM cosmology is assumed). Right : Variation of the (EC) collapsed fraction $f_{\rm ST}$ in the $\Omega_{\rm m}$--$\sigma_8$ plane, for the particular choice of halo mass and redshift indicated. Note the similarity to the dependence of peak height (Fig.~\ref{fig:cosmo_peak_height_om_s8_plane}), although the colour scale here is now inverted, and logarithmic.}
    \label{fig:mass_omega_m}
\end{figure*}

Halo abundance depends on the total mass of material sampled in a survey volume, $M_V$, and on the collapsed fraction $f$ at that scale and redshift.
The survey mass obviously depends on the mean matter density $\rho_m$ and thus on $\Omega_m$ directly, but it also depends on 
$\Omega_m$ indirectly, through the volume sampled for a given solid angle and redshift range. The left panel of Fig.~\ref{fig:mass_omega_m} shows the total mass contained within a survey volume per unit redshift per unit solid angle at two different redshifts, as a function of the cosmological parameter $\Omega_{\rm m}$. While the volume element is a decreasing function of $\Omega_{\rm m}$ for flat $\Lambda$CDM cosmologies (since volume at a given redshift grows as $\Lambda$ increases), the total mass enclosed increases overall, with the greatest $\Omega_{\rm m}$ sensitivity at low redshift.

In the end, however, the total mass $M_V$ has relatively little influence on the overall shape of the abundance contours in the $\Omega_m$--$\sigma_8$ plane, because its variation is of order a factor 2 or less, while the collapsed fraction varies over several orders of magnitude as $\Omega_m$ changes (right panel of Fig.~\ref{fig:mass_omega_m}; see Appendix \ref{sec:Appendix_C} for a discussion of the parametric dependence of the collapsed fraction).

%Fig B1

% Appendix C
\section{Peak Height and collapsed fraction versus
\texorpdfstring{$\Omega_{\rm m}$}{Om} and \texorpdfstring{$\sigma_{\rm 8}$}{s8}}
\label{sec:Appendix_C}

We can write peak height as the product of three factors:
\begin{equation}
\nu = \frac{\delta_c(z)}{\sigma(M)} = \frac{\delta_c(z)}{\sigma_8\Gamma(M)}\,,
\end{equation}
where we have defined $\Gamma(M) \equiv \sigma(M)/\sigma_8$. The redshift evolution of the collapse threshold stems from the linear growth of fluctuations
\begin{equation}
\delta_c(z) = \delta_c(0)\frac{D(0)}{D(z)} = \delta_c(0)\frac{a_0g(0)}{ag(z)}\,,
\end{equation}
where $D$ is the fluctuation amplitude, $g$ is the linear growth factor, and $a$ is the scale factor.
Combining these, 
\begin{equation}
\nu = \frac{(1+z)}{\sigma_8}\frac{1}{\Gamma(M)}\frac{\delta_c(0)}{g(z)/g(0)}\,.
\end{equation}
Thus, while peak height scales simply as $1/\sigma_8$, the dependence on $\Omega_{\rm m}$ is through both the shape of the matter power spectrum $\Gamma(M)$ and the relative growth factor $g(z)/g(0)$. We consider each of these in turn.

\subsubsection{$\Gamma (M)$}
\begin{figure*}
    \centering
    \includegraphics[width=0.3\textwidth]{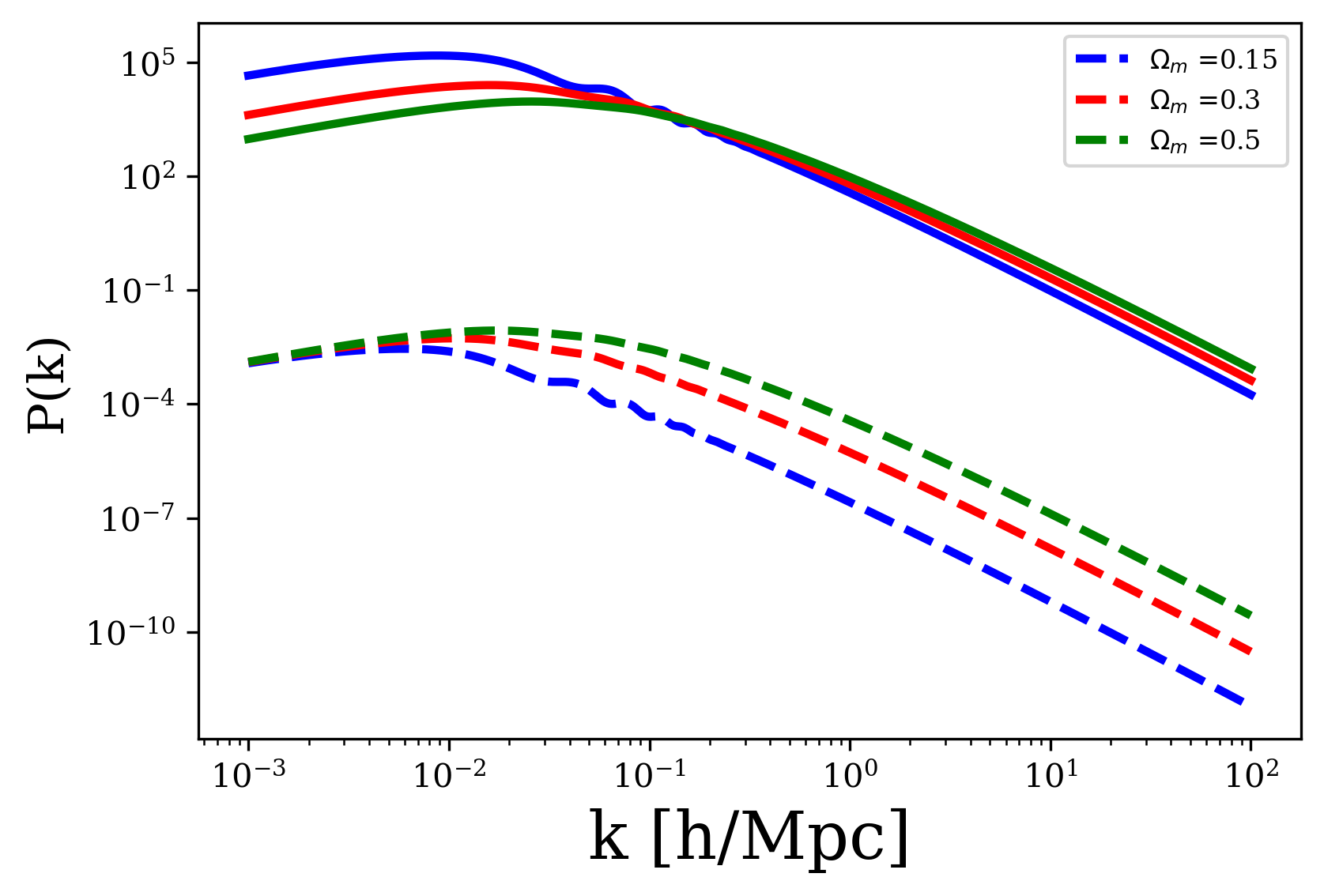} 
    \includegraphics[width=0.3\textwidth]{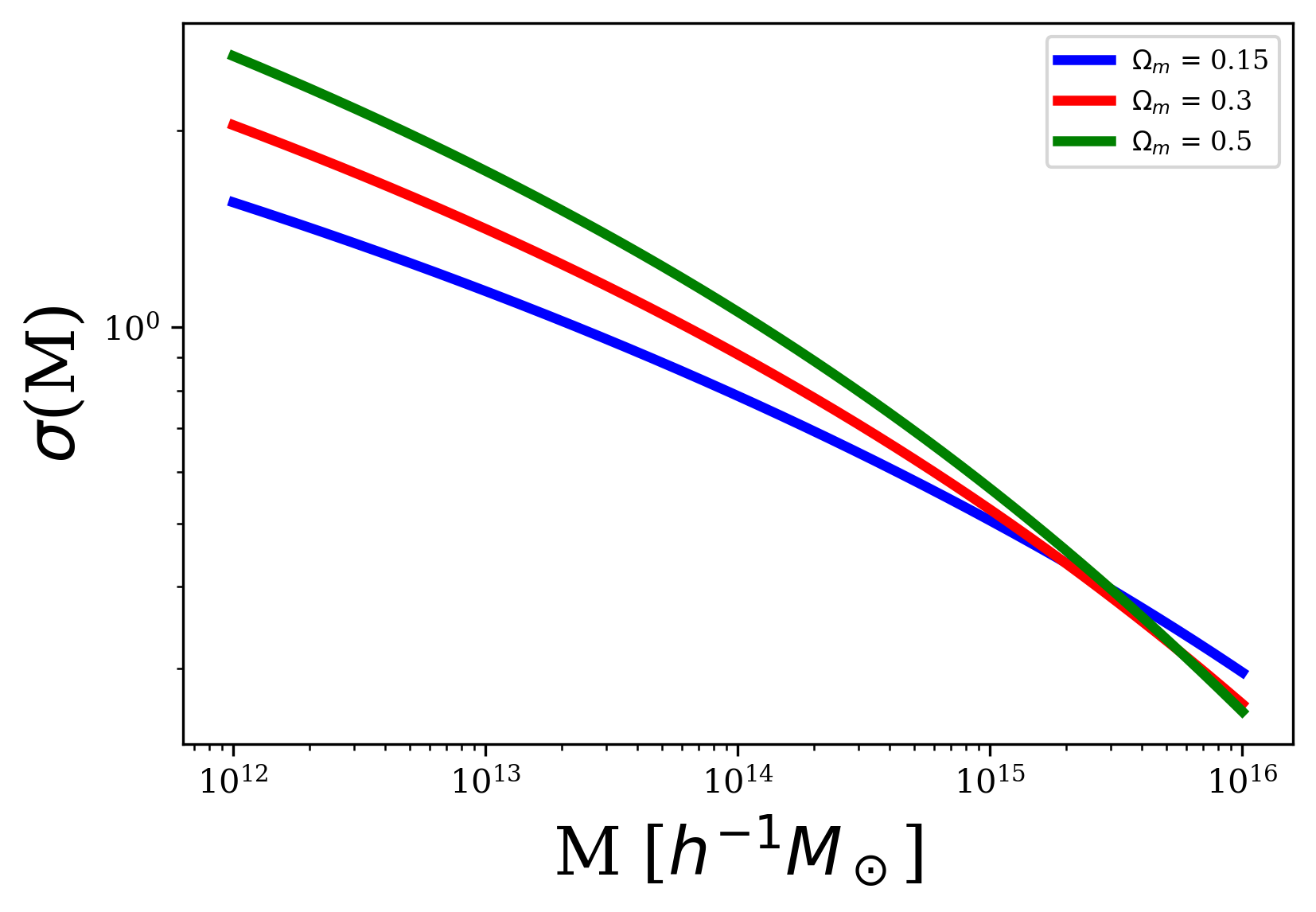}
    \includegraphics[width=0.3\textwidth]{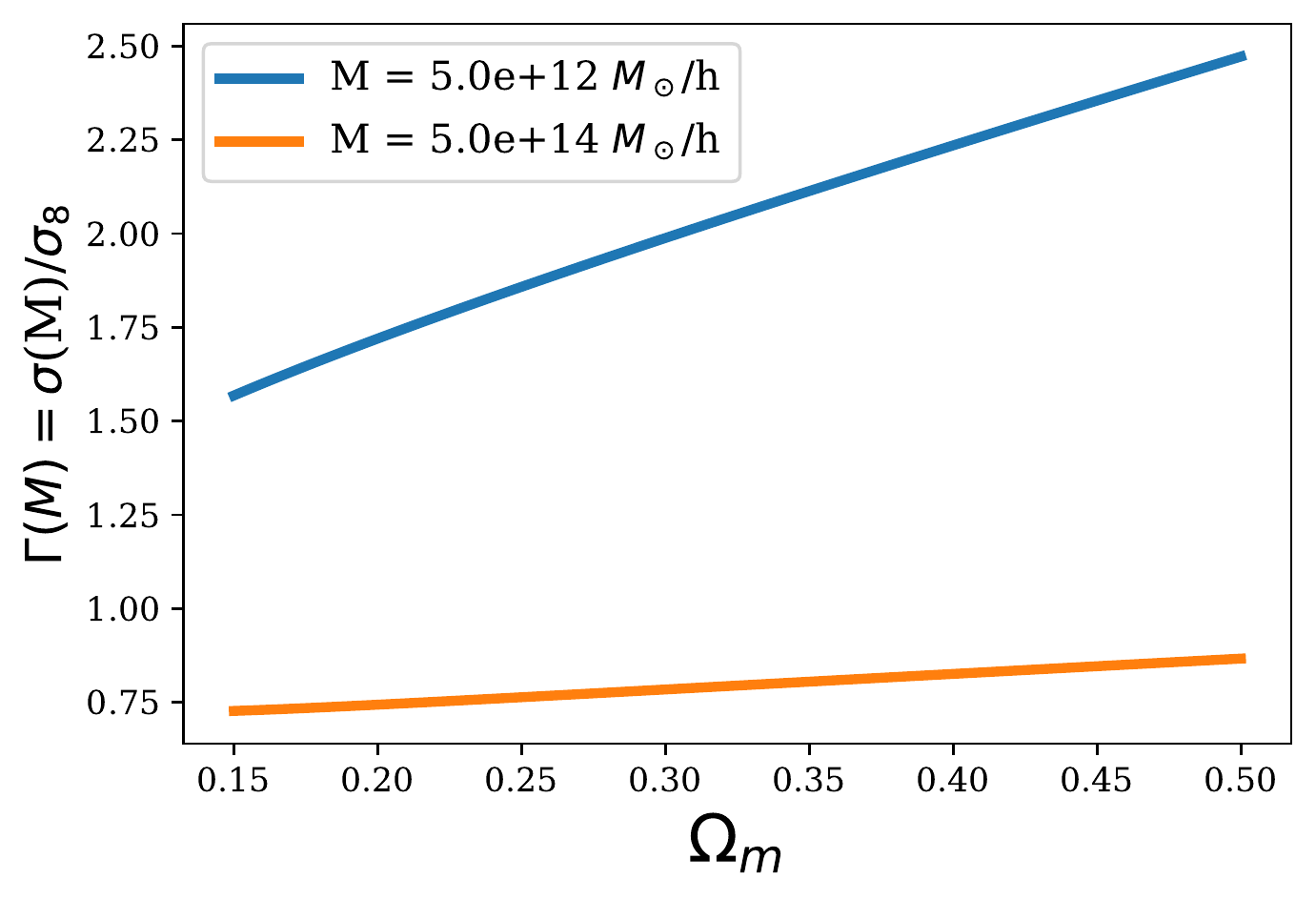} \\
    \includegraphics[width=0.3\textwidth]{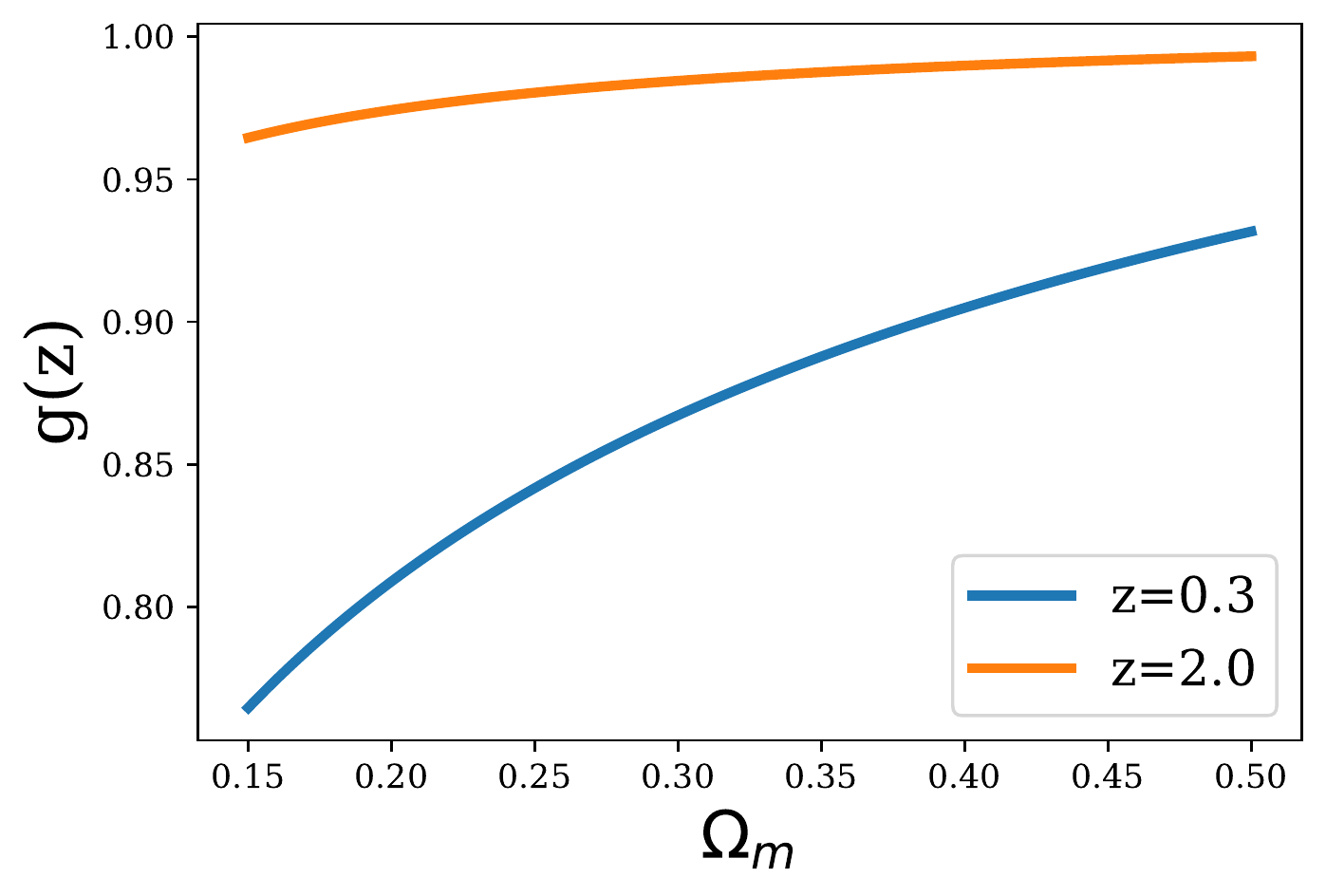} 
    \includegraphics[width=0.3\textwidth]{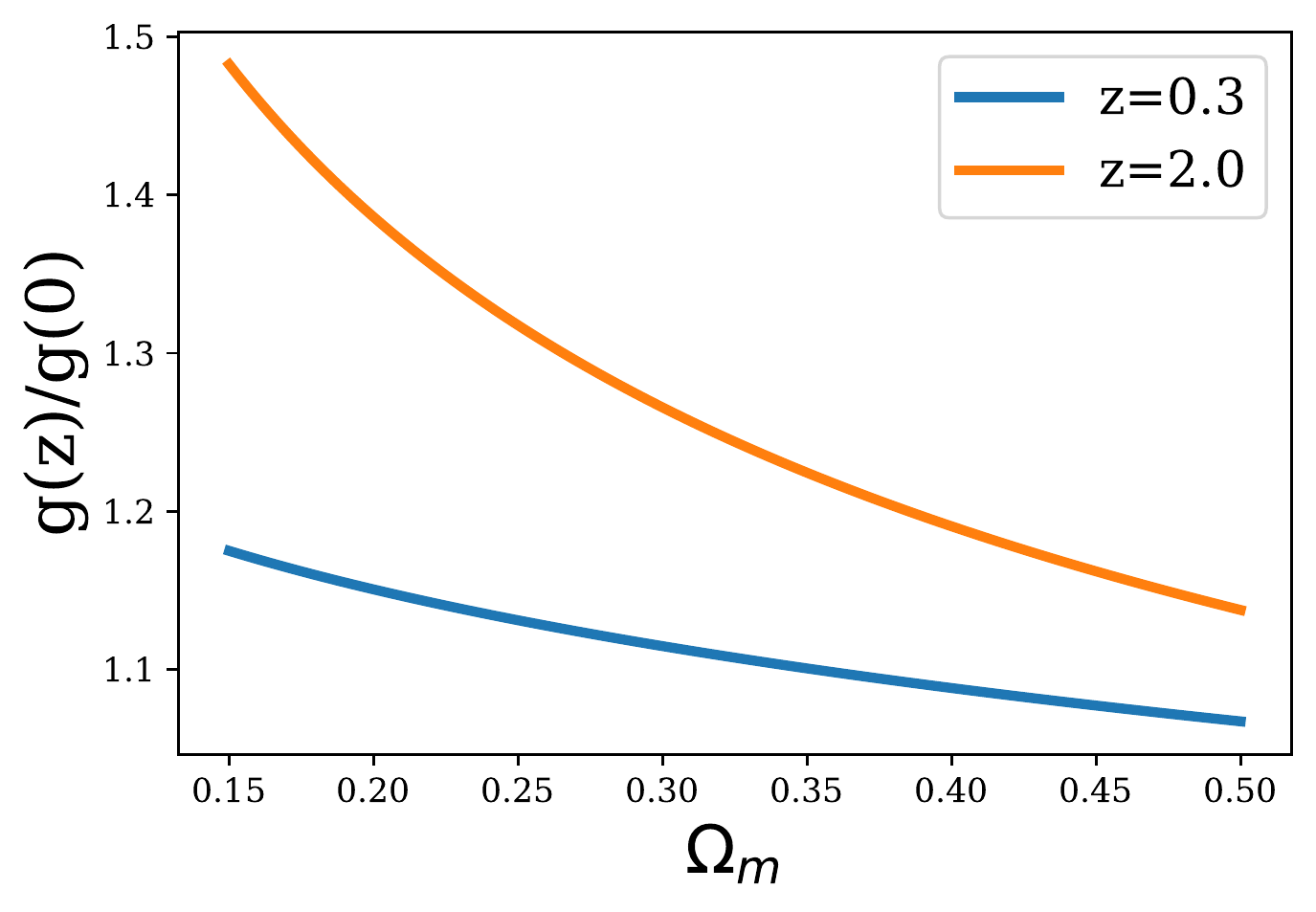}
    \includegraphics[width=0.3\textwidth]{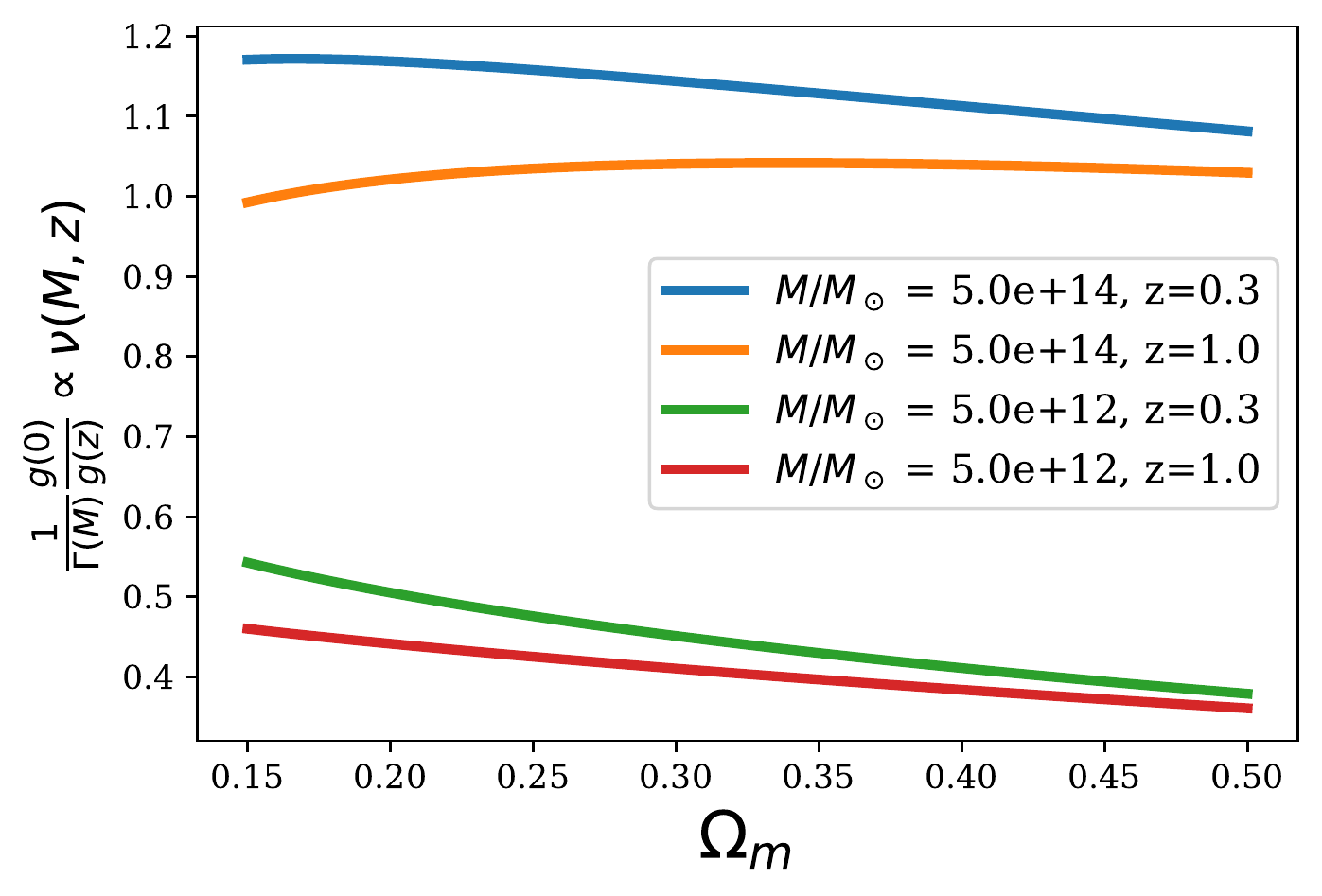}
    \caption{Upper left: Variation of the power-spectrum $P(k)$ with $\Omega_{\rm m}$ at fixed primordial amplitude (dashed lines) and for fixed $\sigma_8$ (solid lines). The upper middle and upper right panels show how $\sigma(M)$ and $\Gamma(M)$ depend on $\Omega_{\rm m}$. The lower left and lower middle panels show how the growth factor and the normalized growth factor vary with $\Omega_{\rm m}$. The lower right panel shows how, as a consequence of these dependencies, the peak height $\nu$ varies with $\Omega_{\rm m}$ at fixed $\sigma_8$, following Eq. \ref{eq:nu_cosmo}.}
    \label{fig:g_ofz_gamma}
\end{figure*}

The function $\Gamma (M)$ describes the shape of the amplitude of fluctuations as a function of mass, $\sigma(M)$, independent of its normalization $\sigma_8$. This shape will depend on the value of the matter density parameter $\Omega_{\rm m}$. More specifically, in cosmologies with larger matter densities,  matter-radiation equality occurs sooner. Growth is suppressed by radiation on all scales below the horizon scale at recombination, but these scales are smaller and spend less time inside the horizon when $\Omega_{\rm m}$ is larger. Thus, for a fixed amplitude $A$ of the primordial power spectrum, small-scale power at recombination will increase with $\Omega_{\rm m}$ (dashed curves in the top left panel of Fig.~\ref{fig:g_ofz_gamma}).

%Fig C1

Fixing $\sigma_8$, the amplitude of fluctuations at scales of $8h^{-1}$Mpc, reduces some of the difference in power (solid curves in the top left panel of Fig.~\ref{fig:g_ofz_gamma}), but the shape of $\sigma(M)$ remains steeper in cosmologies with larger values of $\Omega_{\rm m}$ 
(top middle panel of Fig.~\ref{fig:g_ofz_gamma}).
Thus, over the range of interest, the ratio $\Gamma = \sigma(M)/\sigma_8$ increases with $\Omega_{\rm m}$, especially at lower mass (top left panel of Fig.~\ref{fig:g_ofz_gamma}). We can model this dependence as
\begin{equation}
\Gamma (M) \sim \Omega_{\rm m}^{\beta(M)}\,,
\end{equation}
where $\beta(M)$ decreases with mass, and becomes negative at $M = M_8 \approx 2\times 10^{15} M_\odot/h$ where $\sigma(M) = \sigma_8$.

\subsubsection{Growth factor}
The lower panels of Fig.~\ref{fig:g_ofz_gamma} show how $g(z)$ and $g(z)/g(0)$ vary with $\Omega_{\rm m}$. In a flat $\Lambda$CDM model, the growth factor $g(z)$ at a fixed redshift $z$ is reduced for low $\Omega_{\rm m}$, as dark energy suppresses the growth of fluctuations. The relative amplitude of the effect is greater for very low values of $\Omega_{\rm m}$, or for large redshift ranges, and thus the ratio $g(z)/g(0)$ is {\it largest} for low $\Omega_{\rm m}$ and/or high redshift (middle panel). Mathematically, we can estimate the dependence on the density parameter by using the approximation $g(z) \propto \Omega_{\rm m}^{3/7}(z)$ \citep{Carroll.1992}: 
\begin{equation}
    \frac{g(z)}{g(0)} \propto \left(\frac{\Omega(z)}{\Omega_0}\right)^{3/7}
    \propto\left(\frac{{\bar\rho_m}(z)\rho_c(0)}{{\bar\rho_0}\rho_c(z)}\right)^{3/7}
    \propto \frac{((1+z)^3)^{3/7}}{E(z)^{3/7}} \propto \Omega_{\rm m}^{-\alpha(z)}\,,
\end{equation}
where 
\begin{equation}
E(z) = \frac{H(z)}{H(0)} \sim \sqrt{(1+z)^3\Omega_{\rm m} + \Omega_\Lambda}
\end{equation}
is the Hubble ratio. The index $0<\alpha(z)<1$ is an increasing function of $z$, and approaches the value $3/14$ at high redshift.

\subsubsection{Peak Height}
\begin{figure*}
    \centering
    \includegraphics[width=0.35\textwidth]{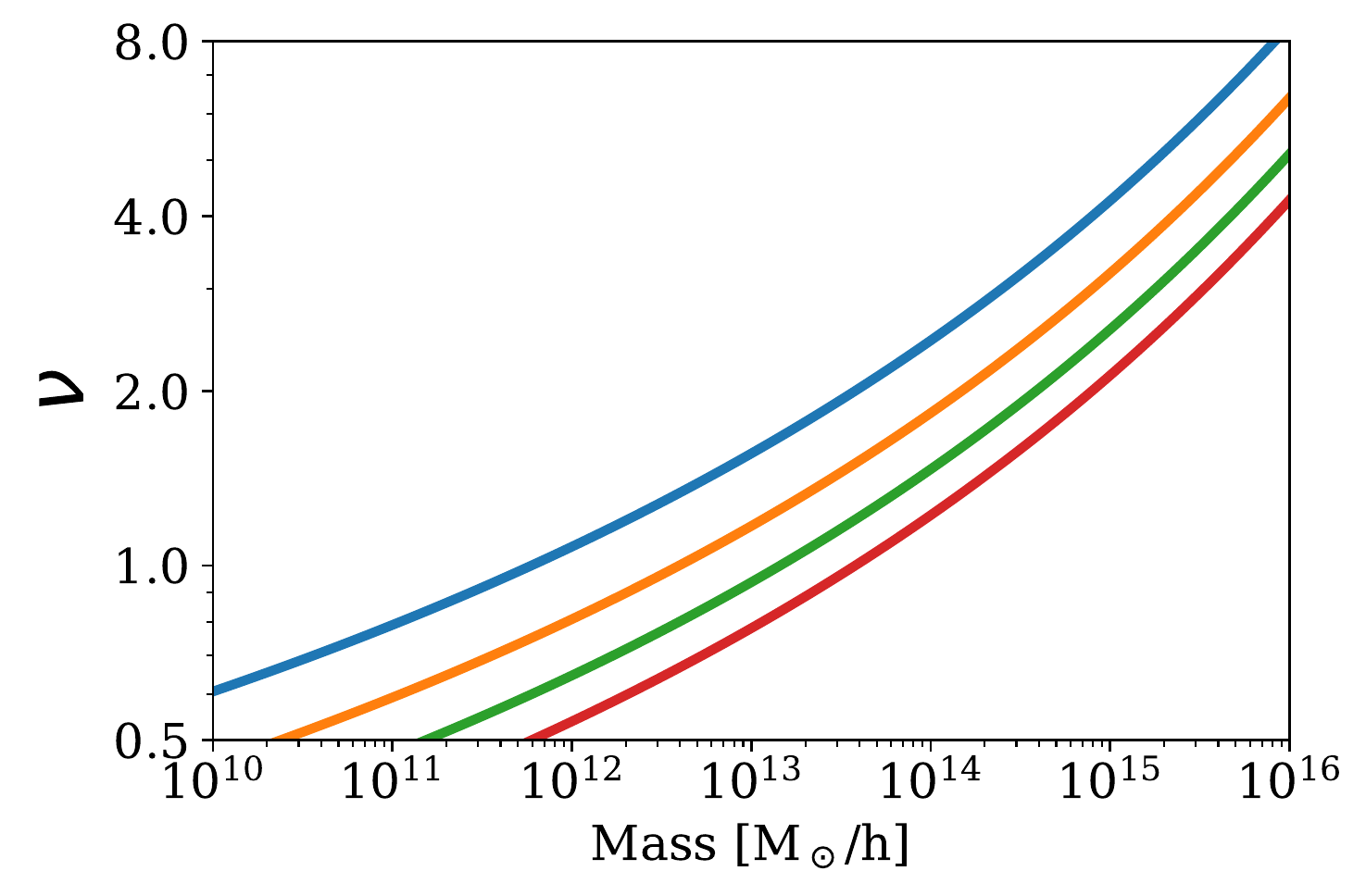}
    \includegraphics[width=0.35\textwidth]{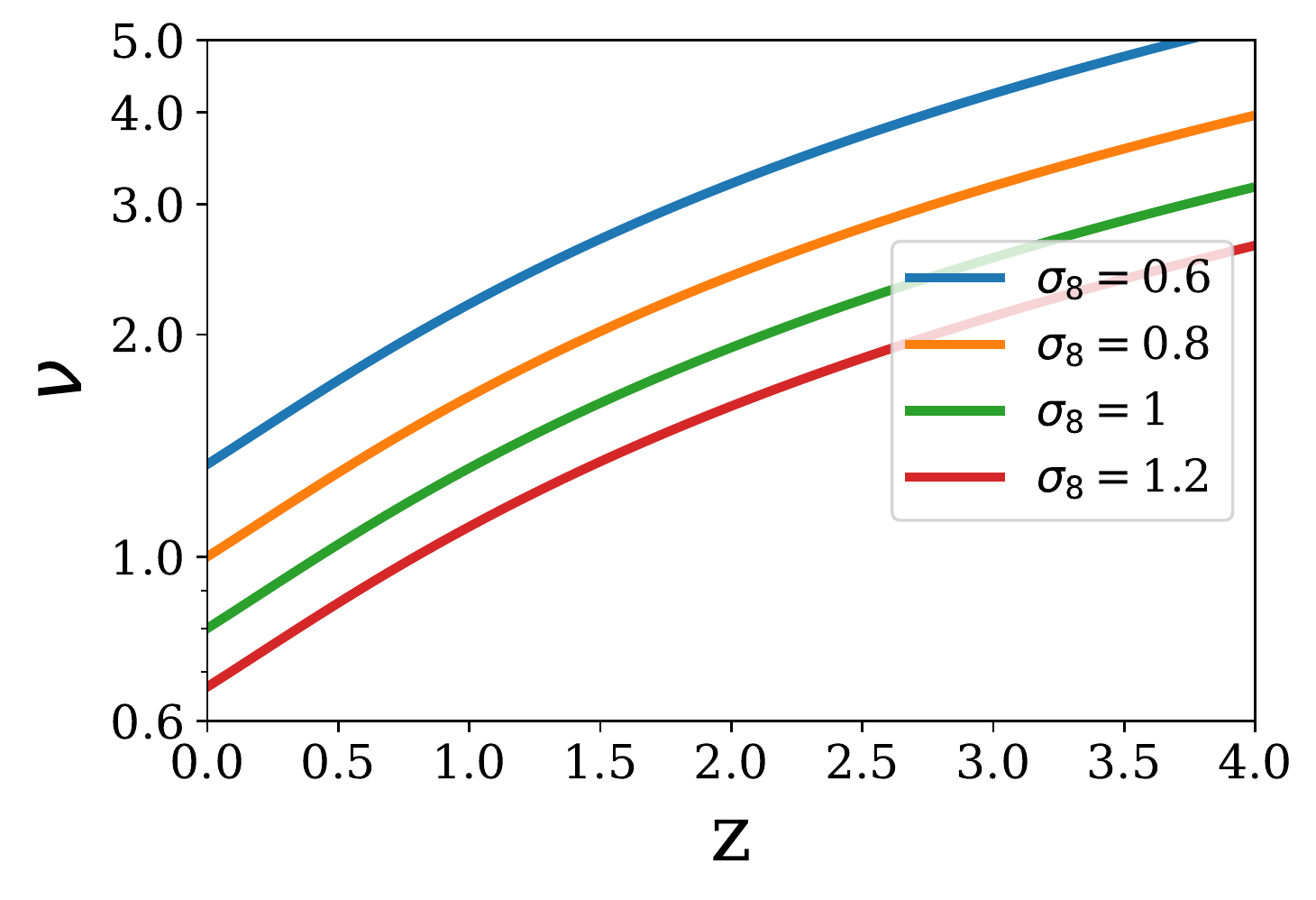}
    \includegraphics[width=0.35\textwidth]{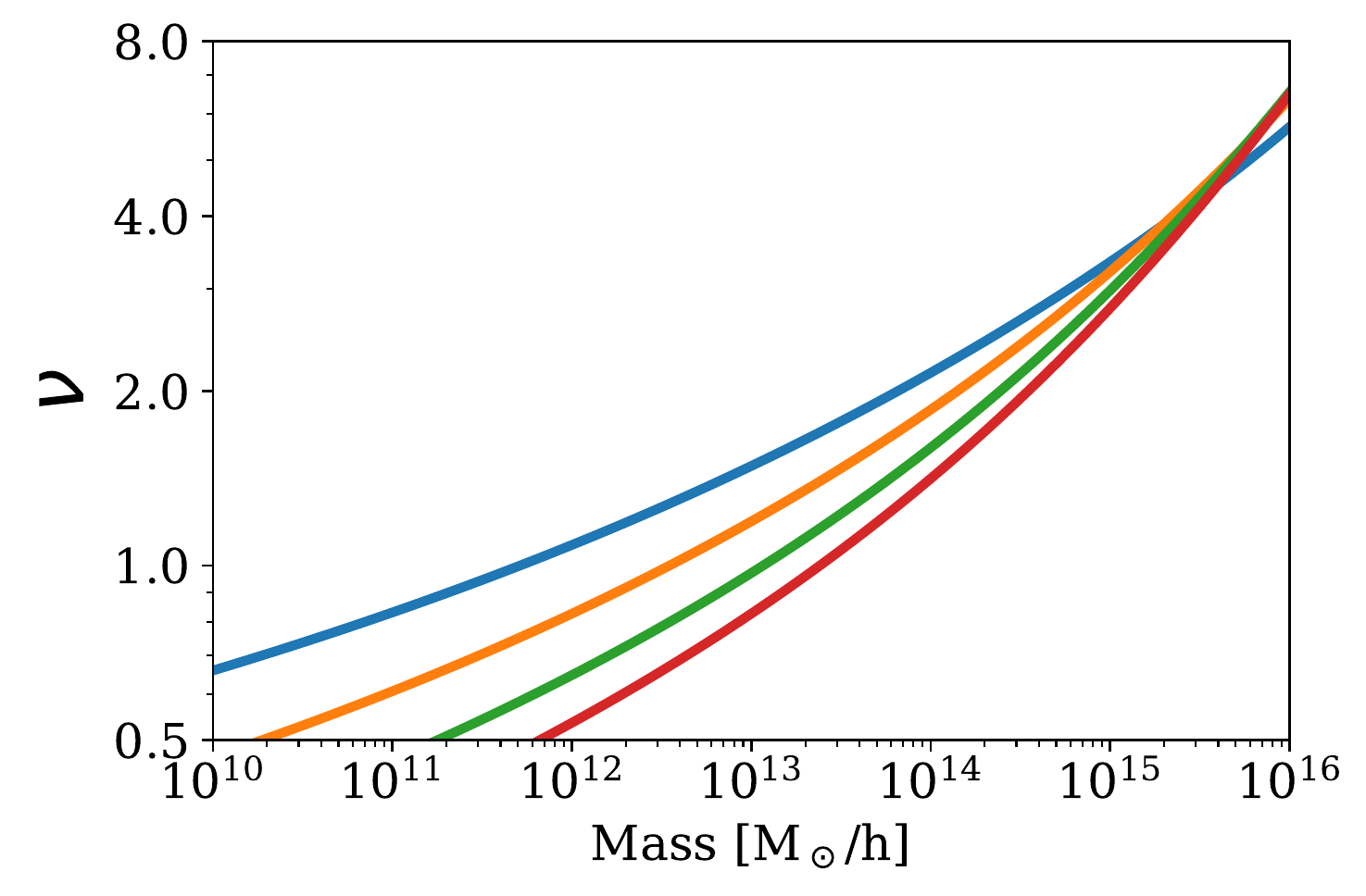}
    \includegraphics[width=0.35\textwidth]{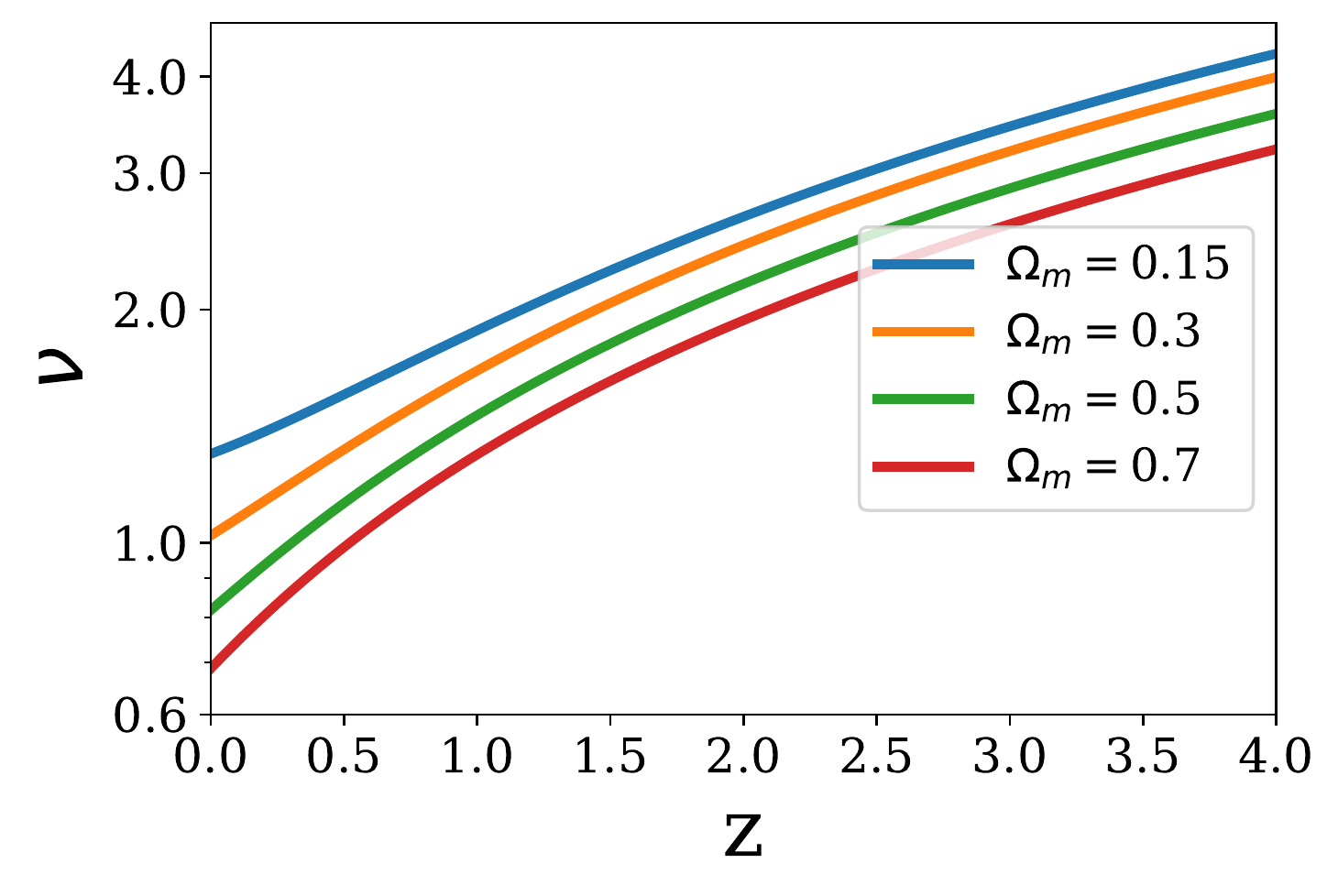}
    \caption{Variation of the peak height $\nu$ with mass (left panels) and redshift (right panels), for different values of the cosmological parameters. The two top panels show the dependence on $\sigma_8$ at fixed $\Omega_{\rm m} = 0.3$, while the bottom panels show the dependence on $\Omega_{\rm m}$ at fixed $\sigma_8 = 0.8$.}
    \label{fig:cosmo_peak_height}
\end{figure*}

Given these results, the overall dependence of the peak height on $\Omega_{\rm m}$ and $\sigma_8$ can be written
\begin{equation}
\label{eq:nu_cosmo}
    \nu(M, z) \propto \frac{1}{\sigma_8}\Omega_{\rm m}^{\alpha(z)}\Omega_{\rm m}^{-\beta (M)}\,,
\end{equation}
where $\alpha >0$ and increases with redshift to the limiting value $3/14$, while $\beta(M)$ is a decreasing function of mass, and is negative at large masses. 
Combining the positive slope of $\Gamma$ with the negative slope of $g(z)/g(0)$, the bottom right-hand panel of Fig.~\ref{fig:g_ofz_gamma} shows how $\nu$ generally decreases with $\Omega_{\rm m}$, but can increase for large masses and high redshifts. This explains the slight $\nu$-dependence seen in Fig.~\ref{fig:cosmo_peak_height_om_s8_plane}. The summary of how the peak height varies with mass and redshift for different values of $\Omega_m$ and $\sigma_8$ is shown in Fig. \ref{fig:cosmo_peak_height}.

%Fig C2

\subsubsection{Collapsed Fraction}

\begin{figure*}
    \centering
    \includegraphics[width=0.45\textwidth]{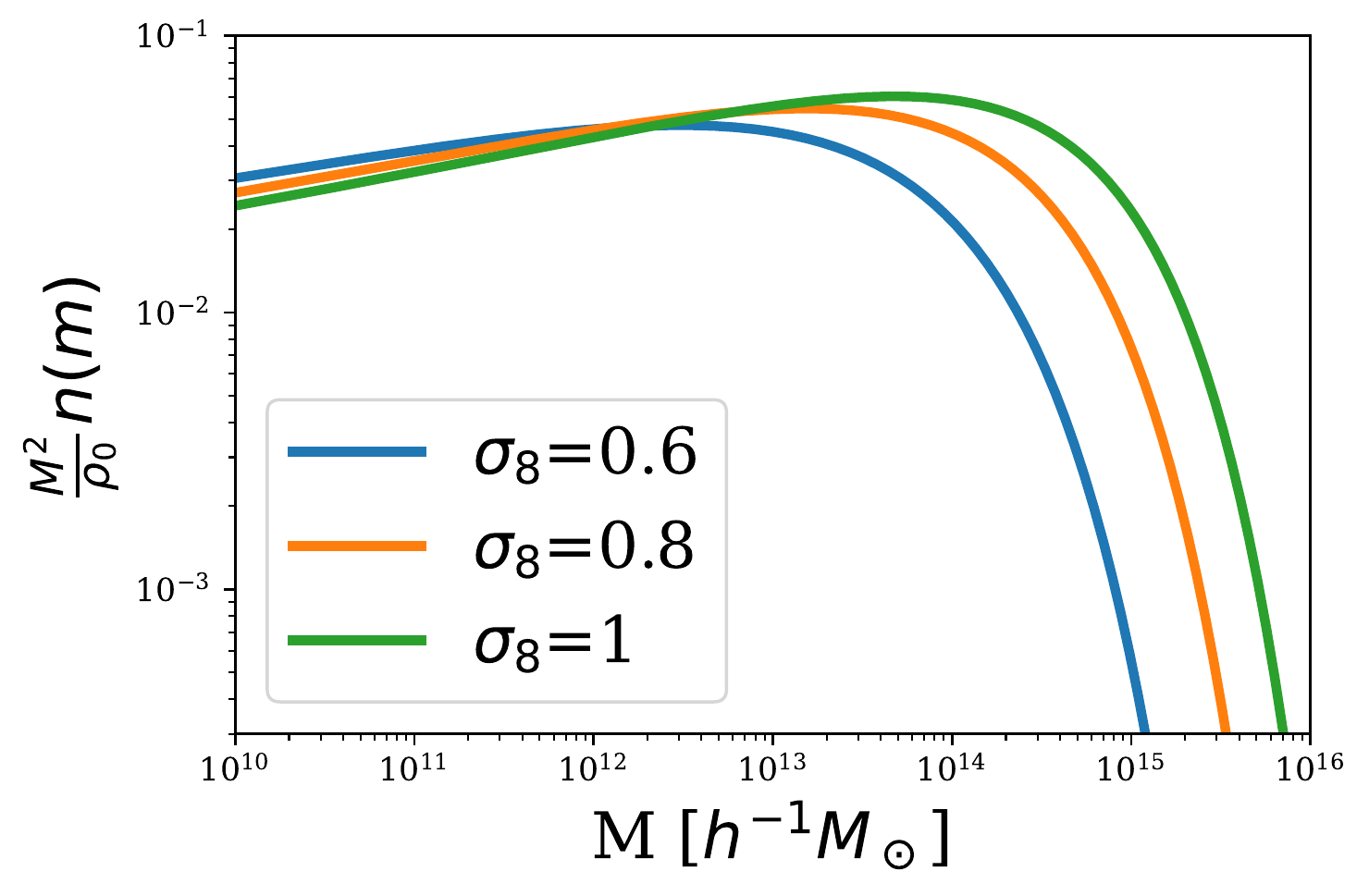}
    \includegraphics[width=0.45\textwidth]{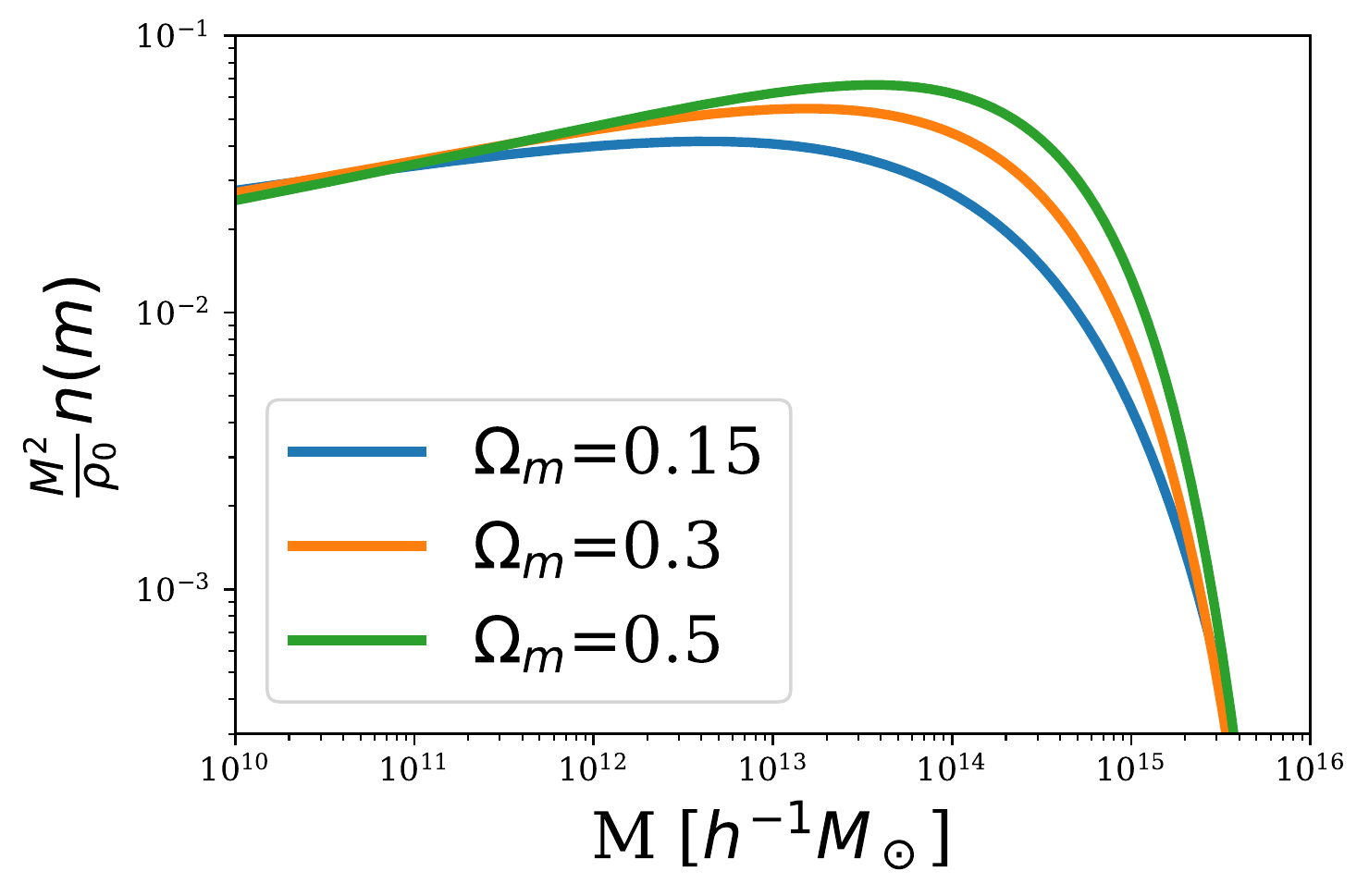}
    \caption{Variation of the collapsed fraction with $\sigma_8$ (left) and $\Omega_{\rm m}$ (right). The collapsed fraction increases slowly for low values of $\nu$, but then drops exponentially at high values; $\sigma_8$ changes the mass at which the transition to exponential behaviour occurs, while $\Omega_{\rm m}$ changes the sharpness of the transition.} 
    \label{fig:dimensionless_hmf}
\end{figure*}

Combining the peak height dependence described above with the functional form of the collapsed fraction as a function of peak height (Eqs. \ref{eq:fps} or \ref{eq:fst}) we obtain the collapsed fraction as a function of mass for different values of $\sigma_8$ and $\Omega_m$, shown in the left and right panels of Fig. \ref{fig:dimensionless_hmf} respectively. It shows two regimes, a power-law increase at low masses, followed by an exponential decrease for cluster mass haloes. The figure also shows why the collapsed fraction sensitivity to $\sigma_8$ and $\Omega_m$ varies with mass, and the transition between the two regimes occurs at different masses for different values of $\sigma_8$.

% Fig C3

% Appendix D
\section{Parametric Dependence of the Formation Time}
\label{sec:Appendix_D}
 
To understand the dependence of formation time on the cosmological parameters $\Omega_{\rm m}$ and $\sigma_8$, we consider the simplest, SC or Press-Schechter (PS) expression for $z_{50}$:
\begin{equation}
    P_{PS}(z_{50}>z|M_0, z_0) \equiv \int_{M_0/2}^{M_0}\frac{M_0}{M}f_{PS}(M, z |M_0, z_0)dM\,,
\end{equation}
where
\begin{equation}
\begin{gathered}
    f_{PS}(M_1, z_1 |M_0, z_0)dM_1 = \frac{1}{\sqrt{2\pi}}\frac{\delta_c(z_1) - \delta_c(z_0)}{(S(M_1)-S(M_0))^{3/2}} \\ \times \exp{\left(-\frac{\left(\delta_c(z_1) - \delta_c(z_0)\right)^2}{2(S(M_1)-S(M_0))}\right)}dS_1\,.
\end{gathered}
\end{equation}

We can express $f_{PS}$ as a function of a single variable 
$$D\nu \equiv \frac{\delta_c(z_1) - \delta_c(z_0)}{\sqrt{S(M_1)-S(M_0)}}\,,$$ as follows:
\begin{equation}
     f_{PS}(M_1, z_1 |M_0, z_0)dM_1 = \sqrt{\frac{2}{\pi}}\exp{\left(-\frac{(D\nu)^2}{2}\right)}|d(D\nu)|\,.   
\end{equation}
Thus, we see that the conditional probability has the same form as the unconditional one, but with the argument $D\nu$ rather than $\nu$. 

The formation redshift distribution is proportional to the average value of the PMF between $M_0/2$ and $M_0$, and since the factor $M_0/M$ does not vary much over this range, it is also approximately equal to the conditional probability evaluated around the middle of the range. This probability $f_{PS}$ in turn goes as $\exp[-(D\nu)^2/2]$, so we expect the parametric dependence of $\langle z_{50}\rangle $ to resemble an inverted, logarithmic version of the dependence for $D\nu$. 

\begin{figure*}
    \centering
    \includegraphics[width=0.4\textwidth]{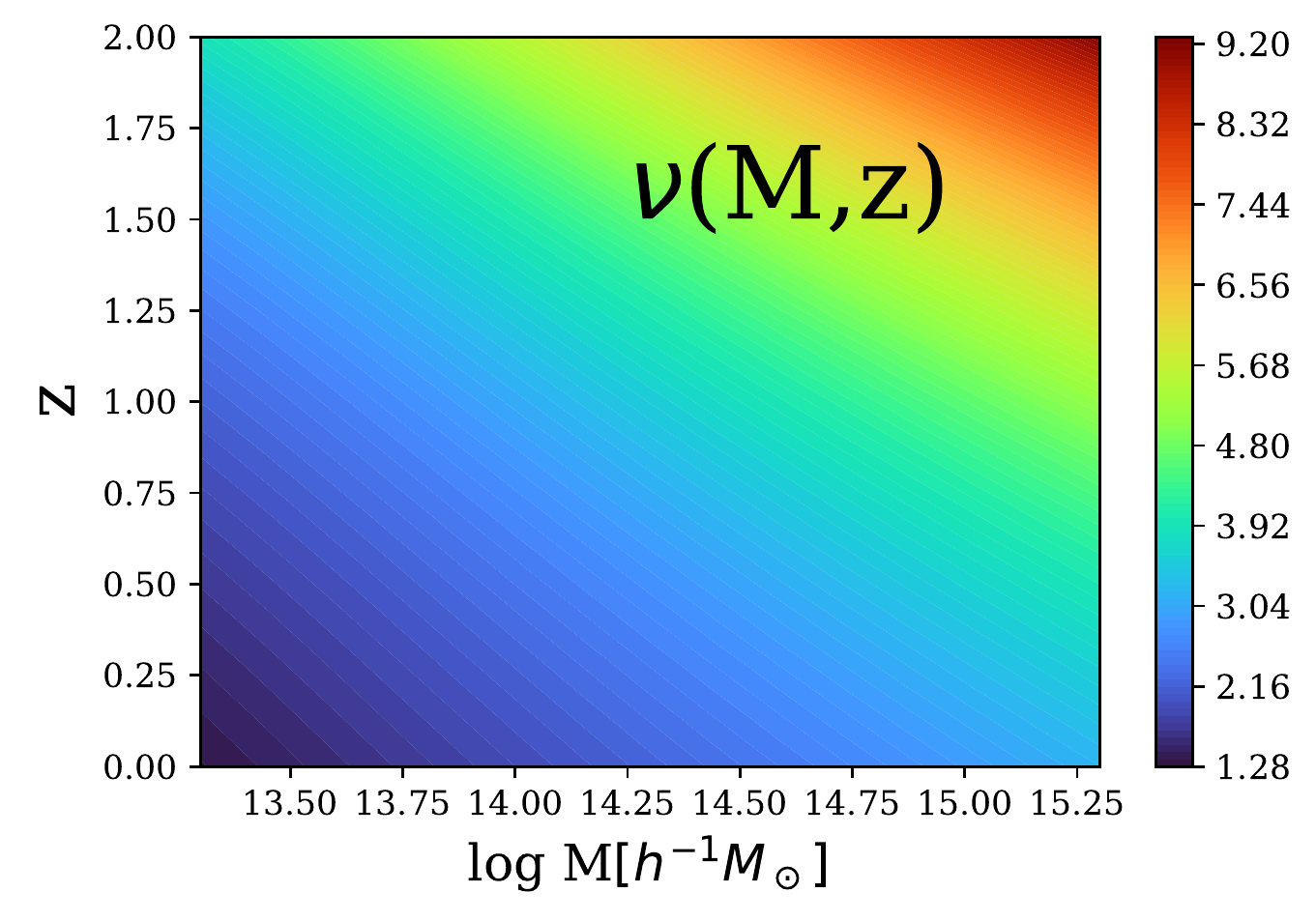}
    \caption{Peak height $\nu$ as a function of mass and redshift, in our fiducial ($\Omega_{\rm m} = 0.3, \Omega_\Lambda = 0.7$) cosmology.}
    \label{fig:nu_m_z}
\end{figure*}

\begin{figure*}
    \centering
    \includegraphics[width=0.67\textwidth]{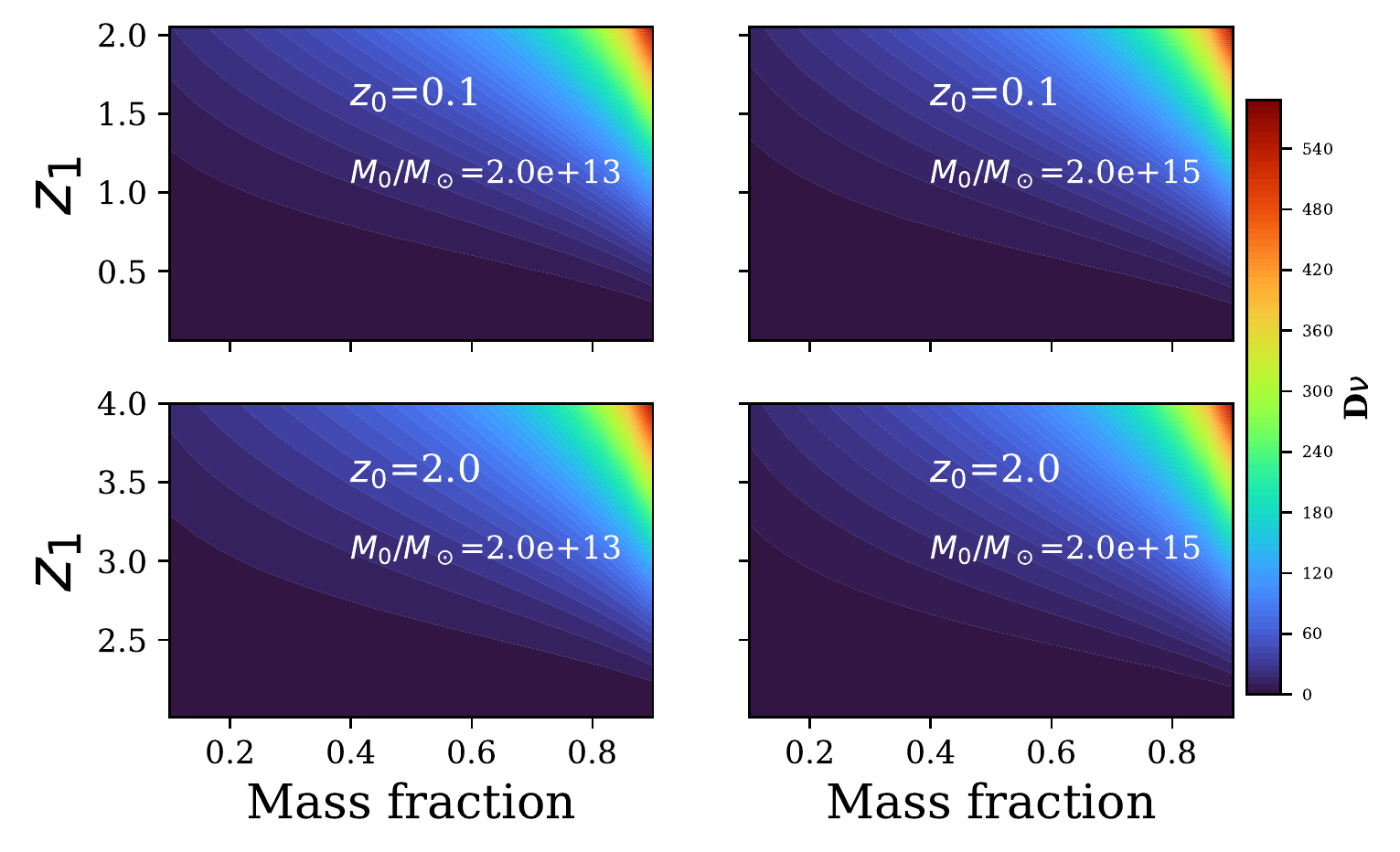}\\
    \includegraphics[width=0.7\textwidth]{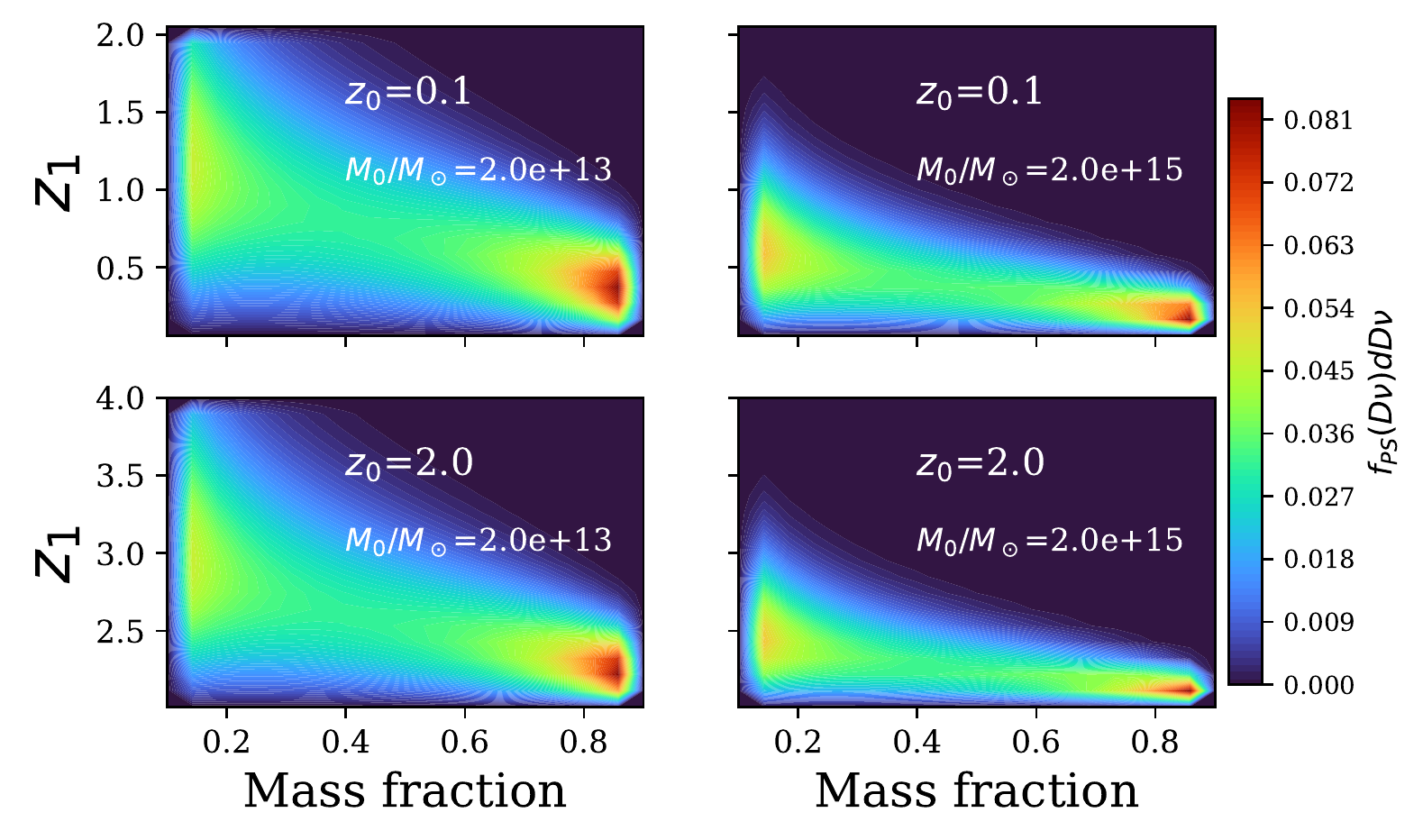}\\
    \caption{Top 4 panels: The variable $D\nu$ as a function of mass fraction $M_1/M_0$ and $z_1$, for the values of $(M_0,z_0)$ indicated, in our fiducial cosmology. Bottom 4 panels: corresponding conditional probability $f_{PS}(D\nu)d(D\nu)$.} 
    \label{fig:dnus_pmf}
\end{figure*}

Fig.~\ref{fig:nu_m_z} shows $\nu$ as a function of mass and redshift, while the top 4 panels of Fig.~\ref{fig:dnus_pmf} show the value of $D\nu$ as a function of $z_1$ and mass fraction $M_1/M_0$, for various values of $M_0$ and $z_0$. We see that the shape of the $D\nu$ contours is generally similar to those for $\nu$, except when $z_1$ is close to $z_0$ (bottom of the plot), or when the mass fraction is close to 1 (right hand side of the plot).  

%Fig D1

The second set of 4 panels in Fig.~\ref{fig:dnus_pmf} shows the value of the conditional probability. As expected, the conditional probability is similar to an inverse, logarithmic mapping of $D\nu$. 

%Fig D2

\begin{figure*}
    \centering
    \includegraphics[width=0.7\textwidth]{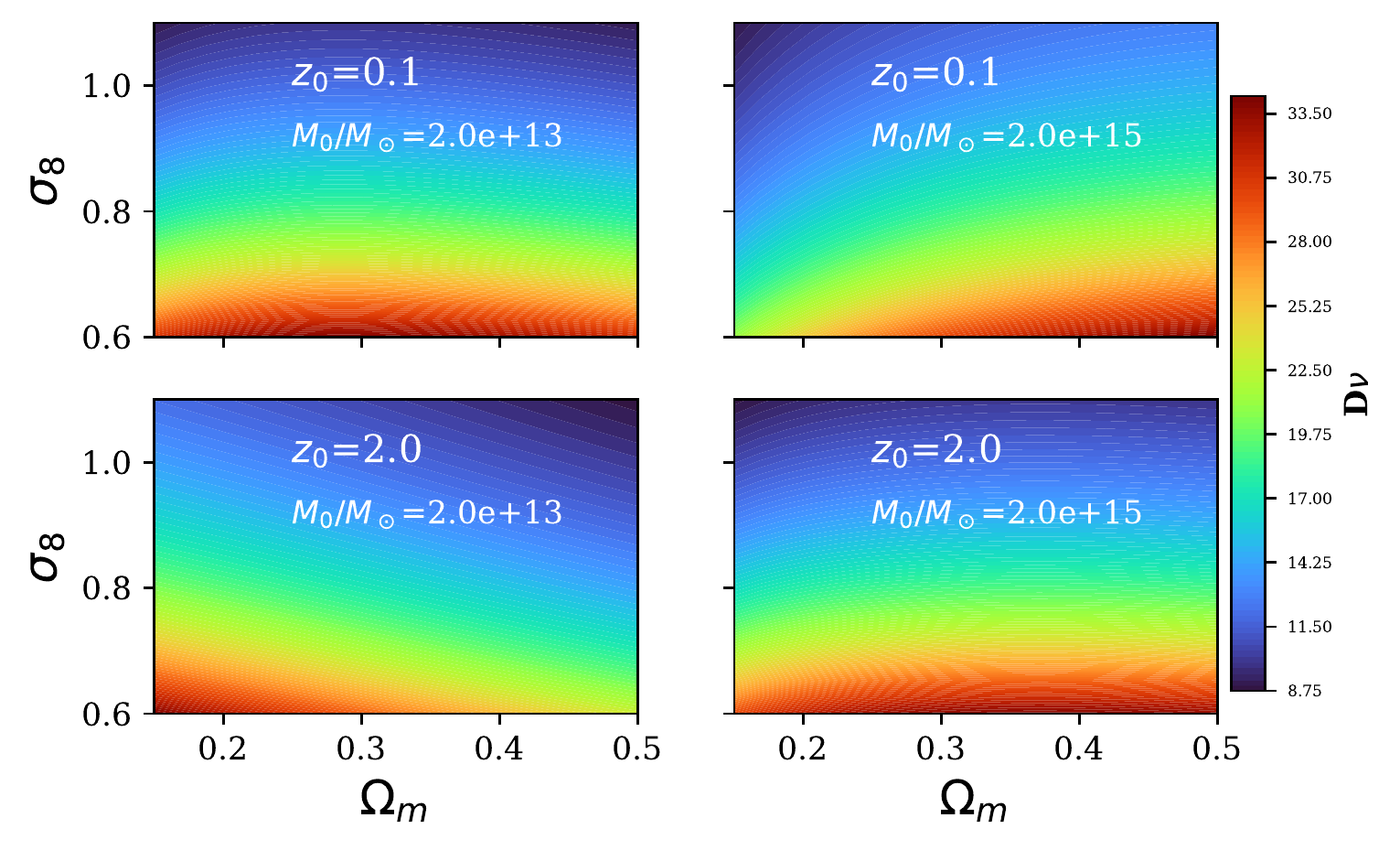}\\
    \includegraphics[width=0.7\textwidth]{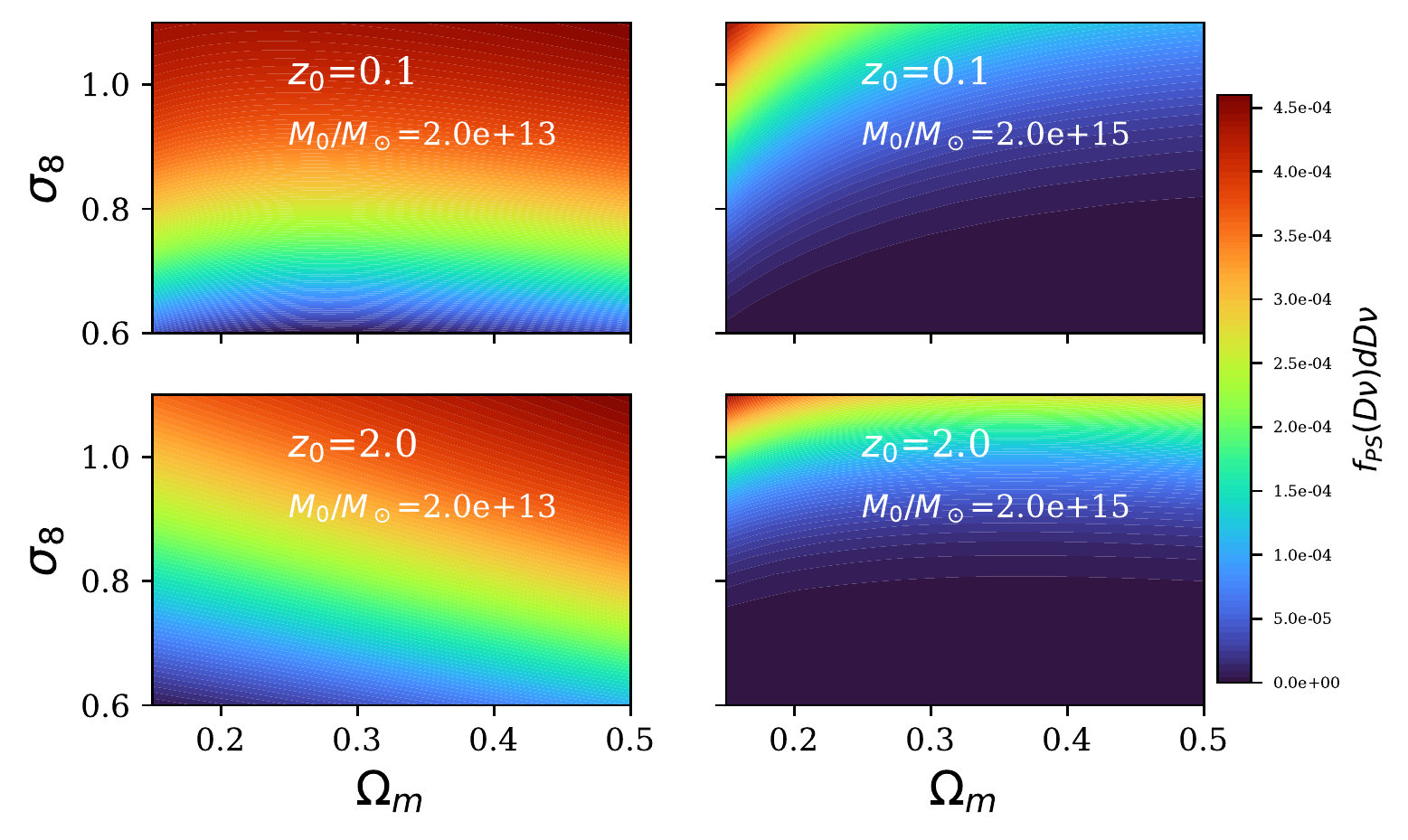}\\
    \caption{Top : $D\nu$ as a function of $\Omega_{\rm m}$ and $\sigma_8$, for the values of $(M_0,z_0)$ indicated, a mass fraction $M_1/M_0 = 0.5$, and $\Delta z = z_1-z_0 = 0.1$. Bottom: The corresponding conditional probability $f_{PS}(D\nu)d(D\nu)$ values as a function of $\Omega_{\rm m}$ and $\sigma_8$.} 
    \label{fig:dnus_pmf_om_s8}
\end{figure*}

Finally, we can consider the behaviour of $D\nu$ and the PMF in the $\Omega_{\rm m}$--$\sigma_8$ plane. The top 4 panels of Fig.~\ref{fig:dnus_pmf_om_s8} show the value of $D\nu$ in this plane, for the values of $(M_0,z_0)$ indicated, a mass fraction $M_1/M_0 = 0.5$, and $\Delta z = z_1-z_0 = 0.1$. The overall pattern is very similar to that of $\nu$ (cf.~Fig.~\ref{fig:cosmo_peak_height_om_s8_plane}).
The bottom four panels show the value of the PMF, for the same choices of  $(z_0, z_1, M_0, M_1)$. Relative to the top panels, we see that the colour scale is inverted and logarithmic, as expected. The overall behaviour explains the shape of the contours in Fig.~\ref{fig:anti-banana}, and their relative orthogonality to abundance contours in the same plane.

%Fig D3

\bsp	% typesetting comment
\label{lastpage}

\end{document}